\documentclass[a4paper,11pt]{article}

\pdfoutput=1

\usepackage{amsmath}
\usepackage{amssymb}
\usepackage{amsbsy}
\usepackage{fixmath}
\usepackage{graphicx}
\usepackage{hyperref}
\usepackage[a4paper,left=0.99in, right=0.99in,top=1.2in, bottom=1.2in]{geometry}
\usepackage{color}
\usepackage{cite}
\numberwithin{equation}{section}

\usepackage{array}
\usepackage{multirow}
\usepackage{setspace}
\usepackage{xspace}

\usepackage{epsfig}
\usepackage{graphicx,color}

\newcommand{\be}{\begin{equation}}
\newcommand{\ee}{\end{equation}}
\newcommand{\GG}{{\cal F}}

\newcommand{\calM}{{\cal M}}
\newcommand{\calA}{{\cal A}}
\newcommand{\GeV}{\ensuremath{\,\mathrm{GeV}}\xspace}

\newcommand{\slashp}{p\hspace{-1.0ex}/}
\newcommand{\eg}{{\it e.g. }}
\newcommand{\ie}{{\it i.e. }}
\newcommand{\cf}{{\it c.f. }}
\newcommand{\ANG}[1]{\langle#1\rangle}
\newcommand{\SQR}[1]{[#1]}
\newcommand{\AL}[1]{\langle#1|}

\newcommand{\AR}[1]{|#1\rangle}
\newcommand{\ts}[1]{\tilde{s}_{#1}}
\newcommand{\that}{{\hat t}}
\newcommand{\shat}{{\hat s}}
\newcommand{\uhat}{{\hat u}}
\newcommand{\tlt}{\bar{t}}
\newcommand{\tls}{\bar{s}}
\newcommand{\tlu}{\bar{u}}
\newcommand{\qbar}{{\bar q}}
\newcommand{\as}{\alpha_s}
\newcommand{\order}[1]{\mathcal{O}\!\left(#1\right)}
\definecolor{red}{rgb}{1,0,0}

\usepackage{hyperref}

\def\be{\begin{equation}}
\def\ee{\end{equation}}
\def\bea{\begin{eqnarray}}
\def\eea{\end{eqnarray}}

\title{ 
    \vskip 2.5cm
    Improved TMD factorization for forward dijet production\\ 
    in dilute-dense hadronic collisions
    \vskip 1cm
}

\author{
P. Kotko$^1$, K. Kutak$^2$, C. Marquet$^3$, E. Petreska$^{3,4}$, S. Sapeta$^5$ and A. van Hameren$^2$ \\\\
$^1$ {\small\it Department of Physics, Penn State University}\\ {\small\it University Park, 16803 PA, USA}\\\\
$^2$ {\small\it The H.\ Niewodnicza\'nski Institute of Nuclear Physics PAN}\\ {\small\it Radzikowskiego 152, 31-342 Krak\'ow, Poland}\\\\
$^3$ {\small\it Centre de Physique Th\'eorique, \'Ecole Polytechnique,}\\
{\small\it CNRS, 91128 Palaiseau, France}\\\\
$^4$ {\small\it Departamento de F\'isica de Part\'iculas and IGFAE,}\\
{\small\it Universidade de Santiago de Compostela, 15782 Santiago de Compostela, Spain}\\\\
$^5$ {\small\it CERN PH-TH, CH-1211, Geneva 23, Switzerland}\\
}

\date{}

\begin{document}
\maketitle

\thispagestyle{empty}

%--------------- preprint numbers ---------------------
\vspace{-44em}
\begin{flushright}
  CERN-PH-TH-2015-045 \\
  CPHT-RR005.0315 \\
  IFJPAN-IV-2015-2
\end{flushright}
\vspace{29em}
%-----------------------------------------------------------------

\vspace{10em}
\begin{abstract}
  
We study forward dijet production in dilute-dense hadronic collisions.
By considering the appropriate limits, we show that both the
transverse-momentum-dependent (TMD) and the high-energy factorization
formulas can be derived from the Color Glass Condensate framework.
Respectively, this happens when the transverse momentum imbalance
of the dijet system, $k_t$, is of the order of either the saturation scale, or
the hard jet momenta, the former being always much smaller than the latter.
We propose a new formula for forward dijets that encompasses
both situations and is therefore applicable regardless of the magnitude of
$k_t$.  That involves generalizing the TMD factorization formula for dijet
production to the case where the incoming small-$x$ gluon is off-shell.  The
derivation is performed in two independent ways, using either Feynman diagram
techniques, or color-ordered amplitudes.
  
\end{abstract}

\newpage
%\tableofcontents

%-----------------------------------------------------------------------------
\section{Introduction}

Forward particle production observables in proton-proton (p+p) and
proton-nucleus (p+A) collisions at the Large Hadron Collider (LHC) offer unique
opportunities to study the dynamics of QCD at small $x$, and in particular the
non-linear regime of parton saturation \cite{Gribov:1984tu}. Indeed, in high-energy hadronic
collisions, forward particle production is sensitive only to high-momentum
partons inside one of the colliding hadrons, which therefore appears dilute. By contrast, for the other hadron or nucleus, it is mainly small-momentum partons, whose density is large, that contribute to the scattering. Such processes, in which a large-$x$ projectile is used as a probe to investigate a small-$x$ target, are sometimes called dilute-dense collisions. Since the high-$x$ part of the projectile wave function is well understood in perturbative QCD, forward particle production is indeed ideal to investigate the small-$x$ part of target wave function. This is true both in p+p and p+A collisions, although using a target nucleus does enhance the dilute-dense asymmetry of such collisions.

The separation between the linear and non-linear regimes of the target wave
function is characterized by a momentum scale $Q_s(x)$, called the saturation
scale, which increases as $x$ decreases. Dilute-dense collisions can be
described from first principles, provided $Q_s\gg\Lambda_{\mathrm{QCD}}$. This
condition is better realized with higher energies (as they open up the phase
space towards lower values of $x$), and with nuclear targets (since, roughly, $Q_s\!\sim\!A^{1/3}$). Over the years, the Color Glass Condensate (CGC) effective theory \cite{Gelis:2010nm} has emerged as the best candidate to approximate QCD in the saturation regime, both in terms of practical applicability and of phenomenological success \cite{Albacete:2014fwa}. In this paper, we focus on forward dijet production in p+A  and 
 p+p collisions. We note that the CGC approach has been very successful in describing forward di-hadron production at RHIC \cite{Albacete:2010pg,Stasto:2011ru,Lappi:2012nh}, in particular it predicted the suppression of azimuthal correlations in d+Au collisions compared to p+p collisions \cite{Marquet:2007vb}, which was observed later experimentally~\cite{Adare:2011sc,Braidot:2010zh}.

With forward dijets at the LHC however, the full complexity of the CGC machinery
is not needed. Indeed, for the di-hadron process at RHIC energies, no particular
ordering of the momentum scales involved is assumed in CGC calculations, while,
at the LHC, the presence of particles with transverse momenta much larger than
the saturation scale clearly must imply some simplifications. On the flip side,
there will be other complications since further QCD dynamics, which is not part
of the CGC framework but which is relevant at large
transverse momenta, must also be considered. There are three important momentum
scales in the forward dijet process: a typical transverse momentum of a hard
jet, $P_t$, whose precise definition will be stated in the next section;
the transverse momentum of the small-$x$ gluons involved in the hard scattering,
$k_t$; and the saturation scale of the small-$x$ target, $Q_s$. Clearly, $P_t$
is always one of the hardest scales, and it is much bigger than $Q_s$, which is
always one of the softest scales. Then, depending on where $k_t$ sits with
respect to these two, three different regimes can be defined.

A first regime, with $Q_s\ll k_t\sim P_t$, corresponds to the domain of
applicability of the so-called \emph{high energy factorization} (HEF) framework
\cite{Catani:1990eg,Deak:2009xt}, in which the description of forward dijets
involves an unintegrated gluon distribution for the small-$x$ target, along with
off-shell hard matrix elements. That is explicitly shown in this work, starting form CGC calculations.
While such a factorization does not occur when non-linear saturation effects are accounted for, we
shall see that taking the $Q_s\ll k_t\sim P_t$ limit is tantamount to restricting the interaction with the
small-x target to a two-gluon exchange, therefore allows to indeed write all the CGC correlators
in terms of a single gluon distribution. Doing so, the matrix elements of the HEF framework are
exactly recovered.

A second regime, with $k_t\sim Q_s\ll P_t$, is where the so-called
\emph{transverse momentum dependent}~(TMD) factorization \cite{Bomhof:2006dp} is valid.
It involves on-shell matrix elements but several unintegrated gluons distributions.
In this regime, non-linear effects are present, and in the large-$N_c$ limit, equivalence with
CGC expressions was shown in \cite{Dominguez:2011wm}. In particular, in that case the
description of forward dijets involves only two independent unintegrated gluons distributions, each of which can
be determined in various other processes \cite{Dominguez:2010xd}. In the present work we shall keep $N_c$
finite, implying, as we show below, that a total of six independent unintegrated gluons distributions are needed.

Finally, the intermediate regime $Q_s\ll k_t\ll P_t$, which is naturally obtained from the two others by taking the
appropriate limits, corresponds to the collinear regime, with on-shell matrix elements and the standard integrated
gluon distribution.

Separately, the HEF and TMD approaches to dijet production have been extensively studied in the literature
\cite{Deak:2009xt,Kutak:2012rf,vanHameren:2014lna,vanHameren:2014ala,vanHameren:2013fla} and
\cite{Bomhof:2006dp,Boer:1999si,Belitsky:2002sm,Boer:2003cm,Collins:2007nk,Vogelsang:2007jk,Rogers:2010dm,Xiao:2010sp},
but little connection has been made between them so far. The first result of
this paper is to reveal that connection, in the context of dilute-dense
collisions, and to show that, in fact, they are both contained in the CGC
description. However, as already mentioned, using the CGC approach is
unnecessarily complicated and one should take advantage of the fact that $P_t\gg Q_s$ to simplify the theoretical formulation. The second result of the paper is precisely to develop a new formula for forward dijets in dilute-dense collisions that encompasses all three situations described above, meaning that it is applicable regardless of the magnitude of $k_t$. As explained below, this is obtained by extending the TMD factorization framework, more precisely by supplementing it with off-shell matrix elements.

Note that the derivation of our new unified formula is performed in two independent ways: first using the standard Feynman diagram technique, and second by exploiting the so-called helicity method that employs color-ordered amplitudes \cite{Mangano:1990by}. With this second method, the gauge invariance of the results is explicit, and the method will also prove very useful in the future, when processes with more particles in the final state are considered. As is the case in the CGC framework, our new formulation contains all the relevant limits, but it has the advantage that it is more amenable to phenomenological implementations than CGC calculations. In addition, it  is 
also better suited to be supplemented with further QCD dynamics relevant at high
$P_t$, such as Sudakov logarithms \cite{Mueller:2012uf,Mueller:2013wwa} or
coherence in the QCD evolution of the gluon density \cite{Ciafaloni:1987ur,Catani:1989sg,Catani:1989yc}. These tasks are left for future work.

The plan of the paper is as follows. In Section 2, we introduce kinematics and notations, and briefly present the HEF and TMD frameworks. In Section 3, we show that the HEF framework can be derived from CGC calculations, when the $Q_s\ll k_t\sim P_t$ limit is considered; namely we explain how the various CGC correlators reduce to a single gluon distribution in that limit, and show that the off-shell matrix elements of the HEF framework are indeed emerging. Section~4 is devoted to the $k_t\sim Q_s\ll P_t$ limit, the derivation of the TMD factorization formula for forward dijets given in \cite{Dominguez:2010xd} is recalled, and extended to the case of finite $N_c$, implying six independent unintegrated gluons distributions instead of two. The hard factors of the TMD framework are computed again in Section 5, but keeping the small-$x$ gluon off-shell, which leads us to our new unified formula for forward dijets in p+A collisions. In Section 6, both the TMD factorization formula and the off-shell hard factors are derived again, but using color-ordered amplitudes, instead of Feynman diagram techniques. Finally, Section 7 is devoted to conclusions and outlook.

%-----------------------------------------------------------------------------
\section{Forward dijets in p+A collisions}

\begin{figure}
  \begin{center}
    \includegraphics[width=0.45\textwidth]{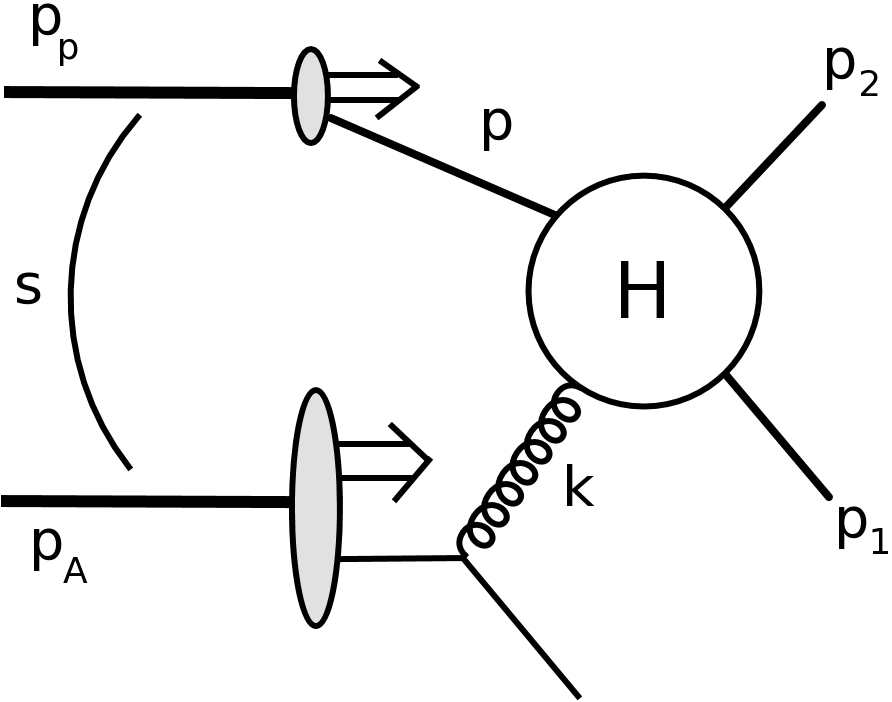}
  \end{center}
  \caption{Inclusive dijet production in p+A collision. The blob $H$ represents
  hard scattering. The solid lines 
  coming out of $H$ represent partons, which can be either quarks or gluons.}
  \label{fig:dijets-pA}
\end{figure}

We shall discuss the process of inclusive dijet production in  the forward
region, in collisions of dilute and dense systems
\begin{equation}
  p (p_p) + A (p_A) \to j_1 (p_1) + j_2 (p_2)+ X\,.
\end{equation}
The process is shown schematically in Fig~\ref{fig:dijets-pA}. The four-momenta
of the projectile and the target are massless and purely longitudinal.
In terms of the light cone variables, $v^\pm = (v^0\pm v^3)/\sqrt{2}$, they take
the simple form
\begin{equation}
  p_p = \sqrt{\frac{s}{2}}(1,0_t,0)\,, \qquad \qquad
  p_A = \sqrt{\frac{s}{2}}(0,0_t,1)\,,
  \label{eq:pp-pA-defs}
\end{equation}
where $s$ is the squared center of mass energy of the p+A system.

The energy (or longitudinal momenta) fractions of the incoming parton (either
a quark or gluon) from the
projectile, $x_1$, and the gluon from the target, $x_2$, can be expressed in
terms of the rapidities and transverse momenta of the produced jets as
\begin{subequations}
\begin{align}
x_1 & = \frac{p_1^+ + p_2^+}{p_p^+}   = \frac{1}{\sqrt{s}} \left(|p_{1t}|
e^{y_1}+|p_{2t}| e^{y_2}\right)\,, \\
x_2 & = \frac{p_1^- + p_2^-}{p_A^-}   = \frac{1}{\sqrt{s}} \left(|p_{1t}| e^{-y_1}+|p_{2t}| e^{-y_2}\right)\,,
\end{align}
\end{subequations}
where $p_{1t}$, $p_{2t}$ are transverse Euclidean two-vectors.
By looking at jets produced in the forward direction, we effectively select
those fractions to be $x_1 \sim 1$ and $x_2 \ll 1$.
Since the target A is probed at low $x_2$, the dominant contributions come from
the subprocesses in which the incoming parton on the target side is a gluon
\begin{equation}
  qg  \to  qg\,,
  \qquad \qquad 
  gg  \to  q\bar q\,,
  \qquad \qquad 
  gg  \to  gg\,.
\end{equation}
In dilute-dense collisions, the large-$x$ partons of the dilute projectile are described in terms of the usual parton distribution functions of collinear factorization $f_{a/p}$, with a scale dependence given by DGLAP evolution equations. By contrast, the small-$x$ gluons of the dense target nucleus are described by a transverse-momentum-dependent distribution, which evolve towards small~$x$ according to non-linear equations.
Moreover, the momentum $k$ of the incoming gluon from the target, besides the longitudinal component $k^-=x_2\sqrt{s/2}$, has in general a non-zero transverse component, $k_T$, which leads to imbalance of transverse momentum of the produced jets
\begin{equation}
  |k_{t}|^2 = |p_{1t}+p_{2t}|^2 = 
  |p_{1t}|^2 + |p_{2t}|^2 + 2|p_{1t}||p_{2t}| \cos\Delta\phi\,,
  \label{eq:ktglue}
\end{equation}
with $k_T^2=-|k_t|^2$. Here, by $k_T$ we mean a
four-vector, as opposed to $k_t = p_{1t} + p_{2t}$, which is a two-dimensional vector
in the transverse plane. They are simply related by: $k_T =
(0,k_t,0)$.
Using the notation defined above, the gluon's four-momentum can be also parametrized as
\begin{equation}
  k=x_2 p_A + k_T\,.
  \label{eq:k-4-vec}
\end{equation}
The Mandelstam variables at the partonic level are defined as

\begin{subequations}
  \label{eq:mandelstam}
  \begin{align}
    \shat & = (p+k)^2 = (p_1 + p_2)^2=\frac{|P_t|^2}{z(1-z)}\,, \\
    \that & = (p_2-p)^2 = (p_1 - k)^2=-\frac{|p_{2t}|^2}{1-z}\,, \\
    \uhat & = (p_1-p)^2 = (p_2 - k)^2=-\frac{|p_{1t}|^2}{z}\,,
  \end{align}
\end{subequations}
with
\begin{equation}
z=\frac{p_1^+}{p_1^+ + p_2^+} \quad\quad \text{and}
\quad\quad P_t=(1-z)p_{1t}-zp_{2t}\ .
\label{eq:zdef}
\end{equation} They sum up to $\shat + \that + \uhat = k_T^2$.

Note that we always neglect the transverse momentum of the high-$x$ partons compared with that of the low-$x$
parton $|k_t|$. This is justified in view of the asymmetry of the problem, $x_1 \sim 1$ and $x_2 \ll 1$, which implies that gluons form
the target have a much bigger average transverse momentum (of the order of $Q_s$) compared to that of the large $x$ partons from the projectile (which of the order of $\Lambda_{QCD}$). And even when the transverse momentum imbalance of the dijet system is of the same order as the jet transverse momenta themselves, implying that both parton distributions are probed in their radiative tail, the small $x_2$ (BFKL) evolution implies a $1/k_t$ behavior on the target side, while DGLAP evolution implies a $1/k_t^2$ behavior on the projectile side.

To take into account small-$x$ effects in dijet production, an approach that has
been broadly used in phenomenological studies involves the so-called high
energy factorization (HEF) formula~\cite{Kutak:2012rf}
\begin{equation}
  \frac{d\sigma^{pA\rightarrow {\rm dijets}+X}}{dy_1dy_2d^2p_{1t}d^2p_{2t}} 
  =
  \frac{1}{16\pi^3 (x_1x_2 s)^2}
  \sum_{a,c,d} 
  x_1 f_{a/p}(x_1,\mu^2)\,
  |\overline{{\cal M}_{ag^*\to cd}}|^2
  {\cal F}_{g/A}(x_2,k_t)\frac{1}{1+\delta_{cd}}\,.
  \label{eq:hef-formula}
\end{equation}
This formula makes use of the unintegrated gluon distribution ${\cal F}_{g/A}$ that is involved in the
calculation of the deep inelastic structure functions. It is determined from fits to DIS data, and then used in
Eq.~(\ref{eq:hef-formula}), along with matrix elements that depend on the transverse momentum imbalance (\ref{eq:ktglue}).
Even though the high energy factorization is not strictly valid for dijet production, there exists a kinematic window, the dilute limit
$Q_s\ll|p_{1t}|,|p_{2t}|,|k_t|$, in which it can be motivated from the CGC approach. We shall demonstrate this explicitly
for all channels in the next section.

A second approach, valid in the regime where the transverse momentum imbalance between the
outgoing particles, Eq.~(\ref{eq:ktglue}), is much smaller than their individual
transverse momenta, is the so-called transverse momentum dependent (TMD) factorization.
This limit,
$|p_{1t}+p_{2t}|\ll|p_{1t}|,|p_{2t}|$, or $|k_t|\ll|P_t|$,
corresponds to the situation of nearly back-to-back dijets. Even though, in
general, there exists no TMD factorization theorem for
jet production in hadron-hadron collisions, such a factorization can be
established in the asymmetric ``dilute-dense'' situation considered here,
where only one of the colliding hadrons is described by a transverse momentum dependent gluon distribution. Again, selecting dijet systems produced in the forward direction implies $x_1 \sim 1$ and $x_2 \ll 1$, which in turn allows us to make that assumption.
The TMD factorization formula reads (so far, this has been obtained in the large-$N_c$ approximation, but this restriction will be lifted in the present work)~\cite{Dominguez:2011wm}

\begin{equation}
\frac{d\sigma^{pA\rightarrow {\rm dijets}+X}}{dy_1dy_2d^2p_{1t}d^2p_{2t}} =
\frac{\alpha _{s}^{2}}{(x_1x_2s)^{2}} \sum_{a,c,d} x_1 f_{a/p}(x_1,\mu^2) \sum_i H_{ag\to cd}^{(i)} \mathcal{F}_{ag}^{(i)}(x_2,k_t) 
\frac{1}{1+\delta_{cd}}\,,
\label{eq:tmd-main}
\end{equation}
where several unintegrated gluon distributions $\mathcal{F}_{ag}^{(i)}$ with
different operator definition are involved and accompanied by different hard
factors $H_{ag\to cd}^{(i)}$. Those hard factors were calculated
in~\cite{Dominguez:2011wm} as if the
small-$x_2$ gluon was on-shell (\ie $|k_t|=0$). The $k_t$ dependence survived only in the gluon distributions. 

By restoring the $k_t$ dependence of the hard factors inside formula
(\ref{eq:tmd-main}), we can make the bridge between the HEF and TMD frameworks
and obtain a unified formulation which encompasses both the dilute and the
nearly back-to-back limit. Note that we follow the conventions used in earlier
papers that dealt with these formalisms, such as Ref.~\cite{Kutak:2012rf} and \cite{Dominguez:2011wm} respectively. Therefore, contrary to the HEF matrix elements $|\overline{{\cal M}_{ag^*\to cd}}|^2$, the hard factors $H_{ag\to cd}^{(i)}$ of the TMD factorization are defined without the $g^4$ factor. In addition, the definition of the gluon distribution also differ by a factor $\pi$. The integrated gluon distribution $x_2 f_{g/A}$ is obtained from $\int dk_t^2\ {\cal F}_{g/A}$ in the HEF formalism, and from $\int d^2k_t\ \mathcal{F}_{ag}^{(i)}$ in the TMD formalism.

Finally, let us point out that, in the frameworks described above, one emits
radiation in the transverse direction that one has no control over, as it is
part of the small-$x$ gluon distributions and therefore is treated fully
inclusively. To be more specific, at this level, transverse momentum conservation is obtained either by several particles of average transverse momentum $Q_s$, or by a third hard jet, depending on the magnitude of $|k_t|$. Due to the small-$x$ evolution, that radiation is ordered in rapidity, therefore it does not contribute to the measured forward dijets systems.

%-----------------------------------------------------------------------------
\section{High energy factorization derived from CGC:  \\
the $\mathbold{|p_{1t}|,|p_{2t}|,|k_t|\gg Q_s}$ limit}
\label{sec:dilute}

We shall demonstrate that the high-energy factorization formula for
double-inclusive particle production, Eq.~(\ref{eq:hef-formula}), is identical
to a result obtained from the CGC formalism in the dilute target approximation.
This is a limit where all the momenta involved in the process are much larger
than the saturation scale: $|p_{1t}|,|p_{2t}|,|k_t|\gg Q_s$. Here, we show explicitly
the equivalence of the HEF and CGC formulas for the ${qg^*\to qg}$
channel and only provide the final results for the two other channels, as the
derivations proceed identically for all of them.
We derive the CGC cross sections for the ${qg^*\to qg}$ and $gg^* \to q\qbar$
channels in the dilute limit following a procedure developed in
Ref.~\cite{Iancu:2013dta} where only the $g g^*\to gg$ sub-process was considered.

\begin{figure}
  \begin{center}
    \includegraphics[width=0.60\textwidth]{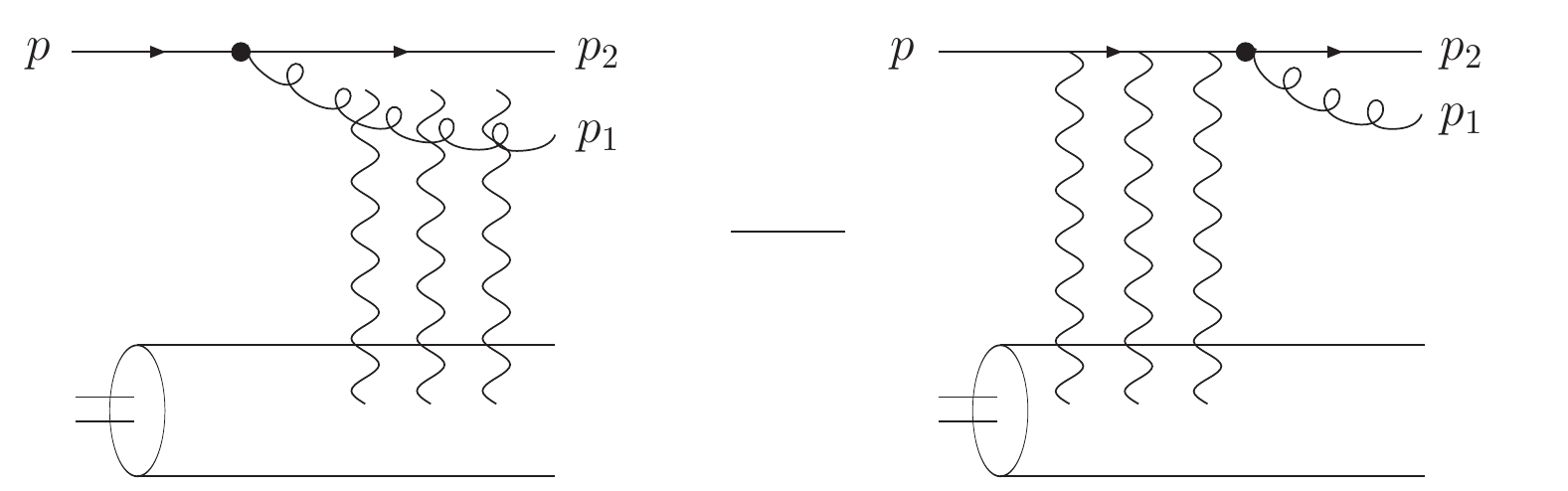}
  \end{center}
  \caption{Amplitude for quark-gluon production in the CGC formalism. Left: the gluon is radiated before the interaction with the target. Right: the gluon is radiated after the interaction with the target. The two terms have a relative minus sign.}
  \label{fig:amplitudeCGC}
\end{figure}

The amplitude for quark-gluon production is schematically presented
in Fig.~\ref{fig:amplitudeCGC} as in Ref.~\cite{Marquet:2007vb}. In the left
diagram, the emission of the gluon from the quark happens before the interaction
with the target, and in the right diagram the emission occurs after the quark
has interacted with the target. There is a relative minus sign between the two
cases as explained in details in Ref.~\cite{Marquet:2007vb}. Multigluon
interactions of quarks and gluons with a target, in the CGC formalism,
enter as Wilson lines in the expression for the amplitude. A quark propagator is
represented as a fundamental Wilson line, while a gluon propagator as an adjoint
Wilson line. As a result, the cross section involves multipoint correlators of
Wilson lines. In particular, the amplitude from Fig.~\ref{fig:amplitudeCGC}, after squaring, has four terms: a correlator of four Wilson lines, $S^{(4)}$, corresponding to interactions happening after the emission of the gluon, both in the amplitude and the complex conjugate, then a correlator of two Wilson lines, $S^{(2)}$,  representing the case when interactions with the target take place before the radiation of the gluon in both amplitude and complex conjugate, and two correlators of three Wilson lines, $S^{(3)}$, for the interference terms. In all the cases the splitting function is the same, and is given by the product of the quark wave functions: $\phi^{\lambda^*}_{\alpha\beta}(p,p^+_1,{\bf{x'}}-{\bf{b'}}) \phi^{\lambda}_{\alpha\beta}(p,p^+_1,\bf{x}-\bf{b})$. The total expression for the inclusive cross section in CGC is then given by the following formula~\cite{Marquet:2007vb}:
\be
\frac{d\sigma(pA\to qgX)}{dy_1 dy_2 d^2p_{1t} d^2p_{2t}} = \alpha_s C_F (1-z) p_1^+ x_1 f_{q/p}(x_1,\mu^2)
\left|\mathcal{M}(p,p_1,p_2)\right|^2\,,
\label{eq:cgc-1}
\ee
where the amplitude squared, $\left|\mathcal{M}(p,p_1,p_2)\right|^2$, has the
form: 
\bea
&&\hspace{-40pt}\left|\mathcal{M}(p,p_1,p_2)\right|^2 = \int \frac{d^2\bf{x}}{(2\pi)^2}\frac{d^2\bf{x'}}{(2\pi)^2}\frac{d^2\bf{b}}{(2\pi)^2}\frac{d^2\bf{b'}}{(2\pi)^2} e^{-ip_{1t}\cdot(\bf{x}-\bf{x'})} e^{-ip_{2t}\cdot(\bf{b}-\bf{b'})}  \nonumber \\
&&\hspace{40pt}\times \sum_{\lambda\alpha\beta} \phi^{\lambda^*}_{\alpha\beta}(p,p^+_1,{\bf{x'}}-{\bf{b'}}) \phi^{\lambda}_{\alpha\beta}(p,p^+_1,\bf{x}-\bf{b})  \nonumber \\
&&\hspace{40pt}\times \left\{S^{(4)}_{qg\bar{q}g}[{\bf{b}},{\bf{x}},{\bf{b'}},{\bf{x'}} ;x_2]- S^{(3)}_{qg\bar{q}}[{\bf{b}},{\bf{x}},{\bf{b'}}+z({\bf{x'}}-{\bf{b'}});x_2] \right . \nonumber \\
&&\hspace{40pt}\left . - S^{(3)}_{qg\bar{q}}[{\bf{b}}+z({\bf{x}}-{\bf{b}}),{\bf{x'}},{\bf{b'}};x_2] + S^{(2)}_{q\bar{q}}[{\bf{b}}+z({\bf{x}}-{\bf{b}}),{\bf{b'}}+z({\bf{x'}}-{\bf{b'}});x_2]\right\}\, ,
\label{eq:cgc-2}
\eea
where $\phi^{\lambda}_{\alpha\beta}$ are mixed-space quark wave functions and $S^{(i)}$ are correlators of Wilson lines explained in details below. 
Following the notation from Fig.~\ref{fig:dijets-pA} and Eq.~(\ref{eq:zdef}), we
use the fraction of the plus components of four-momenta, $z$, with $p_1$ being
the four-momentum of the outgoing gluon and $p_2$, the four-momentum of the
outgoing quark.

The fundamental, ${U}({\bf{x}})$, and adjoint, ${V}({\bf{x}})$, Wilson lines are defined as path-ordered exponentials of the gauge field (written here in the $A^+=0$ gauge):
\be 
{U}({\bf{x}}) = \mathcal{P} \exp \left[ ig \int dx^+ A_a^-(x^+, {\bf{x}}) t^a \right]~~~ {\text{and}} ~~~
{V}({\bf{x}}) = \mathcal{P} \exp \left[ ig \int dx^+ A_a^-(x^+, {\bf{x}}) T^a
\right] \,,
\ee
where $t^a$ and $T^a$ are the generators of the fundamental and adjoint representations of $SU(N)$ respectively. The traces of products of Wilson lines appearing in the cross section are defined in the following way:
\be
S^{(4)}_{qg\bar{q}g}({\bf{b}},{\bf{x}},{\bf{b'}},{\bf{x}})=\frac{1}{C_F N_c}\left<{\text {Tr}} \left({U}({\bf{b}}){U}^\dagger({\bf{b'}})t^dt^c\right)\left[{V}({\bf{x}}){V}^\dagger({\bf{x'}})\right]^{cd}\right>_{x_2}~;
\ee

\be
S^{(3)}_{qg\bar{q}}({\bf{b}},{\bf{x}},{\bf{z'}})=\frac{1}{C_F N_c}\left<{\text {Tr}} \left({U}^\dagger({\bf{z'}})t^c{U}({\bf{b}})t^d\right){V}^{cd}({\bf{x}})\right>_{x_2}~;
\ee

\be
S^{(2)}_{q\bar{q}}({\bf{z}},{\bf{z'}})=\frac{1}{N_c}\left<{\text {Tr}} \left({U}({\bf{z}}){U}^\dagger({\bf{z'}})\right)\right>_{x_2}~.
\ee
The CGC average is taken over the background filed evaluated at $Y=\ln(1/x_2)$. The product of wave functions in the massless limit is: 
\be 
\sum_{\lambda\alpha\beta} \phi^{\lambda^*}_{\alpha\beta}(p,p^+_1,{\bf{u'}}) \phi^{\lambda}_{\alpha\beta}(p,p^+_1,{\bf{u}}) =
\frac{8\pi^2}{p^+_1} \frac{{\bf{u}} \cdot {\bf{u'}}}{|{\bf{u}}|^2  |{\bf{u'}}|^2} (1 + (1-z)^2)~.
\ee
Introducing a change of variables, ${\bf{u}}={\bf{x}}-{\bf{b}}$ and ${\bf{v}}=z{\bf{x}} +(1-z){\bf{b}}$ (and similar for the primed coordinates), we get~\cite{Marquet:2007vb}:
\bea 
&&\hspace{-30pt}\left|\mathcal{M}(p,p_1,p_2)\right|^2 = \int \frac{d^2 {\bf{u}}}{(2\pi)^2}\frac{d^2 {\bf{u'}}}{(2\pi)^2} e^{i P_t \cdot ({\bf{u'}}-{\bf{u}})} 
\sum_{\lambda\alpha\beta} \phi^{\lambda^*}_{\alpha\beta}(p,p_1^+,{\bf{u'}}) \phi^{\lambda}_{\alpha\beta}(p,p_1^+,{\bf{u}})
 \nonumber \\
&&\hspace{-30pt}\times
\int\!\!\! \frac{d^2 {\bf{v}}}{(2\pi)^2}\frac{d^2 {\bf{v'}}}{(2\pi)^2} e^{i k_t
\cdot ({\bf{v'}}-{\bf{v}})}\!\!
\left[ S^{(4)}_{qg\bar{q}g}({\bf{b}},{\bf{x}},{\bf{b'}},{\bf{x'}}) -
S^{(3)}_{qg\bar{q}} ({\bf{b}},{\bf{x}},{\bf{v'}}) - S^{(3)}_{qg\bar{q}}
({\bf{v}},{\bf{x'}},{\bf{b'}}) + S^{(2)}_{q\bar{q}} ({\bf{v}},{\bf{v'}})
\right]\!\!.\,
\eea
The conjugate momentum to ${\bf{u'}}-{\bf{u}}$ is $P_t=(1-z)p_{1t}-zp_{2t}$, and the one corresponding to ${\bf{v'}}-{\bf{v}}$ is the total transverse momentum of the produced particles $k_t=p_{1t}+p_{2t}$. In terms of fundamental Wilson lines only:
\bea
S^{(4)}_{qg\bar{q}g}({\bf{b}},{\bf{x}},{\bf{b'}},{\bf{x'}})= \frac{1}{2C_FN_c}
\left < {\text {Tr}} \left(
{U}({\bf{b}}){U}^\dagger({\bf{b'}}){U}({\bf{x'}}){U}^\dagger({\bf{x}})\right)
{\text{Tr}} \left({U}({\bf{x}}) {U}^\dagger ({\bf{x'}})\right)\right.\\ \nonumber\left. - \frac{1}{N_c} {\text{Tr}}\left(
{U}({\bf{b}}) {U}^\dagger({\bf{b'}})\right) \right>_{x_2} \,,
\label{eq:S4-fund}
\eea
and
\be 
S^{(3)}_{qg\bar{q}} ({\bf{b}},{\bf{x}},{\bf{v'}}) = \frac{1}{2C_FN_c} \left < {\text{Tr}}\left( {U}({\bf{b}}) {U}^\dagger ({\bf{x}})\right) {\text{Tr}}\left( {U}({\bf{x}}) {U}^\dagger ({\bf{v'}})\right) - \frac{1}{N_c} {\text{Tr}}\left( {U}({\bf{b}}) U^\dagger ({\bf{v'}})\right)\right>_{x_2}~.
\label{eq:S3-fund}
\ee

In the dilute target limit we allow for only up to two gluon exchanges between the Wilson line propagators and the nucleus. Accordingly, we expand the Wilson lines to second order in the background field:
\be 
{U}({\bf{x}}) \approx
 1 + ig \int dx^+ A^-(x^+, {\bf{x}}) - \frac{g^2}{2} \int dx^+ dy^+ \mathcal{P} \left\{A^-(x^+,{\bf{x}}) A^-(y^+,{\bf{x}}) \right\} + \mathcal{O} (A^3)~.
\ee
To this order, the expectation values of the four- and three-point correlators are simply expressed in terms of the dipole operator $S^{(2)}_{q\bar{q}} ({\bf{v}},{\bf{v'}})$. The dilute target approximation gives only a leading result in $|{\bf{v}}-{\bf{v'}}|^2 Q_s^2$ for the expectation value of $S^{(2)}_{q\bar{q}} ({\bf{v}},{\bf{v'}})$, which is equivalent to taking the limit $|k_t|\gg Q_s$. Similarly, when all the momenta involved in the process are much larger than the saturation scale, the correlators entering the cross section get the following expressions:
\bea 
&& \hspace{-70pt}  S^{(4)}_{qg\bar{q}g}({\bf{b}},{\bf{x}},{\bf{b'}},{\bf{x'}})  = 1 - g^2 N_c \Gamma _{x_2} ({\bf{x}}-{\bf{x'}}) - g^2 \frac{N_c^2 - 1}{2N_c} \Gamma _{x_2}({\bf{b}}-{\bf{b'}}) \nonumber \\
&& \hspace{30pt} - \frac{g^2 N_c}{2} \left[ \Gamma _{x_2}({\bf{x}}-{\bf{b}}) + \Gamma _{x_2}({\bf{x'}}-{\bf{b'}}) - \Gamma _{x_2}({\bf{x'}}-{\bf{b}}) - \Gamma _{x_2}({\bf{x}}-{\bf{b'}}) \right]~;
\eea
\be 
 S^{(3)}_{qg\bar{q}} ({\bf{b}},{\bf{x}},{\bf{v'}})  = 1 - \frac{g^2 N_c}{2} \Gamma_{x_2} ({\bf{b}}-{\bf{x}}) - \frac{g^2 N_c}{2} \Gamma_{x_2} ({\bf{x}}-{\bf{v'}}) + \frac{g^2}{2 N_c} \Gamma_{x_2} ({\bf{b}}-{\bf{v'}})~;
\ee
\be 
 S^{(2)}_{q\bar{q}} ({\bf{v}},{\bf{v'}}) = 1 - g^2 \frac{N_c^2 - 1}{2N_c} \Gamma _{x_2} ({\bf{v}} - {\bf{v'}})~.
\label{eq:dilutedipole}
\ee
In the above equations:
\be  
\Gamma _{x_2} ({\bf{x}} - {\bf{y}}) = \int dx^+ \left[ \gamma_{x_2} (x^+, {\bf{0}}) - \gamma_{x_2} (x^+, {\bf{r}})\right]~,
\label{eq:Gammadef}
\ee
where ${\bf{r}}={\bf{x}} - {\bf{y}}$ and $\gamma_{x_2} (x^+, {\bf{r}})$ is related to the expectation value of the two-field correlator:
\be 
\left < A^-_a (x^+,{\bf{x}}) A^-_b(y^+,{\bf{y}}) \right>_{x_2} = \delta ^{ab} \delta (x^+ - y^+) \gamma_{x_2} (x^+, {\bf{x}} - {\bf{y}})~.
\ee
Using the expressions for the multi-point functions $S^{(i)}$, we get the following result for the amplitude squared:
\bea 
\left|\mathcal{M}(p,p_1,p_2)\right|^2 = 4 \pi^2 g^2 N_c (1 + (1-z)^2) \frac{1}{p^+_1} \int \frac{d^2 {\bf{u}}}{(2\pi)^2}\frac{d^2 {\bf{u'}}}{(2\pi)^2} e^{i P_t \cdot ({\bf{u'}}-{\bf{u}})} 
\frac{{\bf{u}} \cdot {\bf{u'}}}{|{\bf{u}}|^2  |{\bf{u'}}|^2}
 \nonumber \\
 \times
 \int \frac{d^2 {\bf{v}}}{(2\pi)^2}\frac{d^2 {\bf{v'}}}{(2\pi)^2} e^{i k_t \cdot ({\bf{v'}}-{\bf{v}})}
\left[ \Gamma_{x_2} ({\bf{x}}-{\bf{b'}}) + \Gamma_{x_2} ({\bf{x'}}-{\bf{b}}) + \Gamma_{x_2} ({\bf{x}}-{\bf{v'}}) \right . \nonumber \\ \left .
 +\Gamma_{x_2} ({\bf{v}}-{\bf{x'}})  - 2 \Gamma_{x_2} ({\bf{x}}-{\bf{x'}})   - \frac {N_c^2 -1}{N_c^2} \Gamma_{x_2} ({\bf{b}}-{\bf{b'}}) \right . \nonumber \\ \left .
- \frac {N_c^2 -1}{N_c^2} \Gamma_{x_2} ({\bf{v}}-{\bf{v'}}) - \frac{1}{N_c^2} \Gamma_{x_2} ({\bf{b}}-{\bf{v'}}) - \frac{1}{N_c^2} \Gamma_{x_2} ({\bf{v}}-{\bf{b'}}) \right]~.
\eea
We perform the integrals in the above expression by changing the variables from ${\bf{v}}$ and ${\bf{v'}}$ to ${\bf{r}}$ and ${\bf{B}}$. The integrals over the transverse distances of the type ${\bf{r}}={\bf{v}} - {\bf{v'}}$ are equivalent to the Fourier transform of Eq.~(\ref{eq:Gammadef}) and give the unintegrated
gluon distribution:
\be 
f_{x_2} (k_t) \equiv - k_t^2 \int d^2 {\bf{r}} \, \Gamma_{x_2} ({\bf{r}}) e^{-i k_t \cdot {\bf{r}}} = k_t^2 \int dx^+ \gamma _{x_2} (x^+, k_t)~.
\ee
In our approximation, the correlators do not depend on the impact parameter ${\bf{B}} =( {\bf{v}} + {\bf{v'}})/2$. The integrals over ${\bf{B}}$ factorize and give the transverse area of the target: $\int d^2{\bf{B}} = S_\perp$. Finally, the rest two integrations reduce to:
\be 
\int d^2 {\bf{u}}\, e^{-i P_t \cdot {\bf{u}}} \frac{{\bf{u}}}{|{\bf{u}}|^2} =-2\pi i \frac{{\bf{P_t}}}{|{\bf{P_t}}|^2}.
\ee
In terms of the unintegrated gluon distribution, the amplitude squared then gets the form:
\bea 
&& \hspace{-50pt} \left|\mathcal{M}(p,p_1,p_2)\right|^2 = \frac{2}{(2\pi)^4} g^2 S_\perp N_c \frac{f_{x_2}(k_t)}{k_t^2} (1+ (1-z)^2)
\frac{1}{p^+_1} \nonumber \\
&& \hspace{20pt} \times \left[ \frac{(N_c^2-1)}{2N_c^2} \frac{1}{P_t^2} + \frac{(N_c^2-1)}{2N_c^2} \frac{1}{p_{1t}^2} + \frac{1}{p_{2t}^2}
+ \frac{1}{N_c^2}\frac{P_t \cdot p_{1t}}{P_t^2 p_{1t}^2} + \frac{P_t \cdot p_{2t}}{P_t^2 p_{2t}^2}
+\frac{p_{1t} \cdot p_{2t}}{p_{1t}^2 p_{2t}^2} \right]\, ,
\label{eq:CGCfac-qg}
\eea

We want to show that Eq.~(\ref{eq:CGCfac-qg}) reproduces the HEF formula~(\ref{eq:hef-formula}) with
the appropriate unintegrated parton distribution function and off-shell matrix elements. For this purpose,
we need to find a relation between the unintegrated gluon distribution used in the above equation,
$f_{x_2}(k_t)$, and ${\cal F}_{g/A}(x_2,k_t)$, which appears in the HEF formula~(\ref{eq:hef-formula}).
This is easily done by considering the deep inelastic scattering process, since ${\cal F}_{g/A}(x_2,k_t)$ is precisely
the unintegrated gluon distribution involved in the formulation of the $\gamma^*+A\to X$ total cross section, and is therefore related to the 
$q\bar q$ dipole scattering amplitude in a straightforward manner (see for instance \cite{vanHameren:2014lna,Kutak:2014wga}):
\begin{equation}
{\cal F}_{g/A}(x_2,k_t)=\frac{N_c}{\alpha_s (2\pi)^3}\int d^2{\bf{v}} d^2{\bf{v'}}\
e^{-i k_t \cdot ({\bf{v}}-{\bf{v'}})}\nabla^2_{{\bf{v}}-{\bf{v'}}}\left[1-S^{(2)}_{q\bar{q}} ({\bf{v}},{\bf{v'}})\right]\ .
\label{eq:disgluon}
\end{equation}
In the weak-field limit, using formula~(\ref{eq:dilutedipole}), this gives the relation
\begin{equation}
  f_{x_2}(k_t) = \frac{4\pi^2}{S_\perp (N_c^2-1)} {\cal F}_{g/A} (x_2,k_t)\,.
\end{equation}
Then, the cross section for the $qg$ production channel from
Eq.~(\ref{eq:cgc-1}) can be written in a more compact form

\begin{equation}
\frac{d\sigma(pA\to qgX)}{dy_1 dy_2 d^2p_{1t} d^2p_{2t}} =
\frac{\alpha_s^2}{2\pi}x_1f_{q/p}(x_1,\mu^2)z(1\!-\!z)\hat{P}_{gq}(z)
\left[1+\frac{(1\!-\!z)^2 p_{1t}^{\ 2}}{P_t^2}-\frac{1}{N_c^2}\frac{z^2 p_{2t}^{\ 2}}{P_t^2}\right]
\frac{{\cal F}_{g/A} (x_2,k_t)}{p_{1t}^{\ 2}\ p_{2t}^{\ 2}}~,
\label{eq:sigmaCGCfac-qg}
\end{equation}
where $\hat{P}_{gq}(z)$ is related to the quark-to-gluon splitting function and is given by:
\begin{equation}
\hat{P}_{gq}(z) =  \frac{1+(1\!-\!z)^2}{z}\ .
\end{equation}
It turns out that the above expression for the quark-gluon production cross section is identical
to the result in the HEF formalism, Eq.~(\ref{eq:hef-formula}), containing the off-shell amplitudes
$\overline{\left|\mathcal{M}_{ag^* \to cd}\right|}^2$. The latter have been
calculated in Refs.~\cite{Deak:2009xt},~\cite{vanHameren:2012uj} and~\cite{vanHameren:2013}. 

The equivalence of the CGC and HEF formulas in the dilute limit can be shown in
a similar way for the cross sections of the other two subprocesses, ${gg^* \to
q\bar{q}}$ and ${gg^* \to gg}$. The CGC results for the cross sections in this
limit are:
\begin{equation}
\frac{d\sigma(pA\to q\bar{q}X)}{dy_1 dy_2 d^2 p_{1t} d^2 p_{2t}} =
\frac{\alpha_s^2}{4C_F\pi}x_1f_{g/p}(x_1,\mu^2)z(1\!-\!z)\hat{P}_{qg}(z)
\left[-\frac{1}{N_c^2}+\frac{(1\!-\!z)^2 p_{1t}^{\ 2}+z^2 p_{2t}^{\ 2}}{P_t^2}\right]
\frac{{\cal F}_{g/A} (x_2,k_t)}{p_{1t}^{\ 2}\ p_{2t}^{\ 2}}
\label{eq:sigmaCGCfac-qbarq}
\end{equation}
and~\cite{Iancu:2013dta}
\begin{equation}
\frac{d\sigma(pA\to ggX)}{dy_1 dy_2 d^2 p_{1t} d^2 p_{2t}} =
\frac{\alpha_s^2N_c}{\pi C_F}x_1f_{g/p}(x_1,\mu^2)z(1\!-\!z)\hat{P}_{gg}(z)
\left[1+\frac{(1\!-\!z)^2 p_{1t}^{\ 2}+z^2 p_{2t}^{\ 2}}{P_t^2}\right]
\frac{{\cal F}_{g/A} (x_2, k_t)}{p_{1t}^{\ 2}\ p_{2t}^{\ 2}}\, .
\label{eq:sigmaCGCfac-gg}
\end{equation}
The expressions for $\hat{P}_{qg}(z)$ and $\hat{P}_{gg}(z)$ have the form:
\begin{equation}
\hat{P}_{qg}(z) = z^2+(1\!-\!z)^2 \,,
\quad \quad \quad
\hat{P}_{gg}(z) = \frac{z}{1-z} + \frac{1-z}{z} + z(1-z)\, .
\ee
Again, Eqs.~(\ref{eq:sigmaCGCfac-qbarq}) and~(\ref{eq:sigmaCGCfac-gg}) are
equivalent to the HEF formulas for the corresponding cross sections~\cite{vanHameren:2014lna}.

Therefore, in principle, the HEF formalism should not be employed to include non-linear effects, and one should stick to Balitsky-Fadin-Kuraev-Lipatov (BFKL) evolution \cite{Lipatov:1976zz,Kuraev:1976ge,Balitsky:1978ic}, or Ciafaloni-Catani-Fiorani-Marchesini evolution \cite{Ciafaloni:1987ur,Catani:1989sg,Catani:1989yc}, when evaluating the gluon distribution. In this spirit, most studies are performed using a gluon density evolved with an improved BFKL equation that includes some higher-order corrections \cite{Kwiecinski:1997ee}, but no non-linear effects. However, we note that the HEF framework could be used with the Balitsky-Kovchegov (BK) equation \cite{Balitsky:1995ub,Kovchegov:1999yj} in order to investigate the so-called geometric scaling regime, where saturation effects are felt, even though $Q_s\ll k_t$. The full saturation region, $Q_s\sim k_t$, is however, in principle, out of reach of formula (\ref{eq:hef-formula}). Along these lines, an estimate of saturation effects was obtained in \cite{Kutak:2003bd,Kutak:2004ym}, using the BK equation extended to include the same higher-order corrections as included in the linear case \cite{Kwiecinski:1997ee}.

%-----------------------------------------------------------------------------
\section{TMD factorization for nearly back-to-back jets: \\ the 
$\mathbold{|p_{1t}|,|p_{2t}|\gg |k_t|,Q_s}$ limit}
\label{sec:tmd-on-shell}

In this section we discuss the special case of nearly back-to-back jets,
$|p_{1t}+p_{2t}| \ll |p_{1t}|,|p_{2t}|$, where the differential cross section is
given by formula (\ref{eq:tmd-main}). Several gluon distributions
$\mathcal{F}_{ag}^{(i)}$, with different operator definition, are involved here. Indeed, as explained in \cite{Bomhof:2006dp}, a generic unintegrated gluon distribution of the form
\begin{equation}
\mathcal{F}(x_2,k_t) \stackrel{\text{naive}}{=}
2\int \frac{d\xi^+d^2{\boldsymbol\xi}}{(2\pi )^{3}p_A^{-}}e^{ix_2p_A^{-}\xi ^{+}-ik_t\cdot{\boldsymbol\xi}}
\left\langle A|\text{Tr}\left[
F^{i-}\left(\xi^+,{\boldsymbol\xi}\right)F^{i-}\left( 0\right)
\right]|A\right\rangle\,,
\end{equation}
where $F^{i-}$ are components of the gluon field strength tensor,
must be also supplemented with gauge links, in order to render such a bi-local product of field operators gauge invariant. 

The gauge links are path-ordered exponentials, with the integration path being fixed by the hard part of the process under consideration. Therefore, unintegrated gluon distributions are process-dependent.

In the following, we shall encounter two gauge links $\mathcal{U}^{\left[ +\right] }$ and $\mathcal{U}^{\left[ -\right] }$, as well as the loop
$\mathcal{U}^{\left[\square \right] }=\mathcal{U}^{\left[ +\right] }\mathcal{U}^{\left[ -\right]\dagger}=\mathcal{U}^{\left[ -\right] }\mathcal{U}^{\left[ +\right]\dagger}$. These links are composed of Wilson lines, their simplest expression is obtained in the $A^+=0$ gauge:
\begin{equation} 
\mathcal{U}^{\left[ \pm\right] }= U(0,\pm\infty;{\bf{0}})U(\pm\infty,\xi^+;{\boldsymbol{\xi}})\quad \mbox{with}\quad
U(a,b;{\bf{x}}) = \mathcal{P} \exp \left[ ig \int_a^b dx^+ A_a^-(x^+, {\bf{x}}) t^a \right]\ ,
\end{equation}
but the expressions of the various gluon distributions given below are
gauge-invariant. From now on, $F^{i-}\left(\xi^+,{\boldsymbol\xi}\right)$ is
simply denoted as $F(\xi)$, and the hadronic matrix elements $\langle A | ... | A \rangle\to \langle ... \rangle $. Note however that they are different from the CGC averages $\left< \cdots \right>_{x_2}$ of the previous section. Indeed, the normalization of the hadronic state $\left|A\right>$ is defined as $\left<A'|A\right> = (2\pi)^3 \, 2 p_A^+ \, \delta (p_A^+ - p_A'^{+}) \, \delta^{(2)} \left (p_{At} - p'_{At}\right)$, while the CGC averages are normalized as $\left< 1 \right>_{x_2} = 1$. As explained in \cite{Dominguez:2011wm}, the two can be related by making the replacement $\left< \cdots \right>_{x_2} \to \frac{\left<A\right| \dots \left|A\right>}{  \left<A|A\right>}$.

This approach to dijet production in proton-nucleus collisions was analyzed in
Ref.~\cite{Dominguez:2011wm}. The TMD factorization formula (\ref{eq:tmd-main})
was derived there in the large-$N_c$ limit, and shown to be equivalent to CGC 
calculations (\eg formulas (\ref{eq:cgc-1}) and (\ref{eq:cgc-2}) in
the case of the $qA\to qg$ channel), after taking the limit
$|p_{1t}|,|p_{2t}|\gg |k_t|, Q_s$. In this section, we
derive the TMD factorization formula keeping $N_c$ finite. We obtain corrections to the hard
factors $H_{ag\to cd}^{(i)}$ previously derived, and we calculate new hard factors
corresponding to gluon distributions that were omitted before (as they were
vanishing in the large-$N_c$ limit). The finite $N_c$ extension prevents one to make a further simplification, called correlator factorization, essential to relate the TMD factorization and the CGC formalism, but gives completeness to the main result of this paper, i.e. the new factorization formula we propose below is valid for finite $N_c$. We also check explicitly the gauge invariance of these hard factors  by computing them in a gauge different from the one used in~\cite{Dominguez:2011wm}. 

An important fact to note is that, as a consequence of the
$|k_t|\ll|p_{1t}|,|p_{2t}|$ limit, the $k_t$ dependence in (\ref{eq:tmd-main})
survives only in the gluon distributions, and the hard factors are calculated as
if the small-$x_2$ gluon was on-shell. That is, looking at the hard partonic
interaction represented by the blob $H$ in Fig.~\ref{fig:dijets-pA}, $k^2 =
-|k_t|^2$ is set to zero, and $\shat + \that + \uhat =0$.

%------------------------------------
\subsection[The $qg \to qg$ channel]{The $\mathbold{qg \to qg}$ channel}

\begin{figure}[t]
  \begin{center}
    \includegraphics[width=0.8\textwidth]{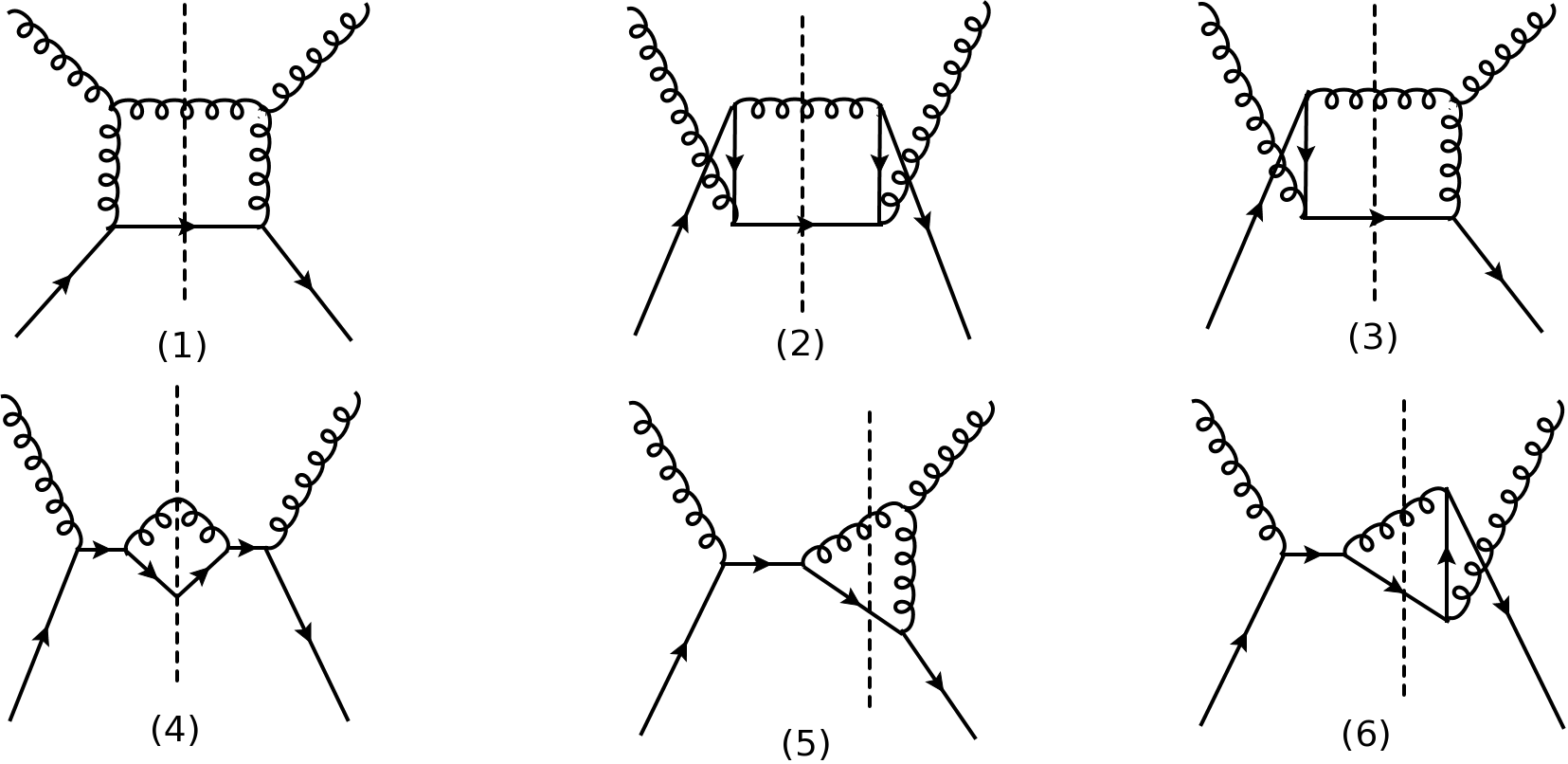}
  \end{center}
  \caption{Diagrams for $qg \to qg$ subprocess. The mirror diagrams of (3), (5)
  and (6) give identical contributions.}
  \label{fig:qg2qg-diag}
\end{figure}

The complete set of independent cut diagrams contributing to this channel is
shown in Fig.~\ref{fig:qg2qg-diag} (mirror images of diagrams (3), (5) and (6)
give identical expressions).

The cross section for a quark-gluon scattering involves only two different TMD gluon distributions as given in Ref.~\cite{Bomhof:2006dp}:

\begin{equation}
\label{eq:fact_gqqg}
\frac{d\sigma^{pA\to qgX}}{d^{2}P_td^{2}k_tdy_{1}dy_{2}}=\frac{\alpha_s^2}{(x_1
x_2 s)^{2}}\ x_1f_{q/p}(x_1, \mu^2)
\sum_{i=1}^2 \mathcal{F}_{qg}^{(i)}H_{qg\to qg}^{(i)}\,,
\end{equation}
with:
\begin{eqnarray}
\mathcal{F}_{qg}^{(1)} &=&2\int \frac{d\xi^+d^2{\boldsymbol\xi}}{(2\pi )^{3}p_A^{-}}e^{ix_2p_A^{-}\xi ^{+}-ik_t\cdot{\boldsymbol\xi}}
\left\langle \text{Tr}\left[ F\left( \xi \right) \mathcal{U}^{\left[ -\right] \dagger }F\left( 0\right) \mathcal{U}^{\left[ +\right] }\right] \right\rangle
= x_2 G^{(2)}(x_2, k_t)\ ,
\label{eq:Fqg1-def}
\\
\mathcal{F}_{qg}^{(2)} &=&2\int \frac{d\xi^+d^2{\boldsymbol\xi}}{(2\pi)^{3}p_A^{-}} e^{ix_2p_A^{-}\xi ^{+}-i k_t \cdot{\boldsymbol\xi}}
\left\langle\text{Tr}\left[ F\left( \xi \right) \frac{\text{Tr}\left[ \mathcal{U}^{\left[\square \right] }\right] }{N_{c}}\mathcal{U}^{\left[ +\right] \dagger}
F\left( 0\right) \mathcal{U}^{\left[ +\right] }\right] \right\rangle\ .
\label{eq:Fqg2-def}
\end{eqnarray}
These are the same gluon distributions as in the large-$N_c$ limit~\cite{Dominguez:2011wm}, no
additional ones are present in this channel. The only difference in the expression~(\ref{eq:fact_gqqg}) when we go to
finite $N_c$ will appear in the hard factor $H_{qg \to qg}^{(1)}$ associated
with $\mathcal{F}_{qg}^{(1)}$. That gluon distribution is sometimes also denoted
$x_2G^{(2)}$, and is called the \emph{dipole distribution}, since it is the one that enters the formulation of the inclusive and semi-inclusive DIS.

In the CGC approach, $x_2G^{(2)}$ can be related to the $q \bar q$
dipole scattering amplitude, and therefore linked to the gluon distribution used
in the HEF formalism: $\mathcal{F}_{g/A} (x_2, k_t)=\pi x_2 G^{(2)}(x_2, k_t)$.
That distribution is not sufficient however to compute the forward dijet cross
section when $|k_t|\sim Q_s$ (\ie the case considered in this section). For completeness, we note that a detailed derivation
of this relation between formula~(\ref{eq:disgluon}), involving a CGC
correlation function, and formula~(\ref{eq:Fqg1-def}), involving matrix elements defining TMDs, can be found in Appendix A of~\cite{Dominguez:2011wm}.

The exact results for the two hard factors read
\begin{eqnarray}
\label{eq:H1-qg-on-shell-def}
H_{qg \to qg}^{(1)}&=&
\frac{1}{2} D_1 - \frac{1}{N_c^2-1} D_2 +D_4 + 2D_5 + 2 D_6\,, \\
\label{eq:H2-qg-on-shell-def}
H_{qg \to qg}^{(2)}&=&\frac{1}{2} D_1 + \frac{N_c}{2C_F} D_2 +2 D_3\,,
\end{eqnarray}
where $D_i$s are the squared and interference diagrams corresponding to the
$qg\to qg$ channel, following the numbering of Fig.~\ref{fig:qg2qg-diag}.
Each term $D_i = C_{u_i} h_i $ represents the product of the color factor,
$C_{u_i}$, and the hard coefficient, $h_i$. What kind of diagrams enter the hard
factors $H_{qg \to qg}^{(i)}$ depends on the type of the gauge links appearing
in each of them. As summarized in table IV of Ref.~\cite{Bomhof:2006dp},
the distribution $\mathcal{F}_{qg}^{(1)}$ is present in diagrams (1), (2), (4),
(5) and (6), while the distribution $\mathcal{F}_{qg}^{(2)}$ appears in diagrams
(1), (2) and (3). The $D_i$ components were computed in
Ref.~\cite{Dominguez:2011wm} (table II) in an axial gauge with the axial vector,
$n$, set to $n=p$, for both the incoming and the outgoing gluon, where $p$ is
the four-momentum of the incoming quark, as defined in Fig.~\ref{fig:dijets-pA}. 
Formulated differently, the polarization vector of each external gluon was chosen such that, besides with the momentum of the gluon, their inner product with $p$ vanishes.
We recovered the same results for $D_i$s in that gauge and performed the same
calculation in a different gauge with the axial vector set to $n=p$ for the
incoming gluon and $n=p_2$ for the outgoing gluon~\footnote{The choice of axial gauge vectors for external gluons corresponds to the choice of the reference momentum for their polarization vectors, see for example \cite{Mangano:1990by}, and is arbitrary for gauge invariant quantities. Thus, the independence on those gauge vectors can be used to confirm that the result is gauge invariant.
}. The results for the hard
factors $H_{qg \to qg}^{(1)}$ and $H_{qg \to qg}^{(2)}$ at finite $N_c$ are
identical in both gauges and they read
\begin{eqnarray}
\label{eq:H1-qg-on-shell}
H_{qg \to qg}^{(1)}&=&
- \frac{\hat{u} ( \hat{s}^2 + \hat{u}^2)}{2\hat{s}\hat{t}^2}
+\frac{1}{2N_c^2}\frac{( \hat{s}^2 + \hat{u}^2)}{\hat{s}\hat{u}}\ ,\\
\label{eq:H2-qg-on-shell}
H_{qg \to qg}^{(2)}&=&
- \frac{\hat{s} ( \hat{s}^2 + \hat{u}^2)}{2\hat{u}\hat{t}^2}\ .
\end{eqnarray}

The hard factors and the TMDs entering the factorization
formula~(\ref{eq:fact_gqqg}) are all gauge invariant. In principle, that leaves
us some freedom and the factorization formula can be rewritten with new hard
factors and the corresponding new gluon distributions formed as linear
combinations of the the old ones.

For reasons that shall be discussed in detail in Section~\ref{sec:HelTMD},
let us define the new hard factors for the $qg \to qg$ subprocess
\begin{equation}
  \label{eq:Kqg2qgon}
  K_{qg\to qg}^{(1)} = H_{qg\to qg}^{(1)} +
                       \frac{1}{N_c^2} H_{qg\to qg}^{(2)} ~~~~~~~~
  {\text{and}} ~~~~~~~
  K_{qg\to qg}^{(2)} = \frac{N_c^2-1}{N_c^2} H_{qg\to qg}^{(2)}\,,
\end{equation}
and the corresponding new gluon TMDs
\begin{eqnarray} 
\Phi_{qg\rightarrow qg}^{\left(1\right)} &=& \mathcal{F}_{qg}^{\left(1\right)}\,, \\
\Phi_{qg\rightarrow qg}^{\left(2\right)} &=& \frac{1}{N_{c}^{2}-1}\left(- \mathcal{F}_{qg}^{\left(1\right)}+
N_c^2\mathcal{F}_{qg}^{\left(2\right)}\right) \,,
\end{eqnarray} 
such that the factorization formula~(\ref{eq:fact_gqqg}) now takes the form
\begin{equation}
  \frac{d\sigma^{pA\to qg X}}{d^{2}P_td^{2} k_t dy_{1}dy_{2}}=
\frac{\alpha_s^2}{(x_1 x_2 s)^2}\  x_1 f_{q/p}(x_1, \mu^2)
  \left[ \Phi_{qg \to qg}^{(1)}K_{qg\to qg}^{(1)} + 
  \Phi_{qg\to qg}^{(2)}K_{qg\to qg}^{(2)} \right]~.
  \label{eq:new-fac-qg2qg}
\end{equation}
The explicit expressions for $K_{qg\to qg}^{(1)}$ and $K_{qg\to qg}^{(2)} $  are
given in Table~\ref{tab:Kfactors-on-shell}.

\begin{table}[t]
\begin{center}
\begin{tabular}{ccc}
\hline \\[-0.5em]
                & $K^{(1)}_{ag \to cd}$ & $K^{(2)}_{ag \to cd}$  \\[0.5em]
\hline \\
$qg \to qg$     & \hspace{10pt} 
$\displaystyle - \frac{\shat^2+\uhat^2}{2 \that^2 \shat \uhat} \left[ \uhat^2 +
\frac{\shat^2-\that^2}{N_c^2}\right]$
&  
$\displaystyle -\frac{C_F}{N_c} \frac{\shat (\shat^2+\uhat^2)}{ \that^2\uhat}$
\\[2em]
$gg \to q\qbar$ & \hspace{10pt} 
$\displaystyle \frac{1}{2 N_c} 
\frac{( \hat{t}^2 + \hat{u}^2)^2}{\hat{s}^2\hat{t}\hat{u}}$ &  \hspace{10pt} 
$\displaystyle -\frac{1}{2C_FN_c^2} \frac{\hat{t}^2 + \hat{u}^2}{\hat{s}^2}$
\\[2em]
$gg \to gg$     &  \hspace{10pt}
$\displaystyle \frac{2 N_c}{C_F} \frac{(\hat{s}^2-\hat{t}\hat{u})^2(\hat{t}^2 +
\hat{u}^2)}{\hat{t}^2\hat{u}^2\hat{s}^2}$
&  
$\displaystyle \frac{2 N_c}{C_F}
\frac{(\hat{s}^2-\hat{t}\hat{u})^2}{\hat{t}\hat{u}\hat{s}^2}$
\\[2em]
\hline 
\end{tabular}
\end{center}
\caption{
  The ``new'' hard factors following from simplified effective TMD factorization
  of Eqs.~(\ref{eq:new-fac-qg2qg}), (\ref{eq:new-fac-gg2qqbar}) and
  (\ref{eq:gg2gg-on-shell2}) in the case with all partons being on shell.
}
\label{tab:Kfactors-on-shell}
\end{table}

%------------------------------------
\subsection[The $gg \to q\qbar$ channel]{The $\mathbold{gg \to q\qbar}$ channel}
\label{sec:gg2qq-on-shell}

\begin{figure}[t]
  \begin{center}
    \includegraphics[width=0.8\textwidth]{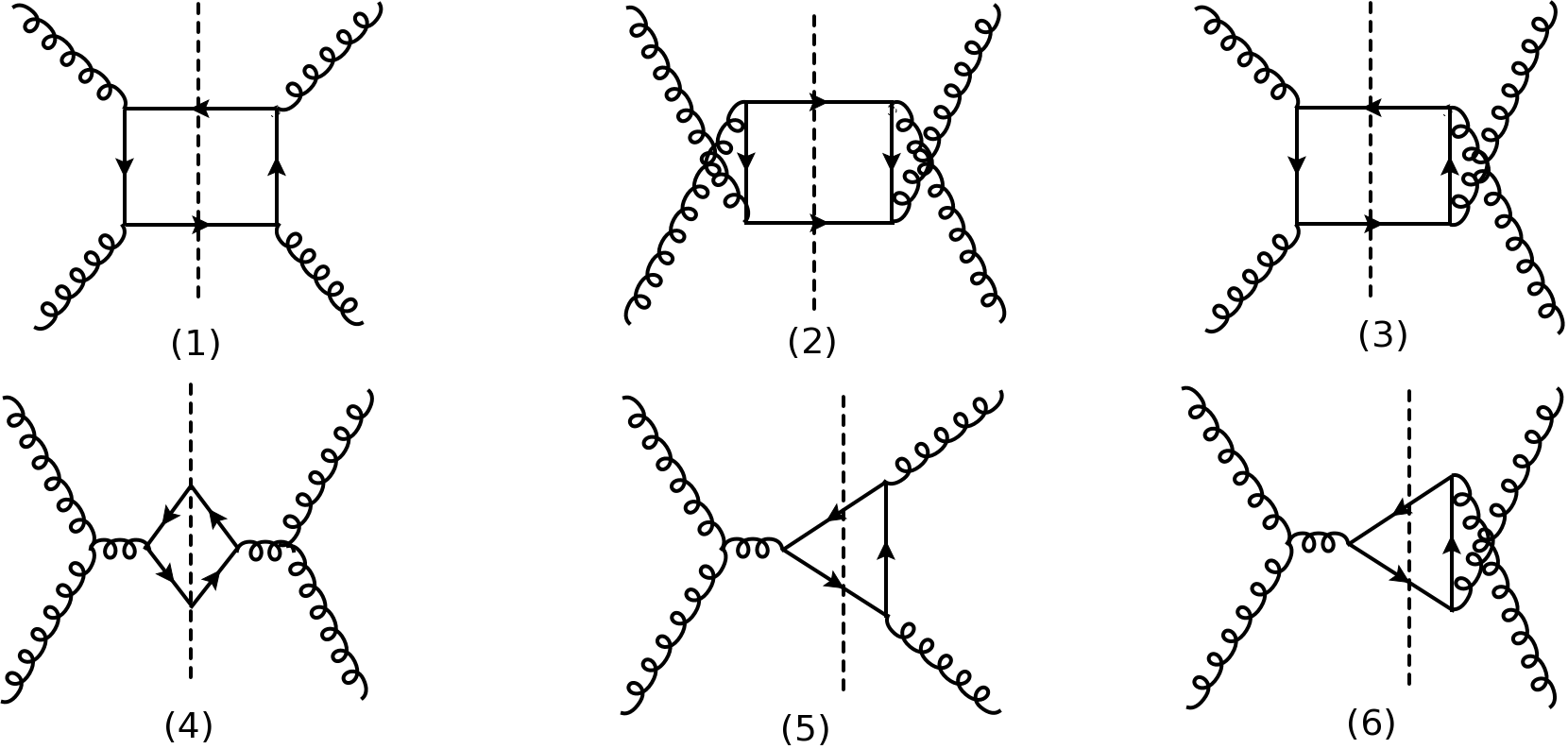}
  \end{center}
  \caption{Diagrams for $gg \to q\qbar$ subprocess. The mirror diagrams of (3),
  (5) and (6) give identical contributions.}
  \label{fig:gg2qqbar-diag}
\end{figure}

The independent cut diagrams contributing to this channel are shown in
Fig.~\ref{fig:gg2qqbar-diag}.

In addition to the two gluon distributions, $\mathcal{F}_{gg}^{(1)}$ and
$\mathcal{F}_{gg}^{(2)}$, used in Ref.~\cite{Dominguez:2011wm}, the result to
all orders in $N_c$ involves a third distribution~\cite{Bomhof:2006dp,Akcakaya:2012si}, 
$\mathcal{F}_{gg}^{(3)}$ (also sometimes denoted $x_2 G^{(1)}$ and called the 
\emph{Weizsacker-Williams gluon distribution}), and the differential cross section reads
\begin{equation}
\label{eq:fact_ggqq}
\frac{d\sigma^{pA\to q\bar{q}X}}{d^{2}P_td^{2} k_t
dy_{1}dy_{2}}=\frac{\alpha_s^2}{(x_1x_2 s)^{2}}\ x_1f_{g/p}(x_1, \mu^2)
\sum_{i=1}^3 \mathcal{F}_{gg}^{(i)}H_{gg\to q\bar{q}}^{(i)}\,,
\end{equation}
with the three gluon TMDs defined as
\begin{eqnarray}
\label{eq:Fgg1}
\mathcal{F}_{gg}^{(1)} &=&2\int \frac{d\xi^+d^2{\boldsymbol\xi}}{(2\pi )^{3}p_A^{-}} e^{ix_2p_A^{-}\xi ^{+}-i k_t \cdot{\boldsymbol\xi}}
\left\langle \text{Tr}\left[ F\left( \xi \right)\frac{\text{Tr}\left[ \mathcal{U}^{\left[\square \right] }\right] }{N_{c}} \mathcal{U}^{\left[ -\right] \dagger }
F\left( 0\right) \mathcal{U}^{\left[ +\right] }\right] \right\rangle\ ,\\
\label{eq:Fgg2}
\mathcal{F}_{gg}^{(2)} &=&2\int \frac{d\xi^+d^2{\boldsymbol\xi}}{(2\pi)^{3}p_A^{-}} e^{ix_2p_A^{-}\xi ^{+}-i k_t \cdot{\boldsymbol\xi}}
\frac{1}{N_c}\left\langle\textrm{Tr}\left[ F\left( \xi \right) \mathcal{U}^{\left[\square\right]\dagger} \right]
\textrm{Tr}\left[ F\left( 0\right) \mathcal{U}^{\left[ \square\right] }\right] \right\rangle\ ,\\
\label{eq:Fgg3}
\mathcal{F}_{gg}^{(3)} &=& 2\int \frac{d\xi^+d^2{\boldsymbol\xi}}{(2\pi )^{3}p_A^{-}} e^{ix_2p_A^{-}\xi ^{+}-i k_t \cdot{\boldsymbol\xi}}
\left\langle \text{Tr}\left[F\left( \xi \right) \mathcal{U}^{\left[+\right] \dagger }F\left( 0\right) \mathcal{U}^{\left[ +\right] }\right] \right\rangle
=x_2 G^{(1)}(x_2, k_t) \ .
\eea
The appropriate hard factors are constructed from the expressions corresponding
to the diagrams (1)-(6) depicted in Fig.~\ref{fig:gg2qqbar-diag}, using the
following formulas
\begin{eqnarray}
\label{eq:ggqq-onshell1-def}
H_{gg \to q \bar{q}}^{(1)}&=&
\frac{N_c}{2C_F} D_1 + \frac{N_c}{2C_F} D_2 +D_4 + 2 D_5 + 2D_6\,, \\
\label{eq:ggqq-onshell2-def}
H_{gg \to q \bar{q}}^{(2)}&=&-2 N_c^2 D_3 -D_4 -2 D_5 -2D_6\,, \\
H_{gg \to q \bar{q}}^{(3)}&=&
-\frac{1}{N_c^2-1} D_1 - \frac{1}{N_c^2-1} D_2 + 2 D_3\,.
\label{eq:ggqq-onshell3-def}
\end{eqnarray}
Again, the components $D_i = C_{u_i} h_i$ were computed
in~\cite{Dominguez:2011wm} (table III) and they were used there to determine the
hard factors $H_{gg \to q \bar{q}}^{(1,2)}$ in the large $N_c$ limit. 
Here, we generalize the results of~\cite{Dominguez:2011wm} to 
the full, finite-$N_c$ case. 
The calculation can be most readily done by exploiting crossing symmetry that
relates the $qg \to qg$ and $gg \to q\qbar$ channels. This allows for
identification of the diagrams between Figs.~\ref{fig:qg2qg-diag} and
\ref{fig:gg2qqbar-diag} and enables one to recycle the $D_i$ expressions
calculated in the previous subsection. For example, the expression corresponding
to the diagram (1) from Fig.~\ref{fig:gg2qqbar-diag}, with the incoming and the
outgoing legs connected, is identical to the already computed expression for the
diagram (4) from Fig.~\ref{fig:qg2qg-diag} (modulo a color averaging factor and
swapping of the momenta $p_1 \leftrightarrow p$). Similarly for all the other
diagrams.  That gives the following set of hard factors for the $gg\to q\qbar$
subprocess:
\begin{eqnarray}
\label{eq:ggqq-onshell1}
H_{gg \to q \bar{q}}^{(1)}&=&
\frac{1}{4C_F} \frac{( \hat{t}^2 + \hat{u}^2)^2}{\hat{s}^2\hat{u}\hat{t}}\ ,\\
\label{eq:ggqq-onshell2}
H_{gg \to q \bar{q}}^{(2)}&=& 
\frac{1}{2C_F} \frac{\hat{t}^2 + \hat{u}^2}{\hat{s}^2}\ ,\\
H_{gg \to q \bar{q}}^{(3)}&=&
- \frac{1}{4 N_c^2 C_F} \frac{\hat{t}^2 + \hat{u}^2}{\hat{t}\hat{u}}\ .
\label{eq:ggqq-onshell3}
\end{eqnarray}
Of the three hard factors, $H_{gg \to q \bar{q}}^{(i)}$, only two are
independent. The third hard factor, $H_{gg \to q \bar{q}}^{(3)}$, can be
expressed as\footnote{The same relation holds of course already at the level of
Eqs.~(\ref{eq:ggqq-onshell1-def})-(\ref{eq:ggqq-onshell3-def}).}
\be 
H_{gg \to q \bar{q}}^{(3)} = - \frac{1}{N_c^2} \left( H_{gg \to q \bar{q}}^{(1)} + H_{gg \to q \bar{q}}^{(2)} \right) \, .
\ee
Therefore, the cross section for quark-antiquark production can be rewritten with only two hard factors and two gluon distributions that are linear combinations of $\mathcal{F}_{gg}^{(1)}$, $\mathcal{F}_{gg}^{(2)}$ and $\mathcal{F}_{gg}^{(3)}$:
\begin{equation}
\frac{d\sigma^{pA\to q\bar{q} X}}{d^{2}P_td^{2} k_t
dy_{1}dy_{2}}=\frac{\alpha_s^2}{(x_1 x_2 s)^{2}}\ x_1f_{g/p}(x_1, \mu^2)
\left[ \Phi_{gg \to q\bar{q}}^{(1)}K_{gg\to q\bar{q}}^{(1)} + \Phi_{gg\to q\bar{q}}^{(2)}K_{gg\to q\bar{q}}^{(2)} \right]~.
\label{eq:new-fac-gg2qqbar}
\end{equation}
In the above, we defined the new gluon TMDs as
\bea 
\Phi_{gg\rightarrow q\overline{q}}^{\left(1\right)} &=& \frac{1}{N_{c}^{2}-1}\left(N_{c}^{2}\mathcal{F}_{gg}^{\left(1\right)}-\mathcal{F}_{gg}^{\left(3\right)}\right)\,, \\
\Phi_{gg\rightarrow q\overline{q}}^{\left(2\right)} &=& -N_{c}^{2}\mathcal{F}_{gg}^{\left(2\right)}+\mathcal{F}_{gg}^{\left(3\right)}\, ,
\eea
and the hard factors $K_{gg\to q\bar{q}}^{(i)}$ as:
\be 
K_{gg\to q\bar{q}}^{(1)} = \frac{N_c^2-1}{N_c^2} H_{gg\to q\bar{q}}^{(1)}~~~
{\text{and}} ~~~
K_{gg\to q\bar{q}}^{(2)} = - \frac{1}{N_c^2} H_{gg\to q\bar{q}}^{(2)} \, .
\label{eq:new-fac-gg2qq}
\ee
The explicit expressions  for the latter are given in Table~\ref{tab:Kfactors-on-shell}.

%------------------------------------
\subsection[The $gg \to gg$ channel]{The $\mathbold{gg \to gg}$ channel}
\label{sec:gg2gg-on-shell}

\begin{figure}[t]
  \begin{center}
    \includegraphics[width=0.8\textwidth]{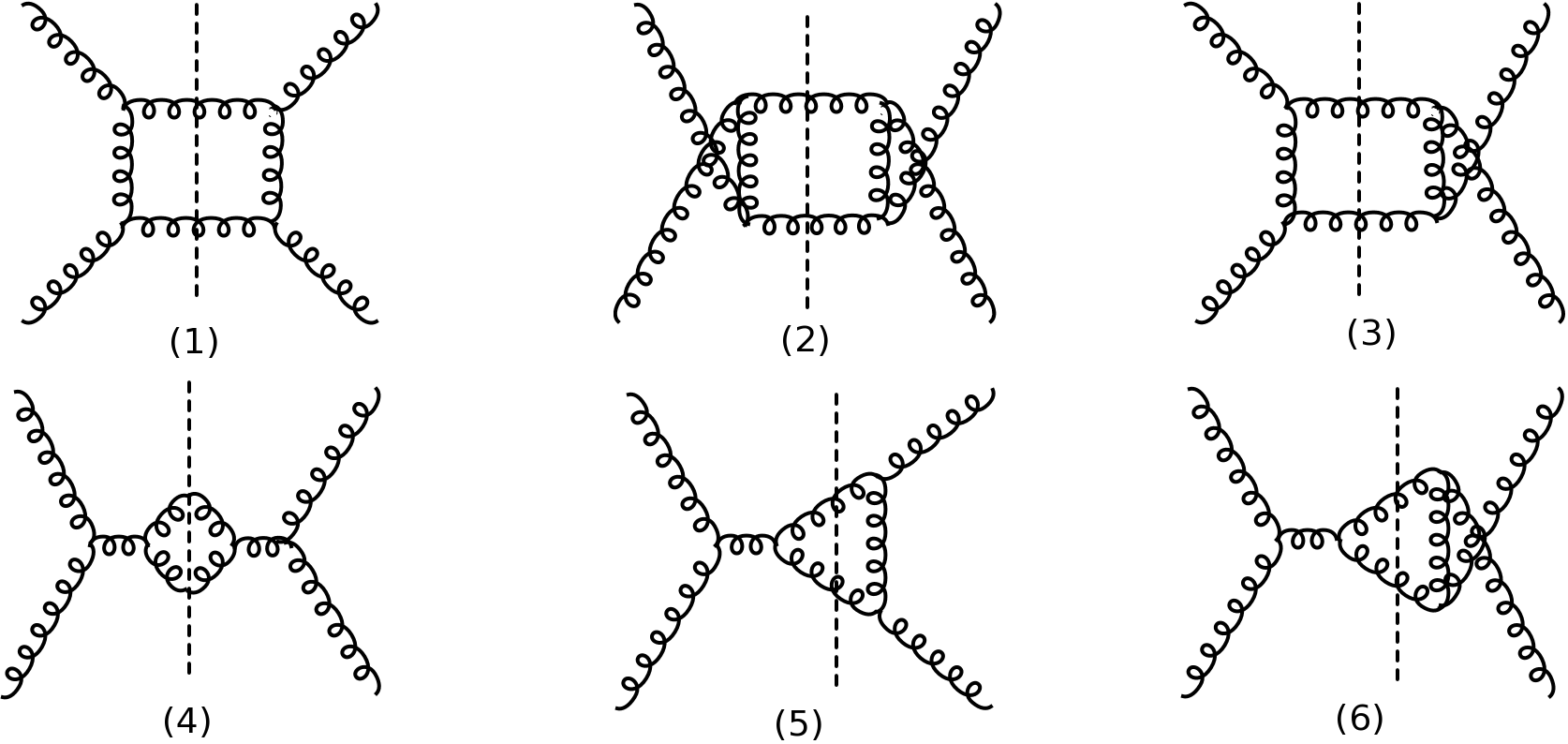}
  \end{center}
  \caption{Set of diagrams for the $gg \to gg$ subprocess involving only 3-gluon
  vertices. The mirror diagrams of (3), (5) and (6) give identical
  contributions.}
  \label{fig:gg2gg-3gdiag}
\end{figure}

\begin{figure}[t]
  \begin{center}
    \includegraphics[width=0.8\textwidth]{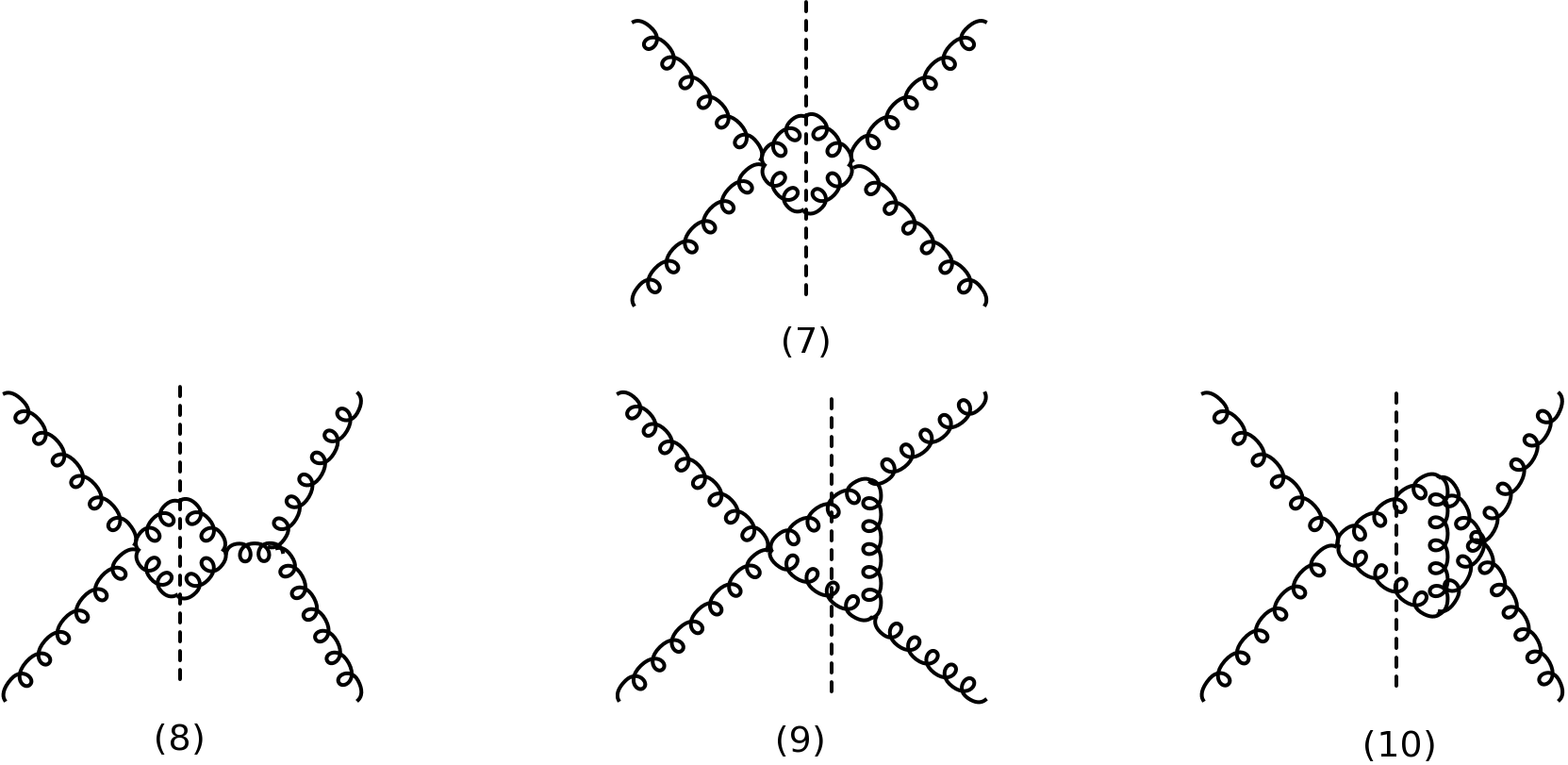}
  \end{center}
  \caption{Set of diagrams for the $gg \to gg$ subprocess involving 4-gluon
  vertex contributions. The mirror diagrams of (8), (9) and (10) give identical
  contributions.}
  \label{fig:gg2gg-4gdiag}
\end{figure}

Finally, the independent cut diagrams  for the $gg \to gg$ channel are given in
Figs.~\ref{fig:gg2gg-3gdiag} and \ref{fig:gg2gg-4gdiag}, and the
corresponding differential cross section for two-gluon production reads:
\begin{equation}
\frac{d\sigma^{pA\to gg X}}{d^{2}P_td^{2} k_t
dy_{1}dy_{2}}=\frac{\alpha_s^2}{(x_1x_2 s)^{2}}\ x_1f_{g/p}(x_1, \mu^2)
\sum_{i=1}^6 \mathcal{F}_{gg}^{(i)}H_{gg\to gg}^{(i)}\,.
\label{eq:gg2gg-on-shell1}
\end{equation}
The $\mathcal{F}_{gg}^{(1,2,3)}$ distributions are the same as the ones
introduced in the previous section in Eqs.~(\ref{eq:Fgg1})-(\ref{eq:Fgg3}).
The remaining three are~\cite{Bomhof:2006dp}:

\begin{eqnarray}
\mathcal{F}_{gg}^{(4)} &\!\!\!=\!\!\!& 2\!\int \frac{d\xi^+d^2{\boldsymbol\xi}}{(2\pi )^{3}p_A^{-}} e^{ix_2p_A^{-}\xi ^{+}-i k_t \cdot{\boldsymbol\xi}}
\left\langle \text{Tr}\left[F\left( \xi \right) \mathcal{U}^{\left[-\right] \dagger }F\left( 0\right) \mathcal{U}^{\left[ -\right] }\right] \right\rangle\,, 
\label{eq:Fgg4}
\\
\mathcal{F}_{gg}^{(5)} &\!\!\!=\!\!\!& 2\!\int \frac{d\xi^+d^2{\boldsymbol\xi}}{(2\pi )^{3}p_A^{-}} e^{ix_2p_A^{-}\xi ^{+}-i k_t \cdot{\boldsymbol\xi}}
\left\langle \text{Tr}\left[F\left( \xi \right) \mathcal{U}^{\left[\square \right] \dagger } \mathcal{U}^{\left[+\right] \dagger }
F\left( 0\right) \mathcal{U}^{\left[\square \right] } \mathcal{U}^{\left[ +\right] }\right] \right\rangle\,, 
\label{eq:Fgg5}
\\
\mathcal{F}_{gg}^{(6)} &\!\!\!=\!\!\!&2\!\int \frac{d\xi^+d^2{\boldsymbol\xi}}{(2\pi )^{3}p_A^{-}} e^{ix_2p_A^{-}\xi ^{+}-i k_t \cdot{\boldsymbol\xi}}
\left\langle \text{Tr}\left[ F\left( \xi \right)\mathcal{U}^{\left[ +\right] \dagger }F\left( 0\right) \mathcal{U}^{\left[ +\right] }\right]
\frac{\text{Tr}\left[ \mathcal{U}^{\left[\square \right] }\right] }{N_{c}} \frac{\text{Tr}\left[ \mathcal{U}^{\left[\square \right] }\right] }{N_{c}}\right\rangle.
\label{eq:Fgg6}
\end{eqnarray}
The associated hard factors are constructed as\footnote{Note that what is
called $H_{gg \to gg}^{(3)}$ in Ref.~\cite{Dominguez:2011wm} is now $H_{gg \to
gg}^{(6)}$. Out of six hard factors, only $H_{gg \to gg}^{(1)}$, $H_{gg \to
gg}^{(2)}$ and $H_{gg \to gg}^{(6)}$ survive in the large-$N_c$ limit.}:
\begin{eqnarray}
\label{eq:H1gggg-def}
H_{gg \to gg}^{(1)}&=&\frac{1}{2}D_1 + \frac{1}{2} D_2 +D_4 + 2D_5 + 2D_6\,,\\
\label{eq:H2gggg-def}
H_{gg \to gg}^{(2)}&=& 2 D_3 - D_4 -2 D_5 -2 D_6\,,\\
\label{eq:H6gggg-def}
H_{gg \to gg}^{(6)}&=& -\frac{N_c^2}{2}  H^{(3)}_{gg \to gg} = 
                       N_c^2 H_{gg \to gg}^{(4)}= N_c^2 H_{gg \to gg}^{(5)}  = 
\frac{1}{2}D_1 + \frac{1}{2} D_2 + 2 D_3\,.
\end{eqnarray}

The calculation of the $gg\to gg$ subprocess requires
inclusion of diagrams with four-gluon vertex.
Therefore, in general, the expressions $D_i$ in the above equations contain
contributions from both, the 3-gluon and 4-gluon vertex diagrams, the latter
shown in Fig.~\ref{fig:gg2gg-4gdiag}. 
The corresponding expressions were computed in~\cite{Dominguez:2011wm},
where they were used to determine the hard factors in the large-$N_c$ limit.
Below, we generalize the result of Ref.~\cite{Dominguez:2011wm} to the
case of finite-$N_c$ , with the help of the exact definitions given in
Eqs.~(\ref{eq:H1gggg-def})-(\ref{eq:H6gggg-def}). The six hard factors read
\begin{eqnarray}
\label{eq:H1gggg}
H_{gg \to gg}^{(1)}&=&
\frac{N_c}{C_F} \frac{(\hat{t}^2 + \hat{u}^2)(\hat{s}^2-\hat{t}\hat{u})^2}{\hat{u}^2\hat{t}^2\hat{s}^2}\ ,\\
\label{eq:H2gggg}
H_{gg \to gg}^{(2)}&=& =\frac{2N_c}{C_F} \frac{(\hat{s}^2-\hat{t}\hat{u})^2}{\hat{u}\hat{t}\hat{s}^2}\ ,\\
\label{eq:H6gggg}
H_{gg \to gg}^{(6)}&=& -\frac{N_c^2}{2}  H^{(3)}_{gg \to gg} = 
                        N_c^2 H_{gg \to gg}^{(4)}= N_c^2 H_{gg \to gg}^{(5)}  = 
\frac{N_c}{C_F} \frac{(\hat{s}^2-\hat{t}\hat{u})^2}{\hat{u}^2\hat{t}^2}\ .
\end{eqnarray}

To get further insight into the above results, we have performed an independent
calculation in a gauge with non-vanishing 4-gluon vertex contribution, with the
axial vectors defined as: 
\begin{equation}
  \begin{array}{llll}
  n = p   & \text{for the gluon } k\,,  &  \qquad
  n = k   & \text{for the gluon } p\,, \\
  n = p_2 & \text{for the gluon } p_1\,, &  \qquad
  n = p_1 &  \text{for the gluon } p_2\,.
  \end{array}
  \label{eq:gauge-gg2gg}
\end{equation}
The contributions to $D_i$s in this gauge, coming from diagrams with 3-gluon
vertices only and depicted in Fig.~\ref{fig:gg2gg-3gdiag}, are given in Table~\ref{tab:facgggg}.
\begin{table}[t]
\begin{center}
\begin{tabular}{ccc}
\hline 
 & $h_i^{(3)}$ & $C_i$ \\
\hline \\
(1) & $\displaystyle \frac{4 \hat{s}^6+4 \hat{t} \hat{s}^5+17 \hat{t}^2 \hat{s}^4+36 \hat{t}^3 \hat{s}^3+24 \hat{t}^4 \hat{s}^2+8 \hat{t}^5
\hat{s}+4 \hat{t}^6}{\hat{s}^4 \hat{t}^2}$ & 
$\displaystyle \frac{N_c}{2C_F}$ \\[15pt]
(2) & $\displaystyle \frac{\hat{s}^6+2 \hat{t} \hat{s}^5+33 \hat{t}^2 \hat{s}^4+60 \hat{t}^3 \hat{s}^3+44 \hat{t}^4 \hat{s}^2+16 \hat{t}^5
\hat{s}+4 \hat{t}^6}{\hat{s}^4 (\hat{s}+\hat{t})^2}$ & 
$\displaystyle \frac{N_c}{2C_F}$ \\[15pt]
(3) & $\displaystyle -\frac{2 \hat{s}^6-9 \hat{t} \hat{s}^5+19 \hat{t}^2 \hat{s}^4+48 \hat{t}^3 \hat{s}^3+4 \hat{t}^4 \hat{s}^2-24 \hat{t}^5
\hat{s}-8 \hat{t}^6}{2 \hat{s}^4 \hat{t} (\hat{s}+\hat{t})}$ & 
$\displaystyle \frac{N_c}{4C_F}$ \\[15pt]
(4) & $\displaystyle \frac{(\hat{s}+2 \hat{t})^2}{\hat{s}^2}$ & 
$\displaystyle \frac{N_c}{2C_F}$ \\[15pt]
(5) & $\displaystyle \frac{(\hat{s}+2 \hat{t}) \left(2 \hat{s}^3-3 \hat{t} \hat{s}^2-2 \hat{t}^2 \hat{s}+2 \hat{t}^3\right)}{2
\hat{s}^3 \hat{t}}$ & 
$\displaystyle \frac{N_c}{4C_F}$ \\[15pt]
(6) & $\displaystyle -\frac{(\hat{s}+2 \hat{t}) \left(\hat{s}^3-7 \hat{t} \hat{s}^2-8 \hat{t}^2 \hat{s}-2 \hat{t}^3\right)}{2
\hat{s}^3 (\hat{s}+\hat{t})}$ & 
$\displaystyle -\frac{N_c}{4C_F}$ \\ \\
\hline
\end{tabular}
\end{center}
\caption{Expressions for the $gg\to gg$ subprocess corresponding to diagrams
(1)-(6) of Fig.~\ref{fig:gg2gg-3gdiag}, hence containing only 3-gluon vertices,
in gauge (\ref{eq:gauge-gg2gg}) with non-vanishing 4-gluon vertex
contributions.}
\label{tab:facgggg}
\end{table}
\vspace{10pt}

\begin{figure}[t]
  \begin{center}
    \includegraphics[width=0.4\textwidth]{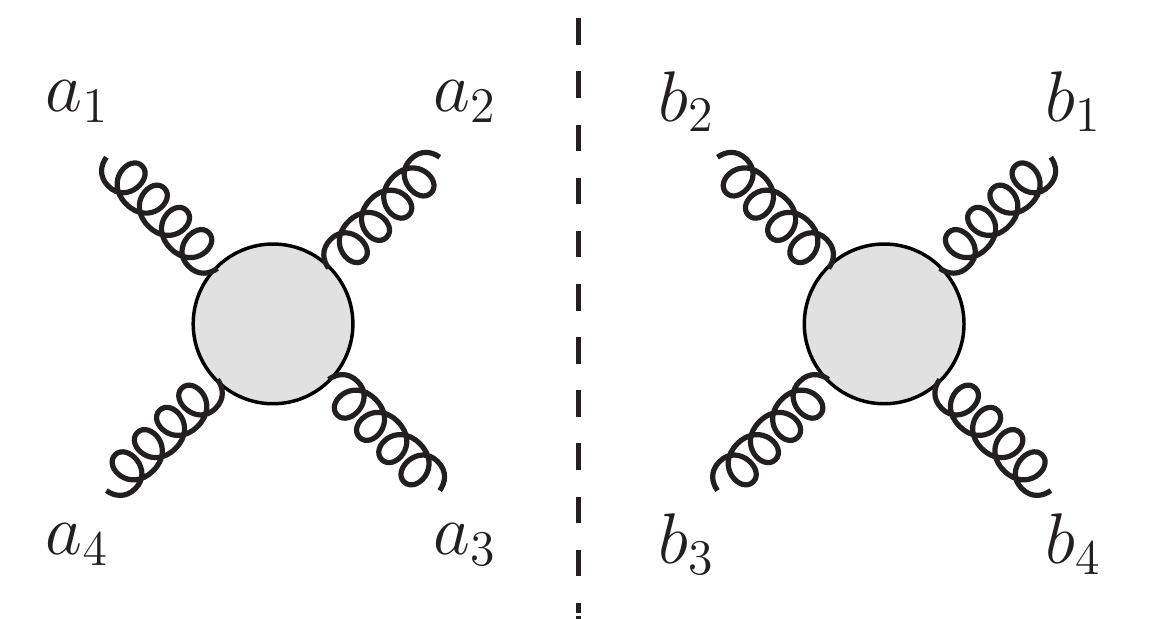}
  \end{center}
  \caption{Color indices for the cut four-gluon squared matrix element.
  }
  \label{fig:4gamp-coa}
\end{figure}

In order to add the 4-gluon vertex contribution and obtain a full
result for the $D_i$ coefficients, let us consider a general
4-gluon amplitude, shown on the left hand side of Fig.~\ref{fig:4gamp-coa}.
A 3-gluon vertex brings a single $SU(N)$ structure constant factor. Each
amplitude in Fig.~\ref{fig:gg2gg-3gdiag} consists of two 3-gluon vertices and
that results in three possible color factor products
\begin{equation}
 c_s \equiv  f^{a_1 c a_4} f^{c a_2 a_3}\,, \quad \quad  
 c_t \equiv  f^{a_1 a_2 c} f^{c a_3 a_4}\,, \quad \quad  
 c_u \equiv  f^{a_1 a_3 c} f^{c a_4 a_2}\,, \quad \quad
 \label{eq:3gv-color-factors}
\end{equation}
for the amplitudes with a gluon exchange in the $t$-, $s$- and $u$-channels,
respectively.  Each of the above amplitudes can now be written as
\begin{equation}
  \calM^{3g}_i = c_i\, \calA^{3g}_i\,,
  \label{eq:3gv-amp}
\end{equation}
where $i$ is either $t$, $s$ or $u$, $c_{i}$ is a color factor from
Eq.~(\ref{eq:3gv-color-factors}), and $\calA_{i}^{3g}$ is a corresponding
kinematic expression. The $3g$ superscript means that only 3-gluon vertices are
involved in the given amplitude.
Similarly, for the conjugate amplitudes, following the notation of
Fig.~\ref{fig:4gamp-coa}, we have
\begin{equation}
 \bar c_s \equiv  f^{b_1 c b_4} f^{c b_2 b_3}\,, \quad \quad  
 \bar c_t \equiv  f^{b_1 b_2 c} f^{c b_3 b_4}\,, \quad \quad  
 \bar c_u \equiv  f^{b_1 b_3 c} f^{c b_4 b_2}\,. \quad \quad
 \label{eq:3gv-color-factors-bar}
\end{equation}
That allows us to identify the color coefficients of the 3-gluon diagrams of
Fig.~\ref{fig:gg2gg-3gdiag} and write them in a compact form
\begin{equation}
  \begin{array}{ccc}
  (1)  \quad \leftrightarrow  \quad c_t \bar c_t\,, &  \qquad
  (2)  \quad \leftrightarrow  \quad c_u \bar c_u\,, &  \qquad
  (3)  \quad \leftrightarrow  \quad c_t \bar c_u\,, \\
  (4)  \quad \leftrightarrow  \quad c_s \bar c_s\,, &  \qquad
  (5)  \quad \leftrightarrow  \quad c_s \bar c_t\,, &  \qquad
  (6)  \quad \leftrightarrow  \quad c_s \bar c_u\,. \\
  \end{array}
  \label{eq:cc-3g}
\end{equation}

The $\order{\as^2}$ contributions from diagrams with 4-gluon vertex are depicted
in Fig.~\ref{fig:gg2gg-4gdiag}, where the first row shows the 4-gluon vertex
amplitude squared, and the second row gives the interference terms with the
three types of $\calM^{3g}$ amplitudes from Eq.~(\ref{eq:3gv-amp}).
A 4-gluon vertex amplitude contains all three color factor products of
Eq.~(\ref{eq:3gv-color-factors}) at once
\begin{equation}
\calM^{4g} = c_t \calA^{4g}_t + c_s \calA^{4g}_s + c_u \calA^{4g}_u\,.
\end{equation}
Therefore, all the contributions from Fig.~\ref{fig:gg2gg-4gdiag} can be
represented in the basis of the color factors defined in
Eq.~(\ref{eq:cc-3g}).
This allows us to distribute all the pieces of diagrams from
Fig.~\ref{fig:gg2gg-4gdiag} over the six $D_i$ expressions,
needed to calculate the hard factors
(\ref{eq:H1gggg-def})-(\ref{eq:H6gggg-def}), according to their color factors.
Hence, the full expressions are 
\begin{eqnarray}
D_1 & = & C_1 \left(h_1^{(3)} + 2 \calA^{4g}_t\calA^{3g}_t + 
                    \calA^{4g}_t \calA^{4g}_t\right)\,,  \\
D_2 & = & C_2 \left(h_2^{(3)} + 2 \calA^{4g}_u\calA^{3g}_u + 
                    \calA^{4g}_u \calA^{4g}_u\right)\,,  \\
D_3 & = & C_3 \left(h_3^{(3)} + \calA^{4g}_t\calA^{4g}_u + 
                  \calA^{4g}_t\calA^{3g}_u + \calA^{4g}_u \calA_t^{3g}\right)
		  \,,\\
D_4 & = & C_4 \left(h_4^{(3)} + 2 \calA^{4g}_s\calA^{3g}_s + 
                    \calA^{4g}_s\calA^{4g}_s\right)\,,  \\
D_5 & = & C_5 \left(h_5^{(3)} + \calA^{4g}_t\calA^{4g}_s + 
          \calA^{4g}_t\calA^{3g}_s + \calA^{4g}_s\calA^{3g}_t\right)\,, \\
D_6 & = & C_6 \left(h_6^{(3)} + \calA^{4g}_s\calA^{4g}_u + 
          \calA^{4g}_u\calA^{3g}_s + \calA^{4g}_s\calA^{3g}_u\right)\,. 
\label{eq:Disgg2gg}
\end{eqnarray}
The results for $D_i$s in the gauge (\ref{eq:gauge-gg2gg}) are summarized in in
Table~\ref{tab:Di}. Plugging those expressions into the hard factor definitions
(\ref{eq:H1gggg-def})-(\ref{eq:H6gggg-def}) leads to the results identical to
Eqs.~(\ref{eq:H1gggg})-(\ref{eq:H6gggg}). 

\begin{table}[t]
\begin{center}
\begin{tabular}{ccc}
\hline 
 & $D_i$ \\
\hline \\
(1) & $\displaystyle \frac{N_c \left(2 \hat{s}^4+2 \hat{s}^3 \hat{t}+3 \hat{s}^2 \hat{t}^2+8 \hat{s} \hat{t}^3+6 \hat{t}^4\right)}{C_F \hat{s}^2 \hat{t}^2 }$  \\[15pt]
(2) & $\displaystyle \frac{N_c \left(\hat{s}^4+4 \hat{s}^3 \hat{t}+15 \hat{s}^2 \hat{t}^2+16 \hat{s} \hat{t}^3+6 \hat{t}^4\right)}{C _F\hat{s}^2 
   (\hat{s}+\hat{t})^2}$
\\[15pt]
(3) & $\displaystyle -\frac{N_c \left(\hat{s}^4+\hat{s}^3 \hat{t}+7 \hat{s}^2 \hat{t}^2+12 \hat{s} \hat{t}^3+6 \hat{t}^4\right)}{2 C_F \hat{s}^2 \hat{t}  (\hat{s}+\hat{t})}$  
\\[15pt]
(4) & $\displaystyle \frac{N_c (\hat{s}+2 \hat{t})^2}{C_F \hat{s}^2 }$
\\[15pt]
(5) & $\displaystyle \frac{N_c (\hat{s}-2 \hat{t}) (\hat{s}+\hat{t}) (\hat{s}+2 \hat{t})}{2 C_F
   \hat{s}^2 \hat{t} }$ 
\\[15pt]
(6) & $\displaystyle -\frac{N_c\ \hat{t} (\hat{s}+2 \hat{t}) (3 \hat{s}+2 \hat{t})}{2 C_F \hat{s}^2  (\hat{s}+\hat{t})}$ 
\\ \\
\hline
\end{tabular}
\end{center}
\caption{Full expressions for the diagrams including three-gluon and four-gluon
vertex contributions in the gauge (\ref{eq:gauge-gg2gg}).}
\label{tab:Di}
\end{table}
\vspace{10pt}

We have already seen that not all of the six hard factors that arise in the
$gg\to gg$ subprocess are independent. As shown in Eq.~(\ref{eq:H6gggg-def}),
the expressions for $H_{gg \to gg}^{(3)}$, $H_{gg \to gg}^{(4)}$, $H_{gg \to
gg}^{(5)}$ and $H_{gg \to gg}^{(6)}$ differ only by numerical factors. On top of
that, when examining further Eqs.~(\ref{eq:H1gggg-def}), (\ref{eq:H2gggg-def})
and (\ref{eq:H6gggg-def}), we see that the hard factors $H_{gg \to gg}^{(1)}$,
$H_{gg \to gg}^{(2)}$ and $H_{gg \to gg}^{(6)}$ are linearly dependent, that is
\begin{equation}
  H_{gg \to gg}^{(6)}  = H_{gg \to gg}^{(1)} + H_{gg \to gg}^{(2)} \, .
  \label{eq:H1H2H6-relation}
\end{equation}
Hence, the cross section for two-gluon production from
Eq.~(\ref{eq:gg2gg-on-shell1}) can be
written in a much simpler, factorized form, with only two hard factors and two
gluon distributions
\begin{equation}
\frac{d\sigma^{pA\to gg X}}{d^{2}P_td^{2} k_t
dy_{1}dy_{2}}=\frac{\alpha_s^2}{(x_1x_2 s)^{2}}\ x_1f_{g/p}(x_1, \mu^2)
\left[ \Phi_{gg\to gg}^{(1)}K_{gg\to gg}^{(1)} + \Phi_{gg\to gg}^{(2)}K_{gg\to gg}^{(2)} \right]~.
\label{eq:gg2gg-on-shell2}
\end{equation}
In this channel, the new gluon TMDs, $\Phi_{gg\to gg}$, are defined as the following linear combinations of 
$\mathcal{F}_{gg}^{(1)}, \mathcal{F}_{gg}^{(2)},\ldots, \mathcal{F}_{gg}^{(6)}$:
\bea
\label{eq:Phigg11}
\Phi_{gg\to gg}^{(1)} &=&  \frac{1}{2}\left(\mathcal{F}_{gg}^{(1)} - \frac{2}{N_c^2} \mathcal{F}_{gg}^{(3)} + \frac{1}{N_c^2} \mathcal{F}_{gg}^{(4)} + \frac{1}{N_c^2} \mathcal{F}_{gg}^{(5)} + \mathcal{F}_{gg}^{(6)}\right) \,, \\
\label{eq:Phigg12}
\Phi_{gg\to gg}^{(2)} &=&  \mathcal{F}_{gg}^{(2)} - \frac{2}{N_c^2} \mathcal{F}_{gg}^{(3)} + \frac{1}{N_c^2} \mathcal{F}_{gg}^{(4)} + \frac{1}{N_c^2} \mathcal{F}_{gg}^{(5)} +  \mathcal{F}_{gg}^{(6)}  \,,
\eea
and the new hard factors are:
\be
K_{gg\to gg}^{(1)} = 2 H_{gg\to gg}^{(1)}\, ,~~~~~~
{\text{and}}~~~~~~K_{gg\to gg}^{(2)} = H_{gg\to gg}^{(2)}\, .
\label{eq:new-fac-gg2gg}
\ee
The explicit expressions  are given in Table~\ref{tab:Kfactors-on-shell}.
We note, that the above simplification occurs naturally when utilizing gauge
invariance from the start, as we will show in section \ref{sec:HelTMD}.

Finally, we point out that, in the large-$N_c$ limit, all the distributions that
were introduced in this section, $\mathcal{F}_{qg}^{(1)}$  $\mathcal{F}_{qg}^{(2)}$, $\mathcal{F}_{gg}^{(1)}$, $\mathcal{F}_{gg}^{(2)}$,
and $\mathcal{F}_{gg}^{(6)}$, can be written in terms of $xG^{(1)}$ and $xG^{(2)}$, and equivalence of formulas
(\ref{eq:new-fac-qg2qg}), (\ref{eq:new-fac-gg2qqbar}) and (\ref{eq:gg2gg-on-shell2}) with CGC results is obtained \cite{Dominguez:2011wm}.

Let use conclude that this part of our work brings two improvements to the
current state of the art for the TMD factorization in forward dijet production.
First of all, we have obtained finite-$N_c$ corrections to the hard factors
of Ref.~\cite{Dominguez:2011wm}. More importantly, however, we have eliminated
the redundancy in the number of gluon distributions needed to write a
factorization formula for this process, which now takes the compact form
\begin{equation}
\frac{d\sigma^{pA\rightarrow {\rm dijets}+X}}{d^{2}P_{t}d^{2}k_{t}dy_{1}dy_{2}}=\frac{\alpha_{s}^{2}}{(x_1 x_2 s)^{2}}
\sum_{a,c,d} x_{1}f_{a/p}(x_{1}, \mu^2)\sum_{i=1}^{2}K_{ag\to cd}^{(i)}\Phi_{ag\rightarrow cd}^{(i)}\ \frac{1}{1+\delta_{cd}}\ ,
\label{eq:gg2gg-new-onshell}
\end{equation}
with only two gluon distributions and two hard factors required in each channel.
Note that, as we shall discuss now, the incoming, small-$x$ gluon is kept on-shell. Eqs.~(\ref{eq:gg2gg-new-onshell})
will be further generalized to the case of the off-shell gluon in Section~\ref{sec:hard-factors-off-shell}.

\subsection{The $\mathbold{|k_t|\gg Q_s}$ limit}
\label{sec:dilute-on-shell}

Finally, let us consider the limit $|k_t|\gg Q_s$. This is the dilute limit considered in Section \ref{sec:dilute}, with the extra requirement that $|k_t|\ll|P_t|$, needed for the validity of those formula. In that limit, the transverse separation between the field operators in the definition of the gluon distribution is restricted to values much smaller than the distance over which the Fourier integrand varies, and the ${\boldsymbol\xi}$ dependence of the gauge links can be neglected. As a result, they simplify, and all the $\mathcal{F}_{ag}^{(i)}$ distributions coincide, except $\mathcal{F}_{gg}^{(2)}$ which vanishes. In terms of the $ \Phi_{ag\to cd}^{(1,2)}$ functions, all six distributions
also reduce to that one gluon distribution, which can therefore be identified with $\mathcal{F}_{g/A}/\pi$.

Then, for all channels, one can easily sum the surviving hard factors. In terms of diagrams, we always obtain $D_1+D_2+2D_3+D_4+2D_5+2D_6$, meaning that we recover the collinear matrix elements.  
Indeed we have (noting that $H_{gg \to gg}^{(3)}+H_{gg \to gg}^{(4)}+H_{gg \to gg}^{(5)}=0$):
\begin{eqnarray}
H_{qg \to qg}^{(1)}+H_{qg \to qg}^{(2)}=K_{qg \to qg}^{(1)}+K_{qg \to qg}^{(2)}
&=&\frac{\hat{s}^2+\hat{u}^2}{\hat{t}^2}-\frac{C_F}{N_c}\frac{\hat{s}^2+\hat{u}^2}{\hat{s}\hat{u}}=\frac1{g^4}
|\overline{{\cal M}_{qg\to qg}}|^2\,,~~~~~~~\\
H_{gg \to q\bar q}^{(1)}+H_{gg \to q\bar q}^{(3)}=K_{gg \to q\bar q}^{(1)}+K_{gg \to q\bar q}^{(2)}
&=&
\frac{1}{2N_c}\frac{\hat{t}^2+\hat{u}^2}{\hat{t}\hat{u}}-\frac{1}{2C_F}\frac{\hat{t}^2+\hat{u}^2}{\hat{s}^2}=\frac1{g^4}|\overline{{\cal
M}_{gg\to q\bar q}}|^2,~~~~~~~  \\
H_{gg \to gg}^{(1)}+H_{gg \to gg}^{(6)}=K_{gg \to gg}^{(1)}+K_{gg \to gg}^{(2)}
&=&\frac{2N_c}{C_F}\frac{(\hat{s}^2-\hat{t}\hat{u})^3}{\hat{s}^2\hat{t}^2\hat{u}^2}=\frac1{g^4}
|\overline{{\cal M}_{gg\to gg}}|^2\,.~~~~~~~
\end{eqnarray}
Therefore, we recover the HEF formula (\ref{eq:hef-formula}), except that, due
to the $|k_t|\ll|P_t|$ limit, the matrix elements are on-shell: the transverse
momentum of the incoming gluon, $k_t$, survives only in $\mathcal{F}_{g/A}$. In other words, we recover the standard high-$|P_t|$ limit:
\begin{equation}
  \frac{d\sigma^{pA\rightarrow {\rm dijets}+X}}{dy_1dy_2dP^2_t dk^2_{t}}
  =
\sum_{a,c,d}  \frac{1}{1+\delta_{cd}}\  x_1 f_{a/p}(x_1,\mu^2)\,\frac{d\hat\sigma_{ag\to cd}}{d\hat{t}}\ {\cal F}_{g/A}(x_2,k_t)\,,
  \label{eq:collinear-formula}
\end{equation}
with $d\hat\sigma_{ag\to cd}/d\hat{t}= |\overline{{\cal M}_{ag\to cd}}|^2 /
[16\pi (x_1x_2 s)^2]$, and where ${\cal F}_{g/A}(x_2,k_t)$ can be identified with $\partial/\partial k_t^2\ x_2 f_{g/A}(x_2,k_t^2)$, the derivative of the integrated gluon distribution.

In the following section, we shall restore the $k_t$ dependence of the hard
factors. This will extend our formulas such that they recover the full HEF
formula when the dilute limit is considered. As a result, we will obtain a
unified description, valid for generic forward dijet system with
$|p_{1t}|,|p_{2t}|\gg Q_s$, without any additional requirement on the magnitude of the transverse momentum imbalance $k_t$.

%-----------------------------------------------------------------------------
\section{Unified description of forward dijets in p+A collisions: \\
TMD factorization with off-shell hard factors}
\label{sec:hard-factors-off-shell}

We shall now generalize the hard factors that enter the TMD factorization
formula (\ref{eq:tmd-main}) to the case with one of the incoming gluons being off
the mass shell, as illustrated in Fig.~\ref{fig:4gamp-off-shell}.
As it has been already stated, the motivation to include the offshellness is to
be able to allow for configurations where the dijets are produced at any
azimuthal angle (of course before application of a jet algorithm that will
suppress very small angles and hence render the results finite). 

As can be seen in Fig.~\ref{fig:matrixelements}
(as an example we chose only purely gluonic matrix element but the same
structure occurs for the other channels), the on-shell matrix element misses
substantial contributions when the jets are produced at small angles near $\Delta\phi=0$
and at small rapidity differences $\Delta Y =|y_1 - y_2| \simeq 0$. In such
configurations, the matrix element develops a structure that is
divergent and it is suppressed only by a jet algorithm, which has to be applied
in order to ensure two-jet configurations \cite{vanHameren:2014lna}.
The matrix elements squared we are after, \ie $g g^*\rightarrow gg$,
$g g^*\rightarrow q\bar q$ and $q g^*\rightarrow qg$, can
be extracted from the high energy limit (or eikonal limit) of $q\,g\rightarrow
q\,g\,g$ and  $q\,g\rightarrow q\,\bar q\,q$ and $q\,q'\rightarrow q\,q'\,g$ \cite{vanHameren:2013}. In
this approach the quark $q$ is an auxiliary line to which the initial state
off-shell gluon $g^*$ couples eikonally.
 
The high energy factorization is a direct procedure where one uses the
standard Feynman rules for all vertices and color factors, and fixes the light-cone
gauge for the on-shell gluons, using a gauge vector given by the longitudinal
component of the off-shell, initial-state gluon's momentum.  In particular, if we apply the
high energy factorization to the process we are after, we set the gauge
vector to $n=p_A$, where $p_A$ is the target four-momentum, as defined in
Fig.~\ref{fig:dijets-pA} and Eq.~(\ref{eq:pp-pA-defs}).
\begin{figure}[t]
  \begin{center}
    \includegraphics[width=0.22\textwidth]{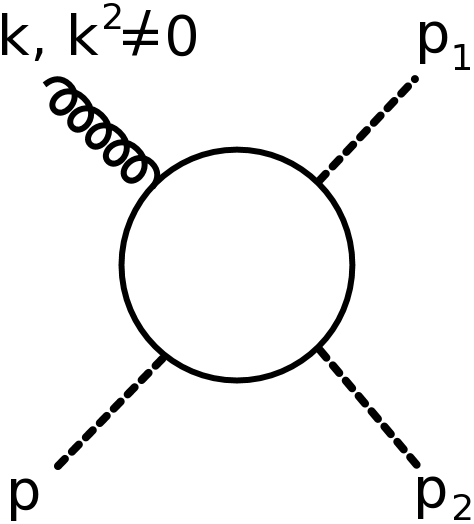}
  \end{center}
  \caption{Four-parton amplitude with the incoming, small-$x$, off-shell gluon.}
  \label{fig:4gamp-off-shell}
\end{figure}
Furthermore, the prescription is to associate with the off-shell gluon a
longitudinal polarization vector, called \emph{nonsense
polarization} \cite{Gribov:1984tu}, of the form~\footnote{
The $\sqrt 2$ factor in Eq.~(\ref{eq:pol-off-shell}) follows from a convention.
It allows for use of the on-shell-like factor $\frac12$ in averaging over
polarization, while calculating matrix elements squared, even in the case of
the off-shell gluon, where the actual number of polarizations in the high energy
limit is 1.
}
\begin{equation}
  \label{eq:pol-off-shell}
\epsilon_\mu^0=\frac{i \sqrt 2\, x_2}{|k_t|} p_{A\, \mu}\,.
\end{equation}
As elaborated in Ref.~\cite{Catani:1990eg}, longitudinally polarized gluons
provide the dominant contribution to the cross section in the high energy limit.
In the square amplitude, this leads to the polarization tensor of the form \cite{Catani:1990eg}
\begin{equation}
\epsilon^0_\mu \epsilon^{0\,*}_\nu\,
%  \bigg |_\text{off-shell gluon}
  = \frac{-2\, x_2^2}{k^2}\, p_{A\, \mu}\, p_{A\,\nu}\,,
\end{equation}
In the above, $x_2=k_\mu p^\mu/p_{\!A \nu} p^\nu$, which follows directly from the definition in Eq.~(\ref{eq:k-4-vec}).
The sum over polarizations of the on-shell gluons takes the standard form,  with the gauge vector given by $p_{A}$
\begin{equation}
  \sum_{\lambda=\pm} \epsilon^{\lambda}_\mu \epsilon^{\lambda *}_\nu\,
%  \bigg |_\text{on-shell gluon} = 
 =  g_{\mu \nu} - \frac{p_{\!A \mu} q_\nu + q_\mu p_{\!A \nu}}{q^\rho p_{\!A \rho}}\,,
\end{equation}
where, depending on the channel, $q=p,\, p_1$ or $p_2$, \cf
Eq.~(\ref{eq:gauge-gg2gg}).

Let us note that the procedure outlined above defines the hard process in a
gauge invariant manner only when a special choice for polarization vectors of
the on-shell gluons is taken. In an arbitrary gauge, for internal and external
gluon lines, more sophisticated methods have to be used, see \eg
 \cite{vanHameren:2013,Lipatov:1995pn,Antonov:2004hh,Kotko:2014aba,vanHameren:2014iua}.

\begin{figure}
  \begin{center}
    \includegraphics[width=0.4\textwidth]{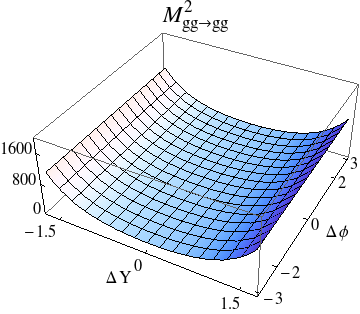}
  \includegraphics[width=0.4\textwidth]{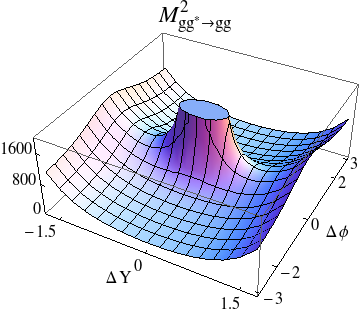}
  \end{center}
  \caption{
  Matrix elements squared for $gg \to gg$ scattering with $p_{t1}=p_{t2}=4\,
  \GeV$ and $\alpha_s=0.2$. Left: the on-shell case. Right: the off-shell case.
  $\Delta Y$ and $\Delta \phi$ are, respectively, the differences in rapidity
  and azimuthal angle of the two outgoing gluons.
  }
  \label{fig:matrixelements}
\end{figure}

To present our results in a compact form, with direct relation to the on-shell
formulas from Section~\ref{sec:tmd-on-shell}, in addition to the 
standard Mandelstam variables given by Eqs.~(\ref{eq:mandelstam}), which now,
however, sum up to $\shat + \that + \uhat = k_T^2$,
we introduce
 their barred versions, defined only with the longitudinal component
of the off-shell gluon
\begin{subequations}
  \begin{align}
  \tls & \ = \ (x_2 p_{\! A} +p)^2 \ =
         \   \ \frac{|P_t|^2}{z(1-z)}+|k_t|^2=x_1x_2s\,,\\
  \tlt & \ = \ (x_2 p_{\! A} -p_1)^2=-z\tls\,, \\
  \tlu & \ = \ (x_2 p_{\! A} -p_2)^2=-(1-z)\tls\,,
  \end{align}
  \label{eq:gen-mandelstam}
\end {subequations}
which are related via the equation
\begin{equation}
  \tls + \tlt + \tlu = 0\,.
\end{equation}
In the on-shell limit, $k_T^2 \to 0$, the variables defined above recover the
standard Mandelstam variables from Eq.~(\ref{eq:mandelstam})
\begin{equation}
  \lim_{|k_t| \to 0} (\tls - \shat) = 0\,, \qquad \quad
  \lim_{|k_t| \to 0} (\tlt - \that) = 0\,, \qquad \quad
  \lim_{|k_t| \to 0} (\tlu - \uhat) = 0\,.
  \label{eq:gen-mandelstam-limit}
\end{equation}

As a consistency check, we have verified that, for all three subprocesses, the
off-shell amplitudes that shall be used to build the hard factors in the
remaining part of this section are identical to
those first calculated in Ref.~\cite{Deak:2009xt}.

From this point onwards, we shall discuss our results only in terms of the new
$K^{(i)}$ hard factors and the new factorization formulas from
Eqs.~(\ref{eq:new-fac-qg2qg}), (\ref{eq:new-fac-gg2qqbar}) and
(\ref{eq:gg2gg-on-shell2}). The results for the old hard factors, $H^{(i)}$, in
the off-shell case are given in Appendix~\ref{app:off-shell-expr} for
completeness.

%------------------------------------
\subsection[The $qg^* \to qg$ channel]{The $\mathbold{qg^* \to qg}$ channel}

The off-shell hard factors for this channel are obtained using definitions given
in Eq.~(\ref{eq:Kqg2qgon}) and then Eqs.~(\ref{eq:H1-qg-on-shell-def}) and (\ref{eq:H2-qg-on-shell-def}).  The
corresponding $D_i$ expressions are collected in Appendix
\ref{app:off-shell-expr} in Table~\ref{tab:Di-off-shell-qg2qg}.
The two hard factors read
\begin{eqnarray}
  K^{(1)}_{qg^*\to qg} & = & 
  - \frac{\tls^2 + \tlu^2}{2 \tlt \that \shat \uhat}
    \Bigg[\tlu \uhat + \frac{\tls \shat - \tlt \that}{N_c^2}\Bigg]\,,\\
  K^{(2)}_{qg^*\to qg} & = & 
  - \frac{C_F}{ N_c} \,
  \frac{\tls\left(\tls^2 + \tlu^2\right)}{ \tlt \that \uhat }\,.
\end{eqnarray}

In the limit $|k_t| \to 0$, simplification given by
Eq.~(\ref{eq:gen-mandelstam-limit}) occurs and
the above formulas manifestly recover the on-shell
results from Table~\ref{tab:Kfactors-on-shell}.

%------------------------------------
\subsection[The $gg^* \to q\bar q$ channel]
{The $\mathbold{gg^* \to q\bar q}$ channel}

The off-shell hard factors are obtained using definitions given in 
Eq.~(\ref{eq:new-fac-gg2qq})
and then 
Eqs.~(\ref{eq:ggqq-onshell1-def}), 
(\ref{eq:ggqq-onshell2-def}) and (\ref{eq:ggqq-onshell3-def}).  The
corresponding $D_i$ expressions are collected in Appendix
\ref{app:off-shell-expr} in Table~\ref{tab:Di-off-shell-gg2qq}. 
The two hard factors take the following compact form 
\begin{eqnarray}
  \label{eq:K1gg2qqoff}
  K_{gg^*\rightarrow q\qbar}^{(1)}&=&
    \frac{1}{2N_c} \frac{\tlt^2+\tlu^2}{ \tls \shat \that \uhat}
    \left[\tlu \uhat + \tlt \that \right]\,, \\
  \label{eq:K2gg2qqoff}
  K_{gg^*\rightarrow q\qbar}^{(2)}&=&
    \frac{1}{4 N_c^2 C_F} \frac{\tlt^2+\tlu^2}{ \tls \shat \that \uhat }
    \left[\tlu \uhat + \tlt \that - \tls \shat \right]\,.
\end{eqnarray}
Again, following Eq.~(\ref{eq:gen-mandelstam-limit}), it is manifest that the
above hard factors reduce to those given in Table~\ref{tab:Kfactors-on-shell},
in the limit $|k_t|\to 0$.

%------------------------------------
\subsection[The $gg^* \to gg$ channel]{The $\mathbold{gg^* \to gg}$ channel}
\label{Sec:hardfactgggg}

In the gauge chosen for our calculation, all the squared diagrams  and
interference terms that involve a 4-gluon vertex are identically zero.
The corresponding  $D_i$s are given in Table~\ref{tab:Di-off-shell-gg2gg}
of Appendix \ref{app:off-shell-expr}.
Using the combinations from Eqs.~(\ref{eq:H1gggg-def})-(\ref{eq:H6gggg-def})
and then the definition from Eq.~(\ref{eq:new-fac-gg2gg}) 
leads to the following set of the off-shell hard factors
\begin{eqnarray}
  \label{eq:Kgg2ggoff1}
  K^{(1)}_{gg^*\to gg} & = & 
  \frac{2 N_c}{C_F}\, 
  \frac{(\tls^2-\tlt \tlu)^2} {\tlt\that \tlu\uhat \tls\shat }\,
  \left[\tlu\uhat + \tlt\that \right]\,, \\
  \label{eq:Kgg2ggoff2}
  K^{(2)}_{gg^*\to gg} & = & 
  -\frac{N_c}{C_F}\, 
  \frac{(\tls^2-\tlt \tlu)^2} {\tlt\that \tlu\uhat \tls\shat }\,
  \left[\tlu\uhat + \tlt\that - \tls\shat \right]\,.
\end{eqnarray}
The on-shell limit is again manifest, with the above equations reducing to
those from Table~\ref{tab:Kfactors-on-shell} as $|k_t|\to 0 $.
%

%-----------------------------------------------------------------------------
\section{Helicity method for TMD amplitudes}

\label{sec:HelTMD}

In the preceding sections, the hard factors accompanying the gluon
densities $\mathcal{F}^{(i)}_{ag}$ were calculated from the squared diagrams presented in Figs.~\ref{fig:qg2qg-diag}-\ref{fig:gg2gg-4gdiag}.
This procedure has certain  drawbacks, especially when one would like
to consider more complicated processes. For multiparticle processes,
the color decompositions and helicity method~\cite{Mangano:1990by,
Dixon:2013uaa} are now considered as the most effective ways to deal with them.
Moreover, it is not obvious how the gauge invariance comes into play for the
separate diagrams from Figs.~\ref{fig:qg2qg-diag}-\ref{fig:gg2gg-4gdiag}
contributing to the hard factors. In the color decomposition method, the
so-called \emph{color ordered amplitudes} are gauge invariant from the start and one
can use them directly to construct hard factors.

In view of the above, and to cross-check the results from Section
\ref{sec:hard-factors-off-shell}, we will give an alternative procedure to obtain the factorization
formulas with off-shell gluon. To this end, we shall need TMD gluon densities corresponding
to color decomposition of amplitudes and the color-ordered amplitudes
themselves.

%------------------------------------
\subsection{Color decompositions}

Let us recall some basic facts about the color decompositions.
We refer to \cite{Mangano:1990by,Dixon:2013uaa} for more details.

We first consider a gluon amplitude $\mathcal{M}^{a_{1}\ldots a_{N}}\left(\varepsilon_{1}^{\lambda_{1}},\ldots,\varepsilon_{N}^{\lambda_{N}}\right)$,
where $a_{1},\ldots,a_{N}$ are the external, adjoint color quantum numbers,
the $\varepsilon_{i}^{\lambda_{i}}$ is a polarization vector for
a gluon $i$ having momentum $k_{i}$ and helicity $\lambda_{i}=\pm$.
The fundamental color decomposition reads 
\begin{equation}
\mathcal{M}^{a_{1}\ldots
a_{N}}\left(\varepsilon_{1}^{\lambda_{1}},\ldots,\varepsilon_{N}^{\lambda_{N}}\right)=\sum_{\sigma\in
S_{N-1}}\mathrm{Tr}\left(t^{a_{1}}t^{a_{\sigma_{2}}}\ldots
t^{a_{\sigma_{N}}}\right)\,\mathcal{M}\left(1^{\lambda_{1}},\sigma_{2}^{\lambda_{\sigma{2}}}\ldots,\sigma_{N}^{\lambda_{\sigma{N}}}\right),\label{eq:ColorDecompGlue}
\end{equation}
where the sum is over a set  $S_{N-1}$ of all non-cyclic permutations
of $\left\{ 1,\dots,N\right\} $. The coefficients of the expansion
define color ordered -- or dual -- amplitudes. They possess several
useful properties. First of all, they are gauge invariant. Second,
there are certain relations between dual amplitudes%
.
Indeed,
the following adjoint color decomposition involves only $\left(N-2\right)!$
different amplitudes~\cite{DelDuca:1999rs}
\begin{equation}
\mathcal{M}^{a_{1}\ldots
a_{N}}\left(\varepsilon_{1}^{\lambda_{1}},\ldots,\varepsilon_{N}^{\lambda_{N}}\right)=\sum_{\sigma\in
S_{N-2}}\left(F^{a_{\sigma_2}}\ldots
F^{a_{\sigma_{N-1}}}\right)_{a_{1}a_{N}}\,\mathcal{M}\left(1^{\lambda_{1}},\sigma_{2}^{\lambda_{\sigma_2}}, \ldots,
\sigma_{N-1}^{\lambda_{\sigma_{N-1}}}, N^{\lambda_N}\right),
\label{eq:ColorDecompGlueAdj}
\end{equation}
where $\left(F^{a}\right)_{bc}=f_{abc}$.

Consider now an amplitude involving a quark anti-quark pair $\mathcal{M}^{D_{1}a_{2}\ldots a_{N-1}\overline{D}_{N}}$
%%%%%\left(\lambda_{1},\varepsilon_{2}^{\lambda_{2}},\ldots,\varepsilon_{N-1}^{\lambda_{N-1}},\lambda_{N}\right)$,
where $D_{i}$, $\overline{D}_{j}$ are the color and the anti-color of the
quark and the anti-quark, respectively. The color decomposition reads
\begin{multline}
\mathcal{M}^{D_{1}a_{2}\ldots a_{N-1}\overline{D}_{N}}\left(\lambda_{1},\varepsilon_{2}^{\lambda_{2}},\ldots,\varepsilon_{N-1}^{\lambda_{N-1}},\lambda_{N}\right)=\\
\sum_{\sigma\in S_{N-2}}\left(t^{a_{\sigma_2}}\ldots
t^{a_{\sigma_{N-1}}}\right)_{D_{1}\overline{D}_{N}}\,\mathcal{M}\left(1^{\lambda_{1}},\sigma_{2}^{\lambda_{\sigma_2}}\ldots,\sigma_{N-1}^{\lambda_{\sigma_{N-1}}},N^{\lambda_{N}}\right).\label{eq:ColorDecompQuark}
\end{multline}
Now $\lambda_1$ and $\lambda_N$ are helicities of the quark and the anti-quark. For amplitudes involving more quark anti-quark pairs the decomposition
is more complicated and we refer to \cite{Mangano:1990by} for details.

It is important to note that the above color decompositions work
also for the case when one of the gluons is off-shell.

\subsection{Gluon TMDs for color ordered amplitudes}

Let us now find the gluon TMDs corresponding to the color ordered amplitudes
squared, as defined in the previous subsection. We constraint ourselves
to the $2\rightarrow2$ processes case considered in this paper.

Let us first consider the $g\left(k_{4}\right)g^*\left(k_{1}\right)\rightarrow g\left(k_{3}\right)g\left(k_{2}\right)$
process. For the purpose of this and next subsections we have assigned
a new set of momenta to the partons. This assignment differs from the one
used before but it is more convenient when dealing with color ordered
amplitudes. The correspondence is achieved by the following relations:
$k_{1}\leftrightarrow k$, $k_{2}\leftrightarrow p_{1}$, $k_{3}\leftrightarrow p_{2}$,
$k_{4}\leftrightarrow p$. Moreover, for the off-shell momentum we
adopt a notation 
\begin{equation}
k_{1}=n_{1}+k_{T}~.\label{eq:k1_def}
\end{equation}
The color decomposition of the four gluon amplitude reads 
\begin{multline}
\mathcal{M}_{gg^{*}\rightarrow gg}^{a_{1}a_{2}a_{3}a_{4}}\left(n_{1},\varepsilon_{2}^{\lambda_{2}},\varepsilon_{3}^{\lambda_{3}},\varepsilon_{4}^{\lambda_{4}}\right)=f_{a_{1}a_{2}c}f_{ca_{3}a_{4}}\,\mathcal{M}_{gg^{*}\rightarrow gg}\left(1^{*},2^{\lambda_{2}},3^{\lambda_{3}},4^{\lambda_{4}}\right) \\
+f_{a_{1}a_{3}c}f_{ca_{2}a_{4}}\,\mathcal{M}_{gg^{*}\rightarrow gg}\left(1^{*},3^{\lambda_{3}},2^{\lambda_{2}},4^{\lambda_{4}}\right),\label{eq:Mg*g1}
\end{multline}
where $n_{1}$ is placed for the off-shell gluon instead of a polarization
vector (in fact it plays a similar role). As far as dual amplitudes
are concerned, we indicate the off-shell gluon by a star. 
In Table~\ref{tab:TMDsgggg}. we calculate 
the gluon TMDs  
that correspond to the color structures exposed in (\ref{eq:Mg*g1})
(after squaring). 
They agree with the gluon TMDs calculated in \cite{Bomhof:2006dp} 
and listed in rows 1 and 3 of Table 8 of \cite{Bomhof:2006dp}. That table defines one more gluon TMD (the row 2) which however is redundant. Clearly, the color decomposition (\ref{eq:Mg*g1}) gives all the necessary color structures and already incorporates 
 the gauge invariance. In summary, the two gluon TMD listed in Table~\ref{tab:TMDsgggg} are the only relevant
TMDs and correspond to the two independent gauge
invariant amplitudes squared and their interference.

\begin{table}
\begin{doublespace}
\begin{centering}
\begin{tabular}{c|>{\centering}p{0.4\textwidth}}
\hline 
color-ordered amplitude squared & gluon TMD\tabularnewline
\hline\\[-1.5em]
$\left|\mathcal{M}_{gg^{*}\rightarrow gg}\left(1^{*},2^{\lambda_{2}},3^{\lambda_{3}},4^{\lambda_{4}}\right)\right|^{2}$ & 
\multirow{2}{0.4\textwidth}{$\Phi_{gg\rightarrow gg}^{\left(1\right)}=\frac{1}{2N{}_{c}^{2}}\big(N_{c}^{2}\mathcal{F}_{gg}^{\left(1\right)}-2\mathcal{F}_{gg}^{\left(3\right)}$\\\hfill
$+\mathcal{F}_{gg}^{\left(4\right)}+\mathcal{F}_{gg}^{\left(5\right)}+N_{c}^{2}\mathcal{F}_{gg}^{\left(6\right)}\big)$}\tabularnewline
%\cline{1-1} 
$\left|\mathcal{M}_{gg^{*}\rightarrow gg}\left(1^{*},3^{\lambda_{3}},2^{\lambda_{2}},4^{\lambda_{4}}\right)\right|^{2}$ & 
%\tabularnewline
\\[0.5em]
\hline 
\\[-1.5em]
$\mathcal{M}_{gg^{*}\rightarrow gg}\left(1^{*},2^{\lambda_{2}},3^{\lambda_{3}},4^{\lambda_{4}}\right)\mathcal{M}_{gg^{*}\rightarrow gg}^{*}\left(1^{*},3^{\lambda_{3}},2^{\lambda_{2}},4^{\lambda_{4}}\right)$ & \multirow{2}{0.4\textwidth}{$\Phi_{gg\rightarrow gg}^{\left(2\right)}=\frac{1}{N{}_{c}^{2}}\big(N_{c}^{2}\mathcal{F}_{gg}^{\left(2\right)}-2\mathcal{F}_{gg}^{\left(3\right)}$\\\hfill$+\mathcal{F}_{gg}^{\left(4\right)}+\mathcal{F}_{gg}^{\left(5\right)}+N_{c}^{2}\mathcal{F}_{gg}^{\left(6\right)}\big)$}\tabularnewline
%\cline{1-1} 
$\mathcal{M}_{gg^{*}\rightarrow
gg}^{*}\left(1^{*},2^{\lambda_{2}},3^{\lambda_{3}},4^{\lambda_{4}}\right)\mathcal{M}_{gg^{*}\rightarrow
gg}\left(1^{*},3^{\lambda_{3}},2^{\lambda_{2}},4^{\lambda_{4}}\right)$ & 
%\tabularnewline
\\[0.5em]
\hline 
\end{tabular}
\par\end{centering}
\end{doublespace}
\caption{Gluon TMDs accompanying the color-ordered amplitudes for $gg^{*}\rightarrow gg$
process. It has been assumed that TMDs are real.
The $\GG^{(i)}_{gg}$ distributions are defined in 
Eqs.~(\ref{eq:Fgg1}), (\ref{eq:Fgg2}), (\ref{eq:Fgg3}) and
in Eqs.~(\ref{eq:Fgg4}), (\ref{eq:Fgg5}), (\ref{eq:Fgg6}).
\label{tab:TMDsgggg}}

\end{table}

Now, let us turn to the $g\left(k_{4}\right)g^{*}\left(k_{1}\right)\rightarrow \overline{q}\left(k_{3}\right) q\left(k_{2}\right) $
process. The color decomposition reads
\begin{multline}
\mathcal{M}_{gg^{*}\rightarrow q\overline{q}}^{D_{2}a_{1}a_{4}\overline{D}_{3}}\left(\lambda_{2},n_{1},\varepsilon_{4}^{\lambda_{4}},\lambda_{3}\right)
=\left(t^{a_{1}}t^{a_{4}}\right)_{D_{2}\overline{D}_{3}}\mathcal{M}_{gg^{*}\rightarrow q\overline{q}}\left(2^{\lambda_{2}},1^{*},4^{\lambda_{4}},3^{\lambda_{3}}\right)\\
+\left(t^{a_{4}}t^{a_{1}}\right)_{D_{2}\overline{D}_{3}}\mathcal{M}_{gg^{*}\rightarrow
q\overline{q}}\left(2^{\lambda_{2}},4^{\lambda_{4}},1^{*},3^{\lambda_{3}}\right)\,.
\end{multline}
The gluon TMDs corresponding to the color structures appearing after
squaring this equation are gathered in Table \ref{tab:TMDsggqq}.
They correspond to rows 1 and 5 of Table 7 in \cite{Bomhof:2006dp}. Again,
we have only two independent TMDs that are needed.

\begin{table}[t]
\begin{doublespace}
\begin{centering}
\begin{tabular}{c|c}
\hline 
color-ordered amplitude squared & gluon TMD\tabularnewline
\hline\\[-1.5em]
$\left|\mathcal{M}_{gg^{*}\rightarrow q\overline{q}}\left(2^{\lambda_{2}},1^{*},4^{\lambda_{4}},3^{\lambda_{3}}\right)\right|^{2}$ & \multirow{2}{*}{$\Phi_{gg\rightarrow q\overline{q}}^{\left(1\right)}=\frac{1}{N_{c}^{2}-1}\left(N_{c}^{2}\mathcal{F}_{gg}^{\left(1\right)}-\mathcal{F}_{gg}^{\left(3\right)}\right)$}
\tabularnewline
%\cline{1-1} 
$\left|\mathcal{M}_{gg^{*}\rightarrow q\overline{q}}\left(2^{\lambda_{2}},4^{\lambda_{4}},1^{*},3^{\lambda_{3}}\right)\right|^{2}$ & 
%\tabularnewline
\\[0.5em]
\hline 
\\[-1.5em]
$\mathcal{M}_{gg^{*}\rightarrow q\overline{q}}\left(2^{\lambda_{2}},1^{*},4^{\lambda_{4}},3^{\lambda_{3}}\right)\mathcal{M}_{gg^{*}\rightarrow q\overline{q}}^{*}\left(2^{\lambda_{2}},4^{\lambda_{4}},1^{*},3^{\lambda_{3}}\right)$ & \multirow{2}{*}{$\Phi_{gg\rightarrow q\overline{q}}^{\left(2\right)}=-N_{c}^{2}\mathcal{F}_{gg}^{\left(2\right)}+\mathcal{F}_{gg}^{\left(3\right)}$}\tabularnewline
%\cline{1-1} 
$\mathcal{M}_{gg^{*}\rightarrow q\overline{q}}^{*}\left(2^{\lambda_{2}},1^{*},4^{\lambda_{4}},3^{\lambda_{3}}\right)\mathcal{M}_{gg^{*}\rightarrow q\overline{q}}\left(2^{\lambda_{2}},4^{\lambda_{4}},1^{*},3^{\lambda_{3}}\right)$ & 
%\tabularnewline
\\[0.5em]
\hline 
\end{tabular}
\par\end{centering}
\end{doublespace}
\caption{Gluon TMDs accompanying the color-ordered amplitudes for $gg^{*}\rightarrow q\overline{q}$
process. It has been assumed that correlators are real.
The $\GG^{(i)}_{gg}$ distributions are defined in 
Eqs.~(\ref{eq:Fgg1}), (\ref{eq:Fgg2}) and ~(\ref{eq:Fgg3}).
\label{tab:TMDsggqq}}
\end{table}

For the process $q\left(k_{4}\right)g^{*}\left(k_{1}\right)\rightarrow
q\left(k_{3}\right)g\left(k_{2}\right)$,
the color decomposition reads 
\begin{multline}
\mathcal{M}_{qg^{*}\rightarrow qg}^{D_{3}a_{1}a_{2}\overline{D}_{4}}\left(\lambda_{3},n_{1},\varepsilon_{2}^{\lambda_{2}},\lambda_{4}\right)=
\left(t^{a_{1}}t^{a_{2}}\right)_{D_{3}\overline{D}_{4}}\mathcal{M}_{qg^{*}\rightarrow qg}\left(3^{\lambda_{3}},1^{*},2^{\lambda_{2}},4^{\lambda_{4}}\right)+\\
\left(t^{a_{2}}t^{a_{1}}\right)_{D_{3}\overline{D}_{4}}\mathcal{M}_{qg^{*}\rightarrow qg}\left(3^{\lambda_{3}},2^{\lambda_{2}},1^{*},4^{\lambda_{4}}\right).
\end{multline}
For anti-quarks we need to exchange the indices $3\leftrightarrow4$.
The TMDs corresponding to those processes are given in Table
\ref{tab:TMDsgqqg}. In general, the TMDs for a sub-process with anti-quarks are different than for quarks, but they turn out to be the same assuming that the correlators are real. Again, we end up with only two independent TMDs.

\begin{table}
\begin{doublespace}
\begin{centering}
\begin{tabular}{c|c}
\hline 
color-ordered amplitude squared & gluon TMD\tabularnewline
\hline\\[-1.5em]
$\mathcal{M}_{qg^{*}\rightarrow qg}\left(3^{\lambda_{3}},1^{*},2^{\lambda_{2}},4^{\lambda_{4}}\right)\mathcal{M}_{qg^{*}\rightarrow qg}^{*}\left(3^{\lambda_{3}},2^{\lambda_{2}},1^{*},4^{\lambda_{4}}\right)$ & \multirow{3}{*}{$\Phi_{qg\rightarrow qg}^{\left(1\right)}=\mathcal{F}_{qg}^{\left(1\right)}$}\tabularnewline
%\cline{1-1} 
$\mathcal{M}_{qg^{*}\rightarrow qg}^{*}\left(3^{\lambda_{3}},1^{*},2^{\lambda_{2}},\lambda_{4}\right)\mathcal{M}_{qg^{*}\rightarrow qg}\left(3^{\lambda_{3}},2^{\lambda_{2}},1^{*},4^{\lambda_{4}}\right)$ & \tabularnewline
%\cline{1-1} 
$\left|\mathcal{M}_{qg^{*}\rightarrow qg}\left(3^{\lambda_{3}},2^{\lambda_{2}},1^{*},4^{\lambda_{4}}\right)\right|^{2}$ & 
%\tabularnewline
\\[0.5em]
\hline 
\\[-1.5em]
$\left|\mathcal{M}_{qg^{*}\rightarrow qg}\left(3^{\lambda_{3}},1^{*},2^{\lambda_{2}},4^{\lambda_{4}}\right)\right|^{2}$ & 
\multirow{1}{*}{$\Phi_{qg\rightarrow qg}^{\left(2\right)}=\frac{1}{N_{c}^{2}-1}\left(-\mathcal{F}_{qg}^{\left(1\right)}+N_{c}^{2}\mathcal{F}_{qg}^{\left(2\right)}\right)$}
%\tabularnewline
\\[0.5em]
\hline
\end{tabular}
\par\end{centering}
\end{doublespace}

\caption{Gluon TMDs accompanying the color-ordered amplitudes for $qg^{*}\rightarrow qg$
process. It has been assumed that correlators are real.
The $\GG^{(i)}_{qg}$ distributions are defined in Eqs.~(\ref{eq:Fqg1-def}) and 
Eqs.~(\ref{eq:Fqg2-def}).
\label{tab:TMDsgqqg}}
\end{table}

\subsection{Off-shell color-ordered helicity amplitudes}

In Section \ref{sec:hard-factors-off-shell}, we have calculated the off-shell
hard factors in a specific axial gauge, with $p_A$ chosen as the gauge vector,
and using the high energy projector (\ref{eq:pol-off-shell}).
As shown in Ref.~\cite{Catani:1990eg}, such a procedure yields results which are
gauge invariant within a subclass of axial gauges with the gauge vector 
$n^\mu = a p_p^\mu+ b p_A^\mu$, where $a$ and $b$ are arbitrary complex numbers.
There are also methods to calculate gauge invariant
off-shell amplitudes in any gauge and choice of polarization vectors
\cite{vanHameren:2012uj,vanHameren:2013,Kotko:2014aba,vanHameren:2014iua}. 
In what follows, we shall use those methods and specifically the results of \cite{vanHameren:2013,vanHameren:2014iua}.

Consider first the gluon amplitudes. For the purpose of this section
only we assume all momenta to be outgoing. For the non-vanishing helicity
configurations, in the helicity basis, we have 
\begin{gather}
\mathcal{M}_{g^{*}g\rightarrow gg}\left(1^{*},2^{-},3^{+},4^{+}\right)=2g^{2}\,\rho_1\,\frac{\ANG{1^{*}2}^{4}}{\ANG{1^{*}2}\ANG{23}\ANG{34}\ANG{41^{*}}}~,\label{eq:Mg*ggg-++}\\
\mathcal{M}_{g^{*}g\rightarrow gg}\left(1^{*},2^{+},3^{-},4^{+}\right)=2g^{2}\,\rho_1\,\frac{\ANG{1^{*}3}^{4}}{\ANG{1^{*}2}\ANG{23}\ANG{34}\ANG{41^{*}}}~,\label{eq:Mg*ggg+-+}\\
\mathcal{M}_{g^{*}g\rightarrow gg}\left(1^{*},2^{+},3^{+},4^{-}\right)=2g^{2}\,\rho_1\,\frac{\ANG{1^{*}4}^{4}}{\ANG{1^{*}2}\ANG{23}\ANG{34}\ANG{41^{*}}}~,\label{eq:Mg*ggg++-}
\end{gather}
where we adopted a shorthand notation for the spinor products $\ANG{ij}=\ANG{k_{i}-|k_{j}+}$
with $|k_{i}\pm\rangle=\frac{1}{2}\left(1\pm\gamma_{5}\right)u\left(k_{i}\right)$, and where $\rho_1$ is a, for our purposes irrelevant, phase factor
(see details \eg in \cite{vanHameren:2014iua}). We also defined $\ANG{1^{*}i}=\ANG{n_{1}i}$
with $n_{1}$ being the longitudinal component of $k_{1}$, \cf
Eq.~(\ref{eq:k1_def}). The other remaining helicity configurations can be
obtained from Eqs.~(\ref{eq:Mg*ggg-++})-(\ref{eq:Mg*ggg++-}) using CP invariance
\begin{equation}
\mathcal{M}_{gg^{*}\rightarrow gg}\left(1^{*},2^{+},3^{-},4^{-}\right)=\mathcal{M}_{gg^{*}\rightarrow gg}^{*}\left(1^{*},2^{-},3^{+},4^{+}\right),
\end{equation}
and so on. For the other color ordered amplitude,
$\mathcal{M}_{gg^{*}\rightarrow gg}\left(1^{*},3,2,4\right)$,
we need to exchange $2\leftrightarrow3$ in the denominators.

The above helicity amplitudes can be efficiently evaluated and squared
numerically, however for the purpose of this paper we shall need analytic
expressions. To this end let us introduce $\SQR{ij}=\ANG{k_{i}+|k_{j}-}$,
which, up to an unimportant phase, is a complex conjugate of $\ANG{ij}$.
Moreover, we have the following relation
\begin{equation}
\ANG{ij}\SQR{ji}=\left(k_{i}+k_{j}\right)^{2}\equiv\tilde{s}_{ij}.\label{eq:stilddef}
\end{equation}
For the products involving $n_{1}$ we use the notation
\begin{equation}
\ANG{1^{*}i}\SQR{i1^{*}}=\left(n_{1}+k_{i}\right)^{2}\equiv\tilde{s}_{1^{*}i}.\label{eq:stild*}
\end{equation}
With this, we get for the required amplitudes squared summed and
averaged over helicities 
\begin{gather}
\left|\overline{\mathcal{M}}_{gg^{*}\rightarrow gg}\left(1^{*},2,3,4\right)\right|^{2}=8g^{4}\,\frac{\ts{1^{*}2}^{4}+\ts{1^{*}3}^{4}+\ts{1^{*}4}^{4}}{\ts{1^{*}2}\ts{23}\ts{34}\ts{41^{*}}},\label{eq:Mg*ggg11}\\
\left|\overline{\mathcal{M}}_{gg^{*}\rightarrow gg}\left(1^{*},3,2,4\right)\right|^{2}=8g^{4}\,\frac{\ts{1^{*}2}^{4}+\ts{1^{*}3}^{4}+\ts{1^{*}4}^{4}}{\ts{1^{*}3}\ts{32}\ts{24}\ts{41^{*}}},\label{eq:Mg*gg22}\\
\overline{\mathcal{M}}_{gg^{*}\rightarrow gg}\left(1^{*},2,3,4\right)\overline{\mathcal{M}}_{gg^{*}\rightarrow gg}^{*}\left(1^{*},3,2,4\right)=-8g^{4}\,\frac{\ts{1^{*}2}^{4}+\ts{1^{*}3}^{4}+\ts{1^{*}4}^{4}}{\ANG{1^{*}2}\ANG{34}\SQR{1^{*}3}\SQR{24}\ts{23}\ts{41^{*}}},\label{eq:Mg*gg12}\\
\overline{\mathcal{M}}_{gg^{*}\rightarrow gg}^{*}\left(1^{*},2,3,4\right)\overline{\mathcal{M}}_{gg^{*}\rightarrow gg}\left(1^{*},3,2,4\right)=-8g^{4}\frac{\ts{1^{*}2}^{4}+\ts{1^{*}3}^{4}+\ts{1^{*}4}^{4}}{\SQR{1^{*}2}\SQR{34}\ANG{1^{*}3}\ANG{24}\ts{23}\ts{41^{*}}},\label{eq:Mg*gg21}
\end{gather}
where we have used overlines to indicate helicity summations. The
last two interference terms enter the cross section as a sum. Therefore, we
may simplify it as
\begin{multline}
\overline{\mathcal{M}}_{gg^{*}\rightarrow gg}\left(1^{*},2,3,4\right)\overline{\mathcal{M}}_{gg^{*}\rightarrow gg}^{*}\left(1^{*},3,2,4\right)+\overline{\mathcal{M}}_{gg^{*}\rightarrow gg}^{*}\left(1^{*},2,3,4\right)\overline{\mathcal{M}}_{gg^{*}\rightarrow gg}\left(1^{*},3,2,4\right)\\
=-8g^{4}\,\frac{(\ts{1^{*}2}^{4}+\ts{1^{*}3}^{4}+\ts{1^{*}4}^{4})(\ts{24}\ts{1^{*}3}-\ts{23}\ts{1^{*}4}+\ts{34}\ts{1^{*}2})}{\ts{1^{*}2}\ts{34}\ts{1^{*}3}\ts{24}\ts{23}\ts{41^{*}}},\label{eq:Mg*ggg1221}
\end{multline}
where we have used 
\begin{equation}
\SQR{1^{*}2}\SQR{34}\ANG{1^{*}3}\ANG{24}\,+\,\ANG{1^{*}2}\ANG{34}\SQR{1^{*}3}\SQR{24}
\,=\,\AL{n_1\!-}\slashp_{3}\slashp_{4}\slashp_{2}\AR{n_1-}\,+\,\AL{n_1\!-}\slashp_{2}\slashp_{4}\slashp_{3}\AR{n_1-}~,
\end{equation}
and applied $\slashp_{i}\slashp_{j}=\tilde{s}_{ij}-\slashp_{j}\slashp_{i}$
a few times. The amplitudes for the on-shell limit are simply obtained
by dropping the star in $1^{*}$ so that the spinor and the scalar products
will be with $k_{1}$ instead of $n_{1}$.

Now let us turn to processes with quarks. We will give only amplitudes
for $g\left(k_{4}\right)g^{*}\left(k_{1}\right)\rightarrow \overline{q}\left(k_{3}\right) q\left(k_{2}\right)$
process, as all the other can be obtained by the crossing symmetry
(taking care of the proper color flow when crossing). We have
\begin{gather}
\mathcal{M}_{gg^{*}\rightarrow q\overline{q}}\left(3^{-},1^{*},4^{+},2^{+}\right)={2g^{2}}\,\rho_1\,\frac{\ANG{21^{*}}^{3}\ANG{31^{*}}}{\ANG{21^{*}}\ANG{1^{*}4}\ANG{43}\ANG{32}},\label{eq:Mg*gqq-++}\\
\mathcal{M}_{gg^{*}\rightarrow q\overline{q}}\left(3^{+},1^{*},4^{+},2^{-}\right)={2g^{2}}\,\rho_1\,\frac{\ANG{31^{*}}^{3}\ANG{21^{*}}}{\ANG{21^{*}}\ANG{1^{*}4}\ANG{43}\ANG{32}}.\label{eq:Mg*gqq++-}
\end{gather}
We note that the above formulas have never been published in the literature and are given here for the first time.
 
Similar as before, the two remaining helicity configurations can be
obtained thanks to CP symmetry. For the color ordered amplitudes with
1 and 4 interchanged, we need to make a replacement $1\leftrightarrow4$
in the denominators. The amplitudes squared and summed over helicities
read (the helicity averaging factor is included)
\begin{gather}
\left|\overline{\mathcal{M}}_{gg^{*}\rightarrow q\overline{q}}\left(3,1^{*},4,2\right)\right|^{2}=2g^{4}\,\frac{\ts{1^{*}3}\left(\ts{1^{*}2}^{2}+\ts{1^{*}3}^{2}\right)}{\ts{1^{*}4}\ts{34}\ts{23}},\label{eq:Mg*gqq11}\\
\left|\overline{\mathcal{M}}_{gg^{*}\rightarrow q\overline{q}}\left(3,4,1^{*},2\right)\right|^{2}=2g^{4}\,\frac{\ts{1^{*}2}\left(\ts{1^{*}2}^{2}+\ts{1^{*}3}^{2}\right)}{\ts{1^{*}4}\ts{24}\ts{23}},\label{eq:Mg*gqq22}\\
\overline{\mathcal{M}}_{gg^{*}\rightarrow q\overline{q}}\left(3,1^{*},4,2\right)\overline{\mathcal{M}}_{gg^{*}\rightarrow q\overline{q}}^{*}\left(3,4,1^{*},2\right)=-2g^{4}\,\frac{\ts{1^{*}2}\ts{1^{*}3}\left(\ts{1^{*}2}^{2}+\ts{1^{*}3}^{2}\right)}{\ANG{21^{*}}\ANG{43}\SQR{31^{*}}\SQR{42}\ts{23}\ts{41^{*}}},\label{eq:Mg*gqq12}\\
\overline{\mathcal{M}}_{gg^{*}\rightarrow q\overline{q}}^{*}\left(3,1^{*},4,2\right)\overline{\mathcal{M}}_{gg^{*}\rightarrow q\overline{q}}\left(3,4,1^{*},2\right)=-2g^{4}\,\frac{\ts{1^{*}2}\ts{1^{*}3}\left(\ts{1^{*}2}^{2}+\ts{1^{*}3}^{2}\right)}{\SQR{21^{*}}\SQR{43}\ANG{31^{*}}\ANG{42}\ts{23}\ts{41^{*}}}.\label{eq:Mg*ggqq21}
\end{gather}
The sum of the last two interference terms simplifies to
\begin{multline}
\overline{\mathcal{M}}_{gg^{*}\rightarrow q\overline{q}}\left(3,1^{*},4,2\right)\overline{\mathcal{M}}_{gg^{*}\rightarrow q\overline{q}}^{*}\left(3,4,1^{*},2\right)+\overline{\mathcal{M}}_{gg^{*}\rightarrow q\overline{q}}^{*}\left(3,1^{*},4,2\right)\overline{\mathcal{M}}_{gg^{*}\rightarrow q\overline{q}}\left(3,4,1^{*},2\right)\\
=-2g^{4}\,\frac{\ts{1^{*}2}\ts{1^{*}3}(\ts{1^{*}2}^{2}+\ts{1^{*}3}^{2})(\ts{24}\ts{1^{*}3}-\ts{23}\ts{1^{*}4}+\ts{34}\ts{1^{*}2})}{\ts{1^{*}2}\ts{34}\ts{1^{*}3}\ts{24}\ts{23}\ts{41^{*}}}.\label{eq:Mg*qq1221}
\end{multline}

In order to obtain amplitudes for $q\left(k_{4}\right)g^{*}\left(k_{1}\right)\rightarrow q\left(k_{3}\right)g\left(k_{2}\right)$
we can use the crossing symmetry. Specifically, we can obtain $\left|\overline{\mathcal{M}}_{qg^{*}\rightarrow qg}\left(3,1^{*},2,4\right)\right|^{2}$,
$\left|\overline{\mathcal{M}}_{qg^{*}\rightarrow qg}\left(3,2,1^{*},4\right)\right|^{2}$
and interference terms by making replacement $2\leftrightarrow4$
in Eqs.~(\ref{eq:Mg*gqq22}), (\ref{eq:Mg*gqq11}), (\ref{eq:Mg*qq1221})
respectively.

\subsection{Hard factors from color-ordered amplitudes}

Having computed the color ordered amplitudes it is now straightforward to
calculate the hard factors $K^{(i)}$. Let us note, that it is the $K^{(i)}$ hard
factors that appear naturally within the color-ordered formalism, not the $H^{(i)}$
factors. It also comes naturally that there are two hard factors and two TMDs per each channel, so the  the factorization formulas can be written in a unified form:
\begin{equation}
\frac{d\sigma^{pA\rightarrow {\rm dijets}+X}}{d^{2}P_{t}d^{2}k_{t}dy_{1}dy_{2}}=\frac{\alpha_{s}^{2}}{(x_1 x_2 s)^{2}}
\sum_{a,c,d} x_{1}f_{a/p}(x_{1}, \mu^2)\sum_{i=1}^{2}K_{ag^*\to cd}^{(i)}\Phi_{ag\rightarrow cd}^{(i)}\ \frac{1}{1+\delta_{cd}}\ ,
\label{eq:gg2gg-mod}
\end{equation}
where $a,c,d$ are the contributing partons. The explicit expressions
for the generalized gluon TMDs $\Phi_{ag\rightarrow cd}^{\left(i\right)}$
are listed in Tables \ref{tab:TMDsgggg}-\ref{tab:TMDsgqqg}. The
hard factors $K^{i}$ were already given in Section \ref{sec:hard-factors-off-shell} (we collect them in Table~\ref{tab:Khardfactors} for convenience).
 In the context of this section, they are obtained
by multiplying the left column of Tables \ref{tab:TMDsgggg}-\ref{tab:TMDsgqqg}
by the corresponding color factors and combining the cells that belong
to the same generalized TMD. More precisely, we have

\renewcommand{\arraystretch}{2.}
\begin{table}
\begin{doublespace}
\begin{centering}
\begin{tabular}{c|c|c}
\hline  
\Large $i$ & \Large 1 &  \Large 2  \\
\hline 
$\displaystyle K_{gg^*\to gg}^{(i)}$ & $ \displaystyle
\frac{N_{c}}{C_F}\,\frac{\left(\overline{s}^{4}+\overline{t}^{4}+\overline{u}^{4}\right)\left(\overline{u}\hat{u}+\overline{t}\hat{t}\right)}{\tlt\that\tlu\uhat\tls\shat}$ & $\displaystyle -\frac{N_{c}}{2C_F}\,\frac{\left(\overline{s}^{4}+\overline{t}^{4}+\overline{u}^{4}\right)\left(\overline{u}\hat{u}+\overline{t}\hat{t}-\overline{s}\hat{s}\right)}{\tlt\that\tlu\uhat\tls\shat}$
\\
\hline 
$\displaystyle K_{gg^*\to q\overline{q}}^{(i)}$ & $\displaystyle \frac{1}{2N_{c}}\,\frac{\left(\overline{t}^{2}+\overline{u}^{2}\right)\left(\overline{u}\hat{u}+\overline{t}\hat{t}\right)}{\overline{s}\hat{s}\hat{t}\hat{u}}$ & $\displaystyle \frac{1}{4N_{c}^2 C_F}\,\frac{\left(\overline{t}^{2}+\overline{u}^{2}\right)\left(\overline{u}\hat{u}+\overline{t}\hat{t}-\overline{s}\hat{s}\right)}{\overline{s}\hat{s}\hat{t}\hat{u}}$
\\
\hline 
$\displaystyle K_{qg^*\to qg}^{(i)}$ & $\displaystyle -\frac{\overline{u}\left(\overline{s}^{2}+\overline{u}^{2}\right)}{2\overline{t}\hat{t}\hat{s}}\left(1+\frac{\overline{s}\hat{s}-\overline{t}\hat{t}}{N_{c}^{2}\ \overline{u}\hat{u}}\right)$ & $\displaystyle -\frac{C_F}{N_c}\,\frac{\overline{s}\left(\overline{s}^{2}+\overline{u}^{2}\right)}{\overline{t}\hat{t}\hat{u}}$
\\
\hline 
\end{tabular}
\par\end{centering}
\end{doublespace}
\caption{The hard factors accompanying the gluon TMDs $\Phi_{ag\rightarrow cd}^{\left(i\right)}$.\label{tab:Khardfactors}}
\end{table}

\begin{gather}
g^4\,K_{gg^*\to gg}^{(1)} = \frac{1}{(2N_c C_F)^2}\, \frac{N_c^3 C_F}{2} \left(\left|\overline{\mathcal{M}}_{gg^{*}\rightarrow gg}\left(1^{*},2,3,4\right)\right|^{2}
+ \left|\overline{\mathcal{M}}_{gg^{*}\rightarrow gg}\left(1^{*},3,2,4\right)\right|^{2} \right)\,, \\
g^4\,K_{gg^*\to gg}^{(2)} = \frac{1}{(2N_c C_F)^2}\, \frac{N_c^3 C_F}{4}  \left( 
\overline{\mathcal{M}}_{gg^{*}\rightarrow gg}\left(1^{*},2,3,4\right)
\overline{\mathcal{M}}_{gg^{*}\rightarrow gg}^{*}\left(1^{*},3,2,4\right)
+ \mathtt{c.c.}
%+\overline{\mathcal{M}}_{g^{*}g\rightarrowgg}^{*}\left(1^{*},2,3,4\right)\overline{\mathcal{M}}_{g^{*}g\rightarrow gg}\left(1^{*},3,2,4\right)
\right)\,,
\end{gather}
for pure gluon channel, and
\begin{gather}
g^4\,K_{gg^*\to q\overline{q}}^{(1)} = \frac{1}{(2N_c C_F)^2}\, N_c C_F^2 \left(
\left|\overline{\mathcal{M}}_{gg^{*}\rightarrow q\overline{q}}\left(3,1^{*},4,2\right)\right|^{2}
+ \left|\overline{\mathcal{M}}_{gg^{*}\rightarrow q\overline{q}}\left(3,4,1^{*},2\right)\right|^{2}
\right)\,, \\
g^4\,K_{gg^*\to q\overline{q}}^{(2)} = \frac{1}{(2N_c C_F)^2}\, \frac{-C_F}{2} \left(
\overline{\mathcal{M}}_{gg^{*}\rightarrow q\overline{q}}\left(3,1^{*},4,2\right)\overline{\mathcal{M}}_{gg^{*}\rightarrow q\overline{q}}^{*}\left(3,4,1^{*},2\right)
+ \mathtt{c.c.} \right) \,,
\end{gather}
for $gg^*\to q\overline{q}$ channel. For the $qg^*\to qg$ sub-process we need to use the crossing symmetry as described in the preceding section. We have
\begin{eqnarray}
g^4\, K_{qg^*\to qg}^{(1)} &=& \frac{1}{2C_F N_c^2}  \Big\lbrace
N_c C_F^2 \left( 
-\left|\overline{\mathcal{M}}_{gg^{*}\rightarrow q\overline{q}}\left(3,1^{*},4,2\right)\right|^{2}
\right)_{2\leftrightarrow 4}  \qquad\qquad \qquad\qquad \qquad \nonumber
\\ &&- \frac{C_F}{2} \left(-
\overline{\mathcal{M}}_{gg^{*}\rightarrow q\overline{q}}\left(3,1^{*},4,2\right)\overline{\mathcal{M}}_{gg^{*}\rightarrow q\overline{q}}^{*}\left(3,4,1^{*},2\right)- \mathtt{c.c.} \right)_{2\leftrightarrow 4}\Big\rbrace \,, 
\label{eq:KgqqgCO1}
\\
g^4\, K_{qg^*\to qg}^{(2)} &=& \frac{1}{2C_F N_c^2}\, N_c C_F^2 \left(
-\left|\overline{\mathcal{M}}_{gg^{*}\rightarrow q\overline{q}}\left(3,4,1^{*},2\right)\right|^{2}
\right)_{2\leftrightarrow 4}\,.
\label{eq:KgqqgCO2}
\end{eqnarray}
In all the formulas above, the first color factor comes from color averaging. The minus signs in front of the amplitudes in (\ref{eq:KgqqgCO1}), (\ref{eq:KgqqgCO2}) come from the crossing of a fermion line. Table~\ref{tab:Khardfactors} is easily recovered using the following relations of $\tilde{s}_{ij}$ to the kinematic variables from Section~\ref{sec:hard-factors-off-shell}
\begin{gather}
\tilde{s}_{23}=\tilde{s}_{14}=\hat{s},\,\,\, \tilde{s}_{34}=\tilde{s}_{12}=\hat{t},\,\,\, \tilde{s}_{24}=\tilde{s}_{13} = \hat{u}\,, \\
\tilde{s}_{1^*4}=\bar{s},\,\,\, \tilde{s}_{1^*2}=\bar{t},\,\,\, \tilde{s}_{1^*3}=\bar{u}\,.  
\end{gather}

%-----------------------------------------------------------------------------
\section{Conclusions and outlook}

Dijet production is one of the key processes studied at the LHC.
Requiring the two jets to be produced in the forward direction creates an asymmetric
situation, in which one of the incoming hadrons is probed at
large $x$, while the other is probed at a very small momentum
fraction.
This kinematic regime poses various challenges, one of the biggest questions being
the existence of a theoretically-consistent and, at the same time,
practically-manageable factorization formula. The standard collinear
factorization is not applicable in this case as the dependence on the transverse
momentum of the low-$x$ gluon in the target, $k_t$, cannot be neglected.

In the limit where the jets' transverse momenta $|p_{1t}|, |p_{2t}| \gg |k_t|
\sim Q_s$, with the latter being the saturation scale of the target, an effective
transverse-momentum-dependent factorization formula for forward dijet production has been derived in
Refs.~\cite{Dominguez:2010xd,Dominguez:2011wm} and it has been shown to be consistent with the CGC framework.
On the other side, the high energy factorization approach~\cite{Catani:1990eg,Deak:2009xt} has been also
successfully applied for studying forward dijet production at the LHC. In this paper, we have examined the theoretical status
of the HEF approach in the context of forward dijet production at hadron colliders and reconciled it with the TMD factorization
by creating a unified framework valid in the limit $|p_{1t}|, |p_{2t}| \gg Q_s$ with an arbitrary value of $|k_t|$, as long as it
is allowed by phase space constraints.
In particular, we have shown in Section~\ref{sec:dilute} that the HEF formula
is indeed justified in the kinematic window of $|p_{1t}|, |p_{2t}| \sim |k_t|
\gg Q_s$, where it was explicitly derived from CGC for all $2\to 2$ channels. This limit
corresponds to the dilute target approximation hence no non-linear effects are
expected.

The second major result of our work is an improvement of the effective TMD
factorization for forward dijet production, first derived in
Ref.~\cite{Dominguez:2011wm}, by taking into account in
Section~\ref{sec:tmd-on-shell} all finite-$N_c$ corrections, as well as
generalizing the factorization formula to the case with an off-shell incoming
gluon in Sections~\ref{sec:hard-factors-off-shell}~and~\ref{sec:HelTMD}.
In addition, we were able to simplify the TMD factorization formula by reducing
the number of gluon distributions to two independent TMDs for each channel.
The main results of this part of our study are summarized in
Eq.~(\ref{eq:gg2gg-mod}),  which gives the new TMD factorization formula, as
well as in Table~\ref{tab:Khardfactors}, where we collect all the off-shell hard factors.  
The corresponding gluon distributions are given in
Tables~\ref{tab:TMDsgggg}, \ref{tab:TMDsggqq} and~\ref{tab:TMDsgqqg}.
The above results were obtained with two independent techniques: a traditional
Feynman diagram approach and helicity methods with color ordered
amplitudes.
The improved TMD factorization formula (\ref{eq:gg2gg-mod}) encapsulates both
the result of Ref.~\cite{Dominguez:2011wm} and the HEF framework as its limiting
cases.

The results obtained in this paper open several avenues for future research that we plan to follow. First, a natural next steps will be to use Eq.~(\ref{eq:gg2gg-mod}) for phenomenological studies. That shall require some input for the six gluon TMDs $\Phi_{ag\rightarrow cd}^{(1,2)}(x,k_t)$, which may be difficult in a general case. But in the large-$N_c$ limit, they can all be written in terms of just two functions: $xG^{(1)}(x,k_t)$ and $xG^{(2)}(x,k_t)$, which in turn can be evaluated within certain models, as in \cite{Stasto:2011ru}.

Another line of possible extension of our framework is to supplement it with
high-$|P_t|$ effects such as Sudakov logarithms or coherence in the evolution of
the gluon density. Essentially, this can be done by adding a $\mu^2$ dependence
to the unintegrated gluon distributions~\cite{Collins:1984kg,Ciafaloni:1987ur,Catani:1989sg,Catani:1989yc,Kimber:1999xc,Kimber:2001sc, Collins:2014jpa}.
The equations that combine such effects with the small-$x$ evolution
\cite{Kutak:2011fu,Kutak:2012qk} show a nontrivial interplay between the
non-linearities and the $\mu^2$ dependence and this may, in particular, weaken
the saturation effects. At the linear level, the so-called single step inclusion of
the hard-scale effects (as demonstrated in \cite{vanHameren:2014ala}) helps in
the description of forward-central dijet data, therefore this direction seems to
be relevant in order to provide complete predictions. Furthermore, first
estimates of azimuthal decorrelations of the forward-forward dijets in the HEF
framework, with inclusion of hard scale effects and non-linearities, show that
they are of similar relevance for this process~\cite{Kutak:2014wga}.

Last but not least, it remains to be proved that the large logarithms generated by higher-order corrections can indeed be absorbed into evolution equations for the various parton distributions (and jet fragmentation functions) involved, and potentially for additional soft factors \cite{Ji:2004wu}.
This limitation however is not specific to our work, the same is true at the
level of the TMD and HEF regimes independently. In the former case, it is known
that TMD factorization generically does not apply for dijet production in
hadron-hadron collisions \cite{Collins:2007nk,Rogers:2010dm}. It is nevertheless
expected that, in dilute-dense collisions, initial state interactions originating from a dilute hadron do not interfere with the intrinsic transverse momentum and thus factorization may hold, although there is no formal proof of this statement yet.

In addition, even though it was possible to write formula
(\ref{eq:gg2gg-new-onshell}) in terms of just two TMDs per channel, this
simplification may not survive after small-$x$ evolution is included, as, in
general, the non-linear equations mix the original $\GG^{(i)}_{ag}$
functions. For instance, $xG^{(1)}$ does not obey a closed equation and, contrary
to what happens with $xG^{(2)}$, the large-$N_c$ limit does not help
\cite{Dominguez:2011gc}. We note that any equivalent linear combination of the
gluon distributions, such as (\ref{eq:tmd-main}) and~(\ref{eq:gg2gg-new-onshell}), is equally valid, and it may turn out that some
alternative choice allows one to write the evolution equations directly in terms of
TMDs. By contrast, it is also possible that the inclusion of small-$x$ evolution
can only be achieved within the full complexity of the CGC, meaning that the
$Q_s\sim|k_t|\ll |P_t|$ limit, which allows one to avoid the quadrupole operator in
(\ref{eq:S4-fund}) and express the cross section in terms of gluon
distributions,
may not help when small-$x$ evolution is considered.

In the HEF regime, the issues are different. The $Q_s\ll |k_t|\sim |P_t|$ limit
makes things simpler from the point of view of small-$x$ evolution, since
non-linear effects can be neglected. However, the off-shellness of the hard
process is not neglected and thus the standard power counting of the twist
expansion becomes useless. One must then resort to different methods, such as
those of Ref.~\cite{Fadin:2006bj}. Any progress towards an all-order proof of
either HEF or TMD factorization for forward dijet production in dilute-dense
collisions will naturally carry over to our improved TMD factorization formula
(\ref{eq:gg2gg-mod}) that combines both regimes. In the meantime, our results represent a viable alternative to CGC calculations, equivalent to them in the kinematic regime appropriate for dijets $Q_s\ll |P_t|$ but more practical.

%-----------------------------------------------------------------------------
\section*{Acknowledgments}
The work of K.K. has been supported by Narodowe Centrum Nauki with Sonata Bis grant DEC-2013/10/E/ST2/00656.
P.K. acknowledges the support of the grants DE-SC-0002145 and DE-FG02-93ER40771. 
S.S. acknowledges useful discussions with Gavin Salam and Fabrizio Caola.
P.K., K.K., S.S. and A.vH. are grateful for hospitality to \'Ecole
Polytechnique, where part of this work has been carried out.
K.K. thanks for the hospitality of Penn State University, where part of this research was done.

%-----------------------------------------------------------------------------
\appendix
%-----------------------------------------------------------------------------
\section{Off-shell expressions}
\label{app:off-shell-expr}

In this appendix, we gather all expressions corresponding to the $D_i$ diagrams
from Fig.~\ref{fig:qg2qg-diag}-\ref{fig:gg2gg-4gdiag} in the case where one of
the incoming gluons is off-shell. 
All calculations were preformed in the axial gauge discussed at the beginning of
Section~\ref{sec:hard-factors-off-shell}, with the axial vectors for the
on-shell gluons set according to Eq.~(\ref{eq:gauge-gg2gg}).

For completeness, we also give here the results for the ``old'' hard factors
defined in Eqs.~(\ref{eq:H1-qg-on-shell-def}), (\ref{eq:H2-qg-on-shell-def})
(\ref{eq:ggqq-onshell1-def}), (\ref{eq:ggqq-onshell2-def}),
(\ref{eq:ggqq-onshell3-def}), (\ref{eq:H1gggg-def}), (\ref{eq:H2gggg-def}) and
(\ref{eq:H6gggg-def}), in the case with off-shell incoming gluon.

Table~\ref{tab:Di-off-shell-qg2qg} gives the $D_i$ expressions for the
subprocesses $qg^* \to qg$.  The two hard factors in this channel read
\begin{eqnarray}
  H^{(1)}_{qg^*\to qg}  = 
  - \frac{\tls^2 + \tlu^2}{2 \shat \that \tlu}
    \Bigg[\tlu - \frac{\that \tlt}{N_c^2 \uhat}\Bigg]\,,
\end{eqnarray}    
    
\begin{eqnarray}   
  H^{(2)}_{qg^*\to qg}  = 
  - \frac{\tls\left(\tls^2 + \tlu^2\right)}{2 \uhat \that \tlt}\,.
\end{eqnarray}
In the limit, $|k_t| \to 0$, simplification given by
Eq.~(\ref{eq:gen-mandelstam-limit}) occurs and
the above formulas manifestly recover the on-shell
results from Eqs.~(\ref{eq:H1-qg-on-shell}) and (\ref{eq:H2-qg-on-shell}).

The corresponding $D_i$ results for the $gg^* \to q\qbar$ subprocess are given
in Table~\ref{tab:Di-off-shell-gg2qq}.
The three ``old'', off-shell hard factors for this channel take the form
\begin{eqnarray}
  \label{eq:H1gg2qqoff}
  H_{gg^*\rightarrow q\qbar}^{(1)}&=&
    \frac{1}{4 C_F} \frac{\tlt^2+\tlu^2}{\uhat \that \shat \tls}
    \left[\uhat \tlu + \that \tlt \right]\,, \\
  \label{eq:H2gg2qqoff}
  H_{gg^*\rightarrow q\qbar}^{(2)}&=&
    \frac{1}{4 C_F} \frac{\tlt^2+\tlu^2}{\uhat \that \shat \tls}
    \left[\shat\tls - \that \tlt - \uhat \tlu \right]\,, \\
  \label{eq:H3gg2qqoff}
  H_{gg^*\rightarrow q\qbar}^{(3)}&=&
    -\frac{1}{4 N_c^2 C_F} \frac{\tlt^2+\tlu^2}{\uhat \that}\,.
\end{eqnarray}
Again, following Eq.~(\ref{eq:gen-mandelstam-limit}), it is manifest that the
above hard factors reduce to
Eqs.~(\ref{eq:ggqq-onshell1})-(\ref{eq:ggqq-onshell3}) in the limit $|k_t| \to
0$. 

Finally, 
the $D_i$ expressions for the subprocess $gg^* \to gg$ 
are given in Table~\ref{tab:Di-off-shell-gg2gg}  and
the six hard factors read
\begin{eqnarray}
  \label{eq:Hgg2ggoff1}
  H^{(1)}_{gg^*\to gg} & = & 
  \frac{N_c}{C_F}\, 
  \frac{(\tls^2-\tlt \tlu)^2} {\that\tlt \uhat\tlu \shat \tls}\,
  \left[\that\tlt + \uhat \tlu\right]\,, \\
  \label{eq:Hgg2ggoff2}
  H^{(2)}_{gg^*\to gg} & = & 
  \frac{N_c}{C_F}\, 
  \frac{(\tls^2-\tlt \tlu)^2} {\that\tlt \uhat\tlu \shat \tls}\,
  \left[\shat\tls -\that\tlt - \uhat \tlu\right]\,, \\
  \label{eq:Hgg2ggoff6}
  H_{gg^* \to gg}^{(6)}&=& -\frac{N_c^2}{2}  H^{(3)}_{gg^* \to gg} = 
                        N_c^2 H_{gg^* \to gg}^{(4)}= N_c^2 H_{gg^* \to gg}^{(5)}  = 
  \frac{N_c}{C_F}\, \frac{(\tls^2-\tlt \tlu)^2} {\that\tlt \uhat\tlu}\,.
\end{eqnarray}
The on-shell limit is again manifest, with the above equations reducing to
Eqs.~(\ref{eq:H1gggg}), (\ref{eq:H2gggg}) and~(\ref{eq:H6gggg}) as $|k_t| \to 0
$.
\begin{table}[t]
\begin{center}
\begin{tabular}{cc}
\hline 
 $qg^* \to qg$ & $D_i$ \\
\hline
(1) & $\displaystyle \frac{2 \bar{t} \bar{u}+\bar{t}^2+2 \bar{u}^2}{\hat{t}
\left(\hat{s}+\hat{t}+\hat{u}\right)}$ \\[15pt]
(2) & $\displaystyle \frac{C_F}{N_c}\frac{\bar{u} \left(2 \bar{t} \bar{u}+\bar{t}^2+2 \bar{u}^2\right)}{\hat{u} \bar{t}
   \left(\hat{s}+\hat{t}+\hat{u}\right)}$ \\[15pt]
(3) & $\displaystyle \frac{\left(2 \bar{t} \bar{u}+\bar{t}^2+2 \bar{u}^2\right)
\left(\bar{t} \left(\hat{s}+\hat{t}\right)+\bar{u}
   \left(\hat{s}+\hat{u}\right)\right)}{4 \hat{t} \hat{u} \bar{t}
   \left(\hat{s}+\hat{t}+\hat{u}\right)}$ \\[15pt]
(4) & $\displaystyle -\frac{C_F}{N_c} \frac{\left(\bar{t}+\bar{u}\right) \left(2 \bar{t} \bar{u}+\bar{t}^2+2 \bar{u}^2\right)}{\hat{s} \bar{t}
   \left(\hat{s}+\hat{t}+\hat{u}\right)}$ \\[15pt]
(5) & $\displaystyle -\frac{\left(2 \bar{t} \bar{u}+\bar{t}^2+2 \bar{u}^2\right)
\left(\bar{t} \left(\hat{s}-\hat{t}\right)+\bar{u}
   \left(\hat{s}+\hat{u}\right)\right)}{4 \hat{s} \hat{t} \bar{t}
   \left(\hat{s}+\hat{t}+\hat{u}\right)}$ \\[15pt]
(6) & $\displaystyle \frac{1}{N_c^2} \frac{\left(2 \bar{t} \bar{u}+\bar{t}^2+2 \bar{u}^2\right) \left(\bar{t} \left(\hat{s}+\hat{t}\right)+\bar{u}
   \left(\hat{s}-\hat{u}\right)\right)}{4 \hat{s} \hat{u} \bar{t}
   \left(\hat{s}+\hat{t}+\hat{u}\right)}$ \\
\hline
\end{tabular}
\end{center}
\caption{Expressions for the $qg^*\to qg$ subprocess with off-shell incoming
gluon corresponding to diagrams (1)-(6) of Fig.~\ref{fig:qg2qg-diag} in gauge
described in Section~\ref{sec:hard-factors-off-shell}.}
\label{tab:Di-off-shell-qg2qg}
\end{table}

\begin{table}[p]
\begin{center}
\begin{tabular}{cc}
\hline 
 $gg^* \to q\qbar $ & $D_i$ \\
\hline
(1) & $\displaystyle \frac{1}{N_c} \frac{\left(\bar{s}+\bar{u}\right) \left(2
\bar{s} \bar{u}+\bar{s}^2+2 \bar{u}^2\right)}{2 \hat{t} \bar{s}
   \left(\hat{s}+\hat{t}+\hat{u}\right)}$ \\[15pt]
(2) & $\displaystyle - \frac{1}{N_c} \frac{\bar{u} \left(2 \bar{s}
\bar{u}+\bar{s}^2+2 \bar{u}^2\right)}{2 \hat{u} \bar{s}
   \left(\hat{s}+\hat{t}+\hat{u}\right)}$ \\[15pt]
(3) & $\displaystyle - \frac{1}{N_c^2 C_F}\frac{\left(2 \bar{s}
\bar{u}+\bar{s}^2+2 \bar{u}^2\right) \left(\bar{s}
\left(\hat{s}+\hat{t}\right)+\bar{u} \left(\hat{t}-\hat{u}\right)\right)}{8
\hat{t} \hat{u} \bar{s} \left(\hat{s}+\hat{t}+\hat{u}\right)}$
\\[15pt]
(4) & $\displaystyle -\frac{1}{C_F}\frac{2 \bar{s} \bar{u}+\bar{s}^2+2
\bar{u}^2}{2 \hat{s} \left(\hat{s}+\hat{t}+\hat{u}\right)}$ \\[15pt]
(5) & $\displaystyle -\frac{1}{C_F}\frac{\left(2 \bar{s} \bar{u}+\bar{s}^2+2
\bar{u}^2\right) \left(\bar{s} \left(\hat{s}-\hat{t}\right)-\bar{u}
\left(\hat{t}+\hat{u}\right)\right)}{8 \hat{s} \hat{t} \bar{s}
\left(\hat{s}+\hat{t}+\hat{u}\right)}$ \\[15pt]
(6) & $\displaystyle -\frac{1}{C_F}\frac{\left(2 \bar{s} \bar{u}+\bar{s}^2+2
\bar{u}^2\right) \left(\bar{s} \left(\hat{s}+\hat{t}\right)+\bar{u}
\left(\hat{t}+\hat{u}\right)\right)}{8 \hat{s} \hat{u} \bar{s}
\left(\hat{s}+\hat{t}+\hat{u}\right)}$ \\
\hline
\end{tabular}
\end{center}
\caption{Expressions for the $gg^*\to q\qbar$ subprocess with off-shell incoming
gluon corresponding to diagrams (1)-(6) of Fig.~\ref{fig:gg2qqbar-diag} in gauge
described in Section~\ref{sec:hard-factors-off-shell}.}
\label{tab:Di-off-shell-gg2qq}
\end{table}
\begin{table}[p]
\begin{center}
\begin{tabular}{cc}
\hline 
 $gg^* \to gg$ & $D_i$ \\
\hline
(1) & $\displaystyle \frac{2 N_c}{C_F} \frac{\left(\bar{t}
\bar{u}+\bar{t}^2+\bar{u}^2\right)^2}{\hat{t} \bar{u}
\left(\bar{t}+\bar{u}\right)
   \left(\hat{s}+\hat{t}+\hat{u}\right)}$ \\[15pt]
(2) & $\displaystyle \frac{2 N_c }{C_F} \frac{\left(\bar{t}
\bar{u}+\bar{t}^2+\bar{u}^2\right)^2}{\hat{u} \bar{t} 
\left(\bar{t}+\bar{u}\right)
   \left(\hat{s}+\hat{t}+\hat{u}\right)}$ \\[15pt]
(3) & $\displaystyle \frac{N_c}{2 C_F} \frac{\left(\bar{t}
\bar{u}+\bar{t}^2+\bar{u}^2\right)^2 \left(\bar{t}
\left(\hat{s}+\hat{t}\right)+\bar{u}
   \left(\hat{s}+\hat{u}\right)\right)}{\hat{t} \hat{u} \bar{t} \bar{u}
   \left(\bar{t}+\bar{u}\right)
      \left(\hat{s}+\hat{t}+\hat{u}\right)}$ \\[15pt]
(4) & $\displaystyle -\frac{2 N_c}{C_F} \frac{\left(\bar{t}
\bar{u}+\bar{t}^2+\bar{u}^2\right)^2}{\hat{s} \bar{t} \bar{u}
   \left(\hat{s}+\hat{t}+\hat{u}\right)}$ \\[15pt]
(5) & $\displaystyle - \frac{N_c}{2 C_F} \frac{\left(\bar{t}
\bar{u}+\bar{t}^2+\bar{u}^2\right)^2 \left(\bar{t}
\left(\hat{s}-\hat{t}\right)+\bar{u}
   \left(\hat{s}+\hat{u}\right)\right)}{\hat{s} \hat{t} \bar{t} \bar{u}
   \left(\bar{t}+\bar{u}\right)
      \left(\hat{s}+\hat{t}+\hat{u}\right)}$ \\[15pt]
(6) & $\displaystyle - \frac{N_c}{2 C_F} \frac{\left(\bar{t}
\bar{u}+\bar{t}^2+\bar{u}^2\right)^2 \left(\bar{t}
\left(\hat{s}+\hat{t}\right)+\bar{u}
   \left(\hat{s}-\hat{u}\right)\right)}{\hat{s} \hat{u} \bar{t} \bar{u}
   \left(\bar{t}+\bar{u}\right)
      \left(\hat{s}+\hat{t}+\hat{u}\right)}$ \\
\hline
\end{tabular}
\end{center}
\caption{Expressions for the $gg^*\to gg$ subprocess with off-shell incoming
gluon in gauge described in Section~\ref{sec:hard-factors-off-shell}.
The numbering (1)-(6) corresponds to the color structures as defined 
in Eq.~(\ref{eq:cc-3g}) and each expression contains contributions from diagrams
with both 3- and 4-gluon vertices.
}
\label{tab:Di-off-shell-gg2gg}
\end{table}


\begin{thebibliography}{99}

\bibitem{Gribov:1984tu}
  L.~V.~Gribov, E.~M.~Levin and M.~G.~Ryskin,
  %``Semihard Processes in QCD,''
  Phys.\ Rept.\  {\bf 100} (1983) 1.
  %%CITATION = PRPLC,100,1;%%

\bibitem{Gelis:2010nm}
  F.~Gelis, E.~Iancu, J.~Jalilian-Marian and R.~Venugopalan,
  %``The Color Glass Condensate,''
  Ann.\ Rev.\ Nucl.\ Part.\ Sci.\  {\bf 60} (2010) 463.
%  [arXiv:1002.0333 [hep-ph]].
  %%CITATION = ARXIV:1002.0333;%%

\bibitem{Albacete:2014fwa} 
  J.~L.~Albacete and C.~Marquet,
  %``Gluon saturation and initial conditions for relativistic heavy ion collisions,''
  Prog.\ Part.\ Nucl.\ Phys.\  {\bf 76} (2014) 1.
%  [arXiv:1401.4866 [hep-ph]].
  %%CITATION = ARXIV:1401.4866;%%

\bibitem{Albacete:2010pg}
  J.~L.~Albacete and C.~Marquet,
  %``Azimuthal correlations of forward di-hadrons in d+Au collisions at RHIC in the Color Glass Condensate,''
  Phys.\ Rev.\ Lett.\  {\bf 105} (2010) 162301.
%  [arXiv:1005.4065 [hep-ph]].
  %%CITATION = ARXIV:1005.4065;%%
  
  \bibitem{Stasto:2011ru}
  A.~Stasto, B.~-W.~Xiao and F.~Yuan,
  %``Back-to-Back Correlations of Di-hadrons in dAu Collisions at RHIC,''
  Phys.\ Lett.\ B {\bf 716} (2012) 430.
%  [arXiv:1109.1817 [hep-ph]].
  %%CITATION = ARXIV:1109.1817;%%
  
\bibitem{Lappi:2012nh}
  T.~Lappi and H.~Mantysaari,
  %``Forward dihadron correlations in deuteron-gold collisions with the Gaussian approximation of JIMWLK,''
  Nucl.\ Phys.\ A {\bf 908} (2013) 51.
%  [arXiv:1209.2853 [hep-ph]].
  %%CITATION = ARXIV:1209.2853;%%

\bibitem{Marquet:2007vb}
  C.~Marquet,
  %``Forward inclusive dijet production and azimuthal correlations in p(A) collisions,''
  Nucl.\ Phys.\ A {\bf 796} (2007) 41.
%  [arXiv:0708.0231 [hep-ph]].
  %%CITATION = ARXIV:0708.0231;%%

\bibitem{Adare:2011sc}
  A.~Adare {\it et al.}  [PHENIX Collaboration],
  %``Suppression of back-to-back hadron pairs at forward rapidity in $d+$Au Collisions at $\sqrt{s_{NN}}=200$ GeV,''
  Phys.\ Rev.\ Lett.\  {\bf 107} (2011) 172301.
%  [arXiv:1105.5112 [nucl-ex]].
  %%CITATION = ARXIV:1105.5112;%%

\bibitem{Braidot:2010zh}
  E.~Braidot [STAR Collaboration],
  %``Suppression of Forward Pion Correlations in d+Au Interactions at STAR,''
  arXiv:1005.2378 [hep-ph].
  %%CITATION = ARXIV:1005.2378;%%

\bibitem{Catani:1990eg}
  S.~Catani, M.~Ciafaloni and F.~Hautmann,
  %``High-energy factorization and small x heavy flavor production,''
  Nucl.\ Phys.\ B {\bf 366} (1991) 135.
  %%CITATION = NUPHA,B366,135;%%

\bibitem{Deak:2009xt}
  M.~Deak, F.~Hautmann, H.~Jung and K.~Kutak,
  %``Forward Jet Production at the Large Hadron Collider,''
  JHEP {\bf 0909} (2009) 121.
%  [arXiv:0908.0538 [hep-ph]].
  %%CITATION = ARXIV:0908.0538;%%

\bibitem{Bomhof:2006dp}
  C.~J.~Bomhof, P.~J.~Mulders and F.~Pijlman,
  %``The Construction of gauge-links in arbitrary hard processes,''
  Eur.\ Phys.\ J.\ C {\bf 47} (2006) 147.
%  [hep-ph/0601171].
  %%CITATION = HEP-PH/0601171;%%
  
\bibitem{Dominguez:2011wm}
  F.~Dominguez, C.~Marquet, B.~-W.~Xiao and F.~Yuan,
  %``Universality of Unintegrated Gluon Distributions at small x,''
  Phys.\ Rev.\ D {\bf 83} (2011) 105005.
%  [arXiv:1101.0715 [hep-ph]].
  %%CITATION = ARXIV:1101.0715;%%  

\bibitem{Dominguez:2010xd}
  F.~Dominguez, B.~-W.~Xiao and F.~Yuan,
  %``$k_t$-factorization for Hard Processes in Nuclei,''
  Phys.\ Rev.\ Lett.\  {\bf 106} (2011) 022301.
%  [arXiv:1009.2141 [hep-ph]].
  %%CITATION = ARXIV:1009.2141;%%

\bibitem{Kutak:2012rf}
  K.~Kutak and S.~Sapeta,
  %``Gluon saturation in dijet production in p-Pb collisions at Large Hadron Collider,''
  Phys.\ Rev.\ D {\bf 86} (2012) 094043.
%  [arXiv:1205.5035 [hep-ph]].
  %%CITATION = ARXIV:1205.5035;%%

\bibitem{vanHameren:2014lna}
  A.~van Hameren, P.~Kotko, K.~Kutak, C.~Marquet and S.~Sapeta,
  %``Saturation effects in forward-forward dijet production in p+Pb
  %collisions,''
  Phys.\ Rev.\ D {\bf 89} (2014) 094014.
%  [arXiv:1402.5065 [hep-ph]].
  %%CITATION = ARXIV:1402.5065;%%
  
\bibitem{vanHameren:2014ala}
  A.~van Hameren, P.~Kotko, K.~Kutak and S.~Sapeta,
  %``Small-$x$ dynamics in forward-central dijet decorrelations at the LHC,''
  Phys.\ Lett.\ B {\bf 737} (2014) 335.
%  [arXiv:1404.6204 [hep-ph]].
  %%CITATION = ARXIV:1404.6204;%%

\bibitem{vanHameren:2013fla}
  A.~van Hameren, P.~Kotko and K.~Kutak,
  %``Three jet production and gluon saturation effects in p-p and p-Pb collisions within high-energy factorization,''
  Phys.\ Rev.\ D {\bf 88} (2013) 9,  094001
   [Erratum-ibid.\ D {\bf 90} (2014) 3,  039901].
%  [arXiv:1308.0452 [hep-ph]].
  %%CITATION = ARXIV:1308.0452;%%

\bibitem{Boer:1999si}
  D.~Boer and P.~J.~Mulders,
  %``Color gauge invariance in the Drell-Yan process,''
  Nucl.\ Phys.\ B {\bf 569} (2000) 505.
%  [hep-ph/9906223].
  %%CITATION = HEP-PH/9906223;%% 
  
\bibitem{Belitsky:2002sm}
  A.~V.~Belitsky, X.~Ji and F.~Yuan,
  %``Final state interactions and gauge invariant parton distributions,''
  Nucl.\ Phys.\ B {\bf 656} (2003) 165.
%  [hep-ph/0208038].
  %%CITATION = HEP-PH/0208038;%%
  
\bibitem{Boer:2003cm}
  D.~Boer, P.~J.~Mulders and F.~Pijlman,
  %``Universality of T odd effects in single spin and azimuthal asymmetries,''
  Nucl.\ Phys.\ B {\bf 667} (2003) 201.
%  [hep-ph/0303034].
  %%CITATION = HEP-PH/0303034;%%

\bibitem{Collins:2007nk} 
  J.~Collins and J.~W.~Qiu,
  %``$k_{T}$ factorization is violated in production of high-transverse-momentum particles in hadron-hadron collisions,''
  Phys.\ Rev.\ D {\bf 75} (2007) 114014.
%  [arXiv:0705.2141 [hep-ph]].
  %%CITATION = ARXIV:0705.2141;%%

\bibitem{Vogelsang:2007jk} 
  W.~Vogelsang and F.~Yuan,
  %``Hadronic Dijet Imbalance and Transverse-Momentum Dependent Parton Distributions,''
  Phys.\ Rev.\ D {\bf 76} (2007) 094013.
%  [arXiv:0708.4398 [hep-ph]].
  %%CITATION = ARXIV:0708.4398;%%

\bibitem{Rogers:2010dm} 
  T.~C.~Rogers and P.~J.~Mulders,
  %``No Generalized TMD-Factorization in Hadro-Production of High Transverse Momentum Hadrons,''
  Phys.\ Rev.\ D {\bf 81} (2010) 094006.
%  [arXiv:1001.2977 [hep-ph]].
  %%CITATION = ARXIV:1001.2977;%%

\bibitem{Xiao:2010sp} 
  B.~W.~Xiao and F.~Yuan,
  %``Non-Universality of Transverse Momentum Dependent Parton Distributions at Small-x,''
  Phys.\ Rev.\ Lett.\  {\bf 105} (2010) 062001.
%  [arXiv:1003.0482 [hep-ph]].
  %%CITATION = ARXIV:1003.0482;%%

\bibitem{Mangano:1990by} 
  M.~L.~Mangano and S.~J.~Parke,
  %``Multiparton amplitudes in gauge theories,''
  Phys.\ Rept.\  {\bf 200} (1991) 301.
%  [hep-th/0509223].
  %%CITATION = HEP-TH/0509223;%%

\bibitem{Mueller:2012uf}
  A.~H.~Mueller, B.~-W.~Xiao and F.~Yuan,
  %``Sudakov Resummation in Small-x Saturation Formalism,''
  Phys.\ Rev.\ Lett.\  {\bf 110} (2013) 082301.
%  [arXiv:1210.5792 [hep-ph]].
  %%CITATION = ARXIV:1210.5792;%%
  
\bibitem{Mueller:2013wwa}
  A.~H.~Mueller, B.~-W.~Xiao and F.~Yuan,
  %``Sudakov Double Logarithms Resummation in Hard Processes in Small-x Saturation Formalism,''
  Phys.\ Rev.\ D {\bf 88} (2013) 114010.
%  [arXiv:1308.2993 [hep-ph]].
  %%CITATION = ARXIV:1308.2993;%%

\bibitem{Ciafaloni:1987ur}
  M.~Ciafaloni,
  %``Coherence Effects in Initial Jets at Small q**2 / s,''
  Nucl.\ Phys.\ B {\bf 296} (1988) 49.
  %%CITATION = NUPHA,B296,49;%%
  
\bibitem{Catani:1989sg}
  S.~Catani, F.~Fiorani and G.~Marchesini,
  %``Small x Behavior of Initial State Radiation in Perturbative QCD,''
  Nucl.\ Phys.\ B {\bf 336} (1990) 18.
  %%CITATION = NUPHA,B336,18;%%
  
 \bibitem{Catani:1989yc}
  S.~Catani, F.~Fiorani and G.~Marchesini,
  %``QCD Coherence in Initial State Radiation,''
  Phys.\ Lett.\ B {\bf 234} (1990) 339.
  %%CITATION = PHLTA,B234,339;%%
  
\bibitem{Iancu:2013dta}
  E.~Iancu and J.~Laidet,
  %``Gluon splitting in a shockwave,''
  Nucl.\ Phys.\ A {\bf 916} (2013) 48.
%  [arXiv:1305.5926 [hep-ph]].
  %%CITATION = ARXIV:1305.5926;%%

\bibitem{Kutak:2014wga}
  K.~Kutak,
  %``Hard scale dependent gluon density, saturation and forward-forward dijet production at the LHC,''
  Phys.\ Rev.\ D {\bf 91} (2015) 3,  034021.
%  [arXiv:1409.3822 [hep-ph]].
  %%CITATION = ARXIV:1409.3822;%%

\bibitem{vanHameren:2012uj} 
  A.~van Hameren, P.~Kotko and K.~Kutak,
  %``Multi-gluon helicity amplitudes with one off-shell leg within high energy factorization,''
  JHEP {\bf 1212} (2012) 029.
%  [arXiv:1207.3332 [hep-ph]].
  %%CITATION = ARXIV:1207.3332;%%

\bibitem{vanHameren:2013} 
  A.~van Hameren, P.~Kotko and K.~Kutak,
  %``Helicity amplitudes for high-energy scattering,''
  JHEP {\bf 1301} (2013) 078.
%  [arXiv:1211.0961 [hep-ph]].
  %%CITATION = ARXIV:1211.0961;%%  

\bibitem{Lipatov:1976zz}
  L.~N.~Lipatov,
  %``Reggeization of the Vector Meson and the Vacuum Singularity in Nonabelian Gauge Theories,''
  Sov.\ J.\ Nucl.\ Phys.\  {\bf 23} (1976) 338.
%   [Yad.\ Fiz.\  {\bf 23} (1976) 642].
  %%CITATION = SJNCA,23,338;%%

\bibitem{Kuraev:1976ge}
  E.~A.~Kuraev, L.~N.~Lipatov and V.~S.~Fadin,
  %``Multi - Reggeon Processes in the Yang-Mills Theory,''
  Sov.\ Phys.\ JETP {\bf 44} (1976) 443
%   [Zh.\ Eksp.\ Teor.\ Fiz.\  {\bf 71} (1976) 840]
   [Erratum-ibid.\  {\bf 45} (1977) 199].
  %%CITATION = SPHJA,44,443;%%
  
\bibitem{Balitsky:1978ic}
  I.~I.~Balitsky and L.~N.~Lipatov,
  %``The Pomeranchuk Singularity in Quantum Chromodynamics,''
  Sov.\ J.\ Nucl.\ Phys.\  {\bf 28} (1978) 822.
%   [Yad.\ Fiz.\  {\bf 28} (1978) 1597].
  %%CITATION = SJNCA,28,822;%%

\bibitem{Kwiecinski:1997ee}
  J.~Kwiecinski, A.~D.~Martin and A.~M.~Stasto,
  %``A Unified BFKL and GLAP description of F2 data,''
  Phys.\ Rev.\ D {\bf 56} (1997) 3991.
%  [hep-ph/9703445].
  %%CITATION = HEP-PH/9703445;%%
  
\bibitem{Balitsky:1995ub}
  I.~Balitsky,
  %``Operator expansion for high-energy scattering,''
  Nucl.\ Phys.\ B {\bf 463} (1996) 99.
%  [hep-ph/9509348].
  %%CITATION = HEP-PH/9509348;%%
  %986 citations counted in INSPIRE as of 06 Feb 2014
  
\bibitem{Kovchegov:1999yj}
  Y.~V.~Kovchegov,
  %``Small x F(2) structure function of a nucleus including multiple pomeron exchanges,''
  Phys.\ Rev.\ D {\bf 60} (1999) 034008.
%  [hep-ph/9901281].
  %%CITATION = HEP-PH/9901281;%%

\bibitem{Kutak:2003bd}
  K.~Kutak and J.~Kwiecinski,
  %``Screening effects in the ultrahigh-energy neutrino interactions,''
  Eur.\ Phys.\ J.\ C {\bf 29} (2003) 521.
%  [hep-ph/0303209].
  %%CITATION = HEP-PH/0303209;%%

\bibitem{Kutak:2004ym}
  K.~Kutak and A.~M.~Stasto,
  %``Unintegrated gluon distribution from modified BK equation,''
  Eur.\ Phys.\ J.\ C {\bf 41} (2005) 343.
%  [hep-ph/0408117].
  %%CITATION = HEP-PH/0408117;%%

\bibitem{Akcakaya:2012si}
  E.~Akcakaya, A.~Schfer and J.~Zhou,
  %``Azimuthal asymmetries for quark pair production in pA collisions,''
  Phys.\ Rev.\ D {\bf 87} (2013) 5,  054010.
%  [arXiv:1208.4965 [hep-ph]].
  %%CITATION = ARXIV:1208.4965;%%

 \bibitem{Lipatov:1995pn}
  L.~N.~Lipatov,
  %``Gauge invariant effective action for high-energy processes in QCD,''
  Nucl.\ Phys.\ B {\bf 452} (1995) 369.
%  [hep-ph/9502308].
  %%CITATION = HEP-PH/9502308;%%
 
 \bibitem{Antonov:2004hh}
  E.~N.~Antonov, L.~N.~Lipatov, E.~A.~Kuraev and I.~O.~Cherednikov,
  %``Feynman rules for effective Regge action,''
  Nucl.\ Phys.\ B {\bf 721} (2005) 111.
%  [hep-ph/0411185].
  %%CITATION = HEP-PH/0411185;%%

\bibitem{Kotko:2014aba} 
  P.~Kotko,
  %``Wilson lines and gauge invariant off-shell amplitudes,''
  JHEP {\bf 1407} (2014) 128.
%  [arXiv:1403.4824 [hep-ph]].
  %%CITATION = ARXIV:1403.4824;%%

\bibitem{vanHameren:2014iua}
  A.~van Hameren,
  %``BCFW recursion for off-shell gluons,''
  JHEP {\bf 1407} (2014) 138.
%  [arXiv:1404.7818 [hep-ph]].
  %%CITATION = ARXIV:1404.7818;%% 

\bibitem{Dixon:2013uaa}
  L.~J.~Dixon,
  %``A brief introduction to modern amplitude methods,''
  arXiv:1310.5353 [hep-ph].
  %%CITATION = ARXIV:1310.5353;%%
  
\bibitem{DelDuca:1999rs}
  V.~Del Duca, L.~J.~Dixon and F.~Maltoni,
  %``New color decompositions for gauge amplitudes at tree and loop level,''
  Nucl.\ Phys.\ B {\bf 571} (2000) 51.
%  [hep-ph/9910563].
  %%CITATION = HEP-PH/9910563;%%

\bibitem{Collins:1984kg}
  J.~C.~Collins, D.~E.~Soper and G.~F.~Sterman,
  %``Transverse Momentum Distribution in Drell-Yan Pair and W and Z Boson Production,''
  Nucl.\ Phys.\ B {\bf 250} (1985) 199.
  %%CITATION = NUPHA,B250,199;%%

\bibitem{Kimber:1999xc}
  M.~A.~Kimber, A.~D.~Martin and M.~G.~Ryskin,
  %``Unintegrated parton distributions and prompt photon hadroproduction,''
  Eur.\ Phys.\ J.\ C {\bf 12} (2000) 655.
%  [hep-ph/9911379].
  %%CITATION = HEP-PH/9911379;%%

\bibitem{Kimber:2001sc}
  M.~A.~Kimber, A.~D.~Martin and M.~G.~Ryskin,
  %``Unintegrated parton distributions,''
  Phys.\ Rev.\ D {\bf 63} (2001) 114027.
%  [hep-ph/0101348].
  %%CITATION = HEP-PH/0101348;%%

\bibitem{Collins:2014jpa}
  J.~Collins and T.~Rogers,
  %``Understanding the large-distance behavior of transverse-momentum-dependent
  %parton densities and the Collins-Soper evolution kernel,''
  Phys.\ Rev.\ D {\bf 91} (2015) 7, 074020.
  %[arXiv:1412.3820 [hep-ph]].
  %%CITATION = ARXIV:1412.3820;%%

\bibitem{Kutak:2011fu}
  K.~Kutak, K.~Golec-Biernat, S.~Jadach and M.~Skrzypek,
  %``Nonlinear equation for coherent gluon emission,''
  JHEP {\bf 1202} (2012) 117.
%  [arXiv:1111.6928 [hep-ph]].
  %%CITATION = ARXIV:1111.6928;%%
  
\bibitem{Kutak:2012qk}
  K.~Kutak,
  %``Resummation in nonlinear equation for high energy factorisable gluon density and its extension to include coherence,''
  JHEP {\bf 1212} (2012) 033.
%  [arXiv:1206.5757 [hep-ph]].
  %%CITATION = ARXIV:1206.5757;%%
  
\bibitem{Ji:2004wu}
  X.~d.~Ji, J.~p.~Ma and F.~Yuan,
  %``QCD factorization for semi-inclusive deep-inelastic scattering at low transverse momentum,''
Phys.\ Rev.\ D {\bf 71} (2005) 034005.
%[hep-ph/0404183].
%%CITATION = HEP-PH/0404183;%%

\bibitem{Dominguez:2011gc}
  F.~Dominguez, A.~H.~Mueller, S.~Munier and B.~W.~Xiao,
  %``On the small-x evolution of the color quadrupole and the Weizs\'acker-Williams gluon distribution,''
Phys.\ Lett.\ B {\bf 705} (2011) 106.
%[arXiv:1108.1752 [hep-ph]].
%%CITATION = ARXIV:1108.1752;%%

\bibitem{Fadin:2006bj}
  V.~S.~Fadin, R.~Fiore, M.~G.~Kozlov and A.~V.~Reznichenko,
  %``Proof of the multi-Regge form of QCD amplitudes with gluon exchanges in the NLA,''
Phys.\ Lett.\ B {\bf 639} (2006) 74.
%[hep-ph/0602006].
%%CITATION = HEP-PH/0602006;%%

\end{thebibliography}
\end{document}